\documentstyle[12pt,twoside]{article}
\def\greaterthansquiggle{\raise.3ex\hbox{$>$\kern-.75em\lower1ex\hbox{$\sim$}}}
\def\lessthansquiggle{\raise.3ex\hbox{$<$\kern-.75em\lower1ex\hbox{$\sim$}}}
\newcommand{\beq}{\begin{equation}}
\newcommand{\eeq}{\end{equation}}
\newcommand{\beqa}{\begin{eqnarray}}
\newcommand{\eeqa}{\end{eqnarray}}
\newcommand{\beqan}{\begin{eqnarray*}}
\newcommand{\eeqan}{\end{eqnarray*}}
\newcommand{\ba}{\begin{array}}
\newcommand{\ea}{\end{array}}
\newcommand{\no}{\nonumber}
\newcommand{\ol}{\overline}
\newcommand{\ra}{\rightarrow}
\newcommand{\Ra}{\Rightarrow}
\newcommand{\ve}{\varepsilon}
\newcommand{\vp}{\varphi}
\newcommand{\vt}{\vartheta}
\newcommand{\dg}{\dagger}
\newcommand{\wt}{\widetilde}
\newcommand{\wh}{\widehat}
\newcommand{\D}{{\cal D}}
\newcommand{\Ha}{{\cal H}}
\newcommand{\cL}{{\cal L}}
\newcommand{\T}{{\cal T}}
\newcommand{\U}{{\cal U}}
\newcommand{\V}{{\cal V}}
\newcommand{\W}{{\cal W}}
\newcommand{\st}{\stackrel}
\newcommand{\dfrac}{\displaystyle \frac}
\normalsize
\begin{document}
\bibliographystyle{plain}
\begin{titlepage}
\begin{flushright}
UWThPh-1995-7\\
June 1995
\end{flushright}
\vspace{2cm}
\begin{center}
{\Large \bf
Automorphisms in Gauge Theories and the Definition of
CP and P}\\[50pt]
W. Grimus and M.N. Rebelo* \\
Institut f\"ur Theoretische Physik \\
Universit\"at Wien \\
Boltzmanngasse 5, A--1090 Wien, Austria
\vfill
{\bf Abstract} \\
\end{center}

We study the possibilities to define CP and parity
in general gauge theories by utilizing the intimate connection of these
discrete symmetries with the group of automorphisms of the gauge Lie
algebra. Special emphasis is put on the scalar gauge interactions and the
CP invariance of the Yukawa couplings.
\vfill
\noindent *) Presently supported by Fonds zur F\"orderung der
wissenschaftlichen Forschung, Project No. P08955--PHY. Permanent address:
Dept. de F\'{\i}sica and CFIF-IST, Instituto Superior T\'ecnico, Lisbon,
Portugal (on leave until September 1995).
\end{titlepage}
\tableofcontents

\newpage
\section{Introduction}

\renewcommand{\theequation}{\arabic{section}.\arabic{equation}}
\setcounter{equation}{0}

The discrete transformations charge conjugation (C), parity (P) and CP
played an important r\^ole in the development of particle physics. In
particular, the hypothesis of P (C) non--conservation \cite{lee} and
its subsequent experimental confirmation \cite{amb} and the discovery
of CP violation \cite{chris} constituted both substantial progress in
our understanding of weak interactions and gave incentives to important
theoretical developments culminating in the advent of the Standard
Model (SM) \cite{SM}. It is obvious, nowadays, that the suitable framework
for discussing CP and P is given by spontaneously broken gauge theories
with chiral fields as the basic fermionic degrees of freedom. In the
SM, CP and P non--invariance appear in two totally different ways: parity is
broken because left and right--handed fermionic multiplets
belong to different representations of the gauge group
whereas complex Yukawa couplings entail CP
non--conservation through the Kobayashi--Maskawa mechanism \cite{KM} for
three quark families. Thus the spontaneous breakdown of the gauge group
$SU(2) \times U(1)$ only affects the manifestation of CP and P violation
but has nothing to do with the non--invariance itself. On the other hand,
on a more fundamental level, in an appropriate extension of the SM or in
a Grand Unified Theory (GUT) \cite{GUT} one might prefer to break CP
or both CP and P spontaneously to provide the same footing for the
breaking of continuous and discrete symmetries. This offers the
possibility to relate the breaking of CP (and P) to the breaking of the
gauge group and thus to have a more intimate relationship between CP
(and P) non--conservation and the gauge structure in addition to the
aesthetic point of view addressed to before.

Adopting the premise of spontaneous breakdown of CP and P we are immediately
led to the question when and how CP and P transformations can be defined
in a gauge theory before spontaneous symmetry breaking. This problem is
particularly acute for P when left and right--handed irreducible fermionic
representations do not match. It is the purpose of this paper to provide
a thorough discussion of these questions, i.e. to consider the symmetry
aspects leading to CP, P and C transformations in general gauge theories.
Thereby we are motivated by the striking structural similarity between
CP and P when formulated in a framework where all fermion fields have the
same chirality. Such a formulation is always possible since it is
equivalent, e.g., to the use of the right--handed field
\beq
(\chi_L)^c \equiv C \gamma_0^T \chi_L^*
\eeq
instead of $\chi_L$ (for conventions and the definition of the matrix $C$,
see app. A). In this work we will use right--handed fields for
definiteness. It is well known that the key to an understanding of the
relation between CP and P is given by the automorphisms of the gauge
group $G$ \cite{sla} and that a gauge theory containing only the gauge
bosons and fermions is always CP invariant \cite{sla,smo}. These points
will be worked out and commented upon in detail in the first chapters
of this paper. Furthermore, gauge interactions of scalar fields, the second
focus of this work,
are dealt with in the same way and this part of the interaction is CP
invariant as well. However, complications in the definition of CP and P
transformations arise for Higgs fields in real and pseudoreal irreducible
representations (irreps) of $G$. Finally, the third focus is on
Yukawa interactions $\cL_Y$. In this connection not only the group
structure of the couplings represented by Clebsch--Gordan coefficients
is important but also the couplings pertaining to each group singlet
contained in $\cL_Y$. In the case of replication of irreps of $G$ like the
well--known families of quarks and leptons there is freedom of
defining CP and P with respect to unitary rotations among the ``families''
(see refs. \cite{bin,eck81} for the context of left--right symmetric models).
Choosing one of these ``horizontal rotations'', e.g. in the definition of
CP, and requiring CP invariance of $\cL_Y$ we impose conditions on the
Yukawa coupling constants. In the simplest case
(when the horizontal rotations are
proportional to unit matrices) and with appropriate phase
conventions they have to be real . Thus the Yukawa interactions are not
automatically CP invariant in contrast to the gauge interactions. In
the discussions of the three main subjects of this work special attention
will be paid to the existence of certain bases with respect to the
group structure and to the horizontal structure to provide ``canonical''
forms of CP where its properties are especially transparent.

\paragraph{General remarks on the scope of the paper:} To clarify the range
of validity of this paper
some remarks are in order. We will always deal with compact
Lie groups $G$ as gauge groups which entails unitary representations
(reps) and the real Lie algebra $\cL_c$ of $G$ is compact (see app.~B).
On the other hand, requiring unitary reps on quantum theoretical grounds,
it can be shown that one can confine oneself to compact groups
\cite{comp}. We will not only cover simple groups but also groups of the type
$G' \times G''$ and $G' \times U(1)$ with $G'$ and $G''$ simple.
In physical terms
these cases require two independent gauge coupling constants. More
complicated gauge structures like the SM gauge group can easily be
discussed by an extension of these considerations. Special emphasis
will be put on $G = G' \times G'$ ($G'$ simple) extended by a discrete
element such that only one gauge coupling constant is present. A typical
case would be left--right symmetric models \cite{pati,LR}.

In general there will also be symmetries of the Lagrangian which are not
gauged, whether discrete or continuous. Therefore the total symmetry
group will have the structure $G \times H$ where $H$ is the group of
these additional symmetries. Multiplets identical with respect to the
gauge group may be distinguished by different transformation properties
under $H$. In certain cases this can forbid operations to be discussed
in what follows which involve non--trivial multiplicities of irreps of
$G$. For instance, if in a model there is a complex scalar multiplet
transforming according to a real irrep of $G$ but to a complex irrep
of $H$ then this multiplet cannot be split into two real multiplets.
We leave it with this caveat and concentrate on the gauge group from
now on.

Although we will be mainly concerned with reps of $\cL_c$ it will
be tacitly assumed that these reps can be extended to reps of $G$.
Furthermore, in our discussion of the various terms of the Lagrangian we will
not cover the scalar potential because we think that
in concrete cases it can be treated
with the methods of this paper though its general discussion is involved.
Finally, since for gauge
theories the CPT theorem holds \cite{CPT} (see also ref. \cite{bjo}) and
since CP is discussed extensively time reversal is only mentioned
shortly in this work.

\paragraph{Plan of the paper:} In sect. 2 the main features of CP and P
are worked out in detail by means of three examples: QED, QCD and the
$SO(10)$--GUT. CP--type transformations comprising CP and P are
introduced in sect. 3 where Conditions A and B for invariance of the
Lagrangian containing gauge fields and fermions are derived. Since
Condition A requires that a CP--type transformation on the gauge fields
corresponds to an automorphism of $\cL_c$ sect. 4 is devoted to a
detailed discussion of Aut~$(\cL_c)$. Sects. 5 and 6 treat Condition
B. In sect. 5 CP transformations are introduced as special solutions
of Condition B and the existence of the ``CP basis'' is discussed where all the
generators of $\cL_c$ are either symmetric or antisymmetric matrices.
In sect. 6 the general solution of Condition B is expounded and parity
transformations are defined by characteristic properties of the
associated automorphisms. Several notions of CP--type transformations
are introduced in sects. 5 and 6 to clarify the subject.
They are related to each other in the following way:
$$
\mbox{CP--type transformation} \left\{ \ba{l}
\mbox{ generalized CP } \{ \mbox{ canonical CP} \\[12pt]
\mbox{ parity} \left\{
\ba{l} \mbox{ internal parity} \\ \mbox{ external parity} \ea\right.
\ea \right.
$$
The brace $\{$ signifies that the more general notion is found to the left
of it. In the examples of sect. 2 on the one hand, the structural
similarity between CP and P will become apparent, on the other hand,
their distinctive features will be elucidated as well. QED and QCD
provide examples of external parity whereas the $SO(10)$--GUT, where all
fermionic degrees of freedom are in a 16--dimensional irrep, allows to
define an internal parity. In sect. 7 the whole discussion is carried
over to scalars with emphasis on the special cases of real and
pseudoreal irreps. Sect. 8 is devoted to the invariance of the Yukawa
couplings under CP--type transformations. Furthermore, the general
solutions of the condition imposed on these couplings by a generalized
CP transformation are derived. Transformations of the charge conjugation
type are considered in sect. 9. Finally, in sect. 10 a summary is
presented.

In app. A our conventions concerning the $\gamma$ matrices and the
charge conjugation matrix $C$ can be found. All properties of Lie
algebras necessary for our purposes are collected in app. B. In app. C
spinor reps of $so(N)$ are defined via the Clifford algebra and
app. D describes the Lie algebra isomorphisms $so(4) \cong su(2) \oplus
su(2)$ and $so(6) \cong su(4)$. The remaining appendices E--I give
proofs which are not carried out in the main body of the paper.

\paragraph{Notational remarks:} For convenience we will often leave
out the arguments $x$ and $\wh x = (x^0,-\vec x)$ in CP--type
transformations. Equivalence of reps will be denoted by $\sim$. The
letter $D$ will be used for reps of $\cL_c$ and also of $G$. Its use
should be clear from the context. Yet $- D^T$ refers only to the
complex conjugate of the Lie algebra rep~$D$. Finally
let us collect the abbreviations used in this work: CSA, GUT, ON, SM,
rep and irrep denote Cartan subalgebra, Grand Unified Theory,
orthonormal, Standard Model, representation and irreducible
representation, respectively.

\section{Examples}
\renewcommand{\theequation}{\arabic{section}.\arabic{equation}}
\setcounter{equation}{0}

Before coming to the general discussion it is quite instructive to
consider some examples which enable us to get
a feeling for the main features. In this
light we will discuss now QED, QCD and the GUT based on the spinor
rep of $SO(10)$ (strictly speaking, an irrep of $Spin(10)$, its universal
covering group, see e.g. ref. \cite{corn}).

\paragraph{QED:} Denoting the electron field by $e(x)$ we know that
the Lagrangian is invariant under
\beqa
CP: \quad e(x) &\ra& - C e(\wh x)^* \no \\
P: \quad e(x) &\ra& \gamma^0 e(\wh x)
\eeqa
where $\wh x = (x^0, - \vec x)$. In terms of chiral fields eq. (2.1)
reads
\beqa
CP: \quad e_{L,R}(x) &\ra& - C e_{L,R}(\wh x)^* \no \\
P: \quad e_{L,R} (x) &\ra& \gamma^0 e_{R,L}(\wh x)
\eeqa
expressing the fact that, having started with Dirac fields, CP does not
mix chiralities whereas parity connects fields with different chirality.
Defining
\beq
\chi_{R1} \equiv e_R, \qquad \chi_{R2} \equiv C \gamma_0^T e_L^*
\eeq
we can get yet another view of eq. (2.1) \cite{gri93}, namely
\beqa
CP: \left( \ba{c} \chi_{R1} \\ \chi_{R2} \ea \right) &\ra&
\left( \ba{rc} -1 & 0 \\ 0 & 1 \ea \right) C \left( \ba{c}
\chi_{R1} \\ \chi_{R2} \ea \right)^* \no \\
P: \left( \ba{c} \chi_{R1} \\ \chi_{R2} \ea \right) &\ra&
\left( \ba{cr} 0 & -1 \\ 1 & 0 \ea \right) C \left( \ba{c}
\chi_{R1} \\ \chi_{R2} \ea \right)^* .
\eeqa
With
\beq
U_{CP} \equiv \left( \ba{cr} 1 & 0 \\ 0 & -1 \ea \right), \qquad
U_P \equiv \left( \ba{rc} 0 & 1 \\ -1 & 0 \ea \right)
\eeq
we observe that the form of CP and P is only distinguished by the matrix
$U$ in the CP--type transformation
\beq
\left( \ba{c} \chi_{R1} \\ \chi_{R2} \ea \right) \ra - U C
\left( \ba{c} \chi_{R1} \\ \chi_{R2} \ea \right)^* .
\eeq
How are $U_{CP}$ and $U_P$ characterized with respect to the gauge group
$U(1)_{em}$? The chiral fields
$\chi_{R1,2}$ transform with complex conjugate phases, i.e.
\beq
\chi_{R1} \ra e^{i\alpha} \chi_{R1}, \qquad
\chi_{R2} \ra e^{-i\alpha} \chi_{R2}, \qquad
e = \chi_{R1} + (\chi_{R2})^c \ra e^{i\alpha} e
\eeq
and the chiral field vector consists of two irreps which are exchanged
under $U_P$ whereas $U_{CP}$ acts within the respective irreps. Thus, in the
formulation where the previously mentioned structural similarity
becomes obvious, CP and P can be distinguished by group theoretical
properties of the matrix $U$. Eq. (2.4) has to be supplemented by the
transformation properties of the photon field
\beqa
CP : \quad A_\mu(x) &\ra& - \ve(\mu) A_\mu (\wh x) \no \\
P : \quad A_\mu(x) &\ra& \ve(\mu) A_\mu (\wh x)
\eeqa
where the $\ve (\mu)$ are the signs $+1$ for $\mu = 0$ and $-1$
for $\mu = 1, 2, 3$.

\paragraph{QCD:} We confine ourselves to a single quark flavour $q$.
It is well known that QCD is invariant under the transformations
analogous to eq. (2.2) and with the same reasoning as before we get
for the ``CP--type'' transformation
\beq
\chi_R \ra - U C \chi_R^* \qquad \mbox{with} \qquad
\chi_R = \left( \ba{c} \chi_{R1} \\ \chi_{R2} \ea \right) =
\left( \ba{cc} q_R \\ (q_L)^c \ea \right)
\eeq
and
\beq
U_{CP} = \left( \ba{cr} {\bf 1} & 0 \\ 0 & -{\bf 1} \ea \right), \qquad
U_P = \left( \ba{rc} 0 & {\bf 1} \\ -{\bf 1} & 0 \ea \right).
\eeq
Eq. (2.7) now reads
\beq
\chi_{R1} \ra D \chi_{R1}, \qquad
\chi_{R2} \ra D^* \chi_{R2}, \qquad
q = \chi_{R1} + (\chi_{R2})^c \ra Dq
\eeq
with $D \in SU(3)_c$ and eq. (2.8) is adapted to QCD by
\beqa
CP : W_\mu^a(x) &\ra& \eta_a \ve(\mu) W_\mu^a(\wh x) \no \\
P : W_\mu^a(x) &\ra& \ve(\mu) W_\mu^a(\wh x)
\eeqa
with $a = 1,\ldots,8$. The signs $\eta_a$ depend on the transposition
properties of the generators $\lambda_a/2$ where $\lambda_a$ are the
Gell--Mann matrices \cite{gell} and are obtained by
\beq
- \lambda_a^T = \eta_a \lambda_a .
\eeq
As is well--known this choice of signs in eq.(2.12) also leaves invariant the
pure gauge part of QCD. In sect. 3 (see also app. B) this fact will be
connected with the mapping $\lambda_a \ra \eta_a \lambda_a$ being an
automorphism of the Lie algebra $su(3)$.

\paragraph{$\bf SO(10)$ -- GUT:} A more involved example of CP and P is
provided by a Grand Unified Theory based on $SO(10)$
\cite{SO(10),geo,ross,moh} or actually on its universal covering group
$Spin(10)$. Here, all 15 fundamental fermions of one SM family together with
an additional right--handed neutrino carrying lepton number $+1$ fit
nicely into a 16--dimensional spinor irrep. In order to implement CP
and P a more elaborate analysis of the spinor irrep has to be
performed. For our purpose, it is appropriate to take advantage of
\beq
so(10) \supset so(6) \oplus so(4) \cong su(4) \oplus su(2) \oplus
su(2)
\eeq
since the classification of the states according to $SU(4)_c \times
SU(2)_L \times SU(2)_R$ is well known. Here, $SU(4)_c$ denotes the
colour or Pati--Salam--$SU(4)$ \cite{pati} where lepton number is
treated as ``fourth colour''. Furthermore, eq. (2.14) also represents a
step towards a realistic pattern of spontaneous symmetry breaking. All
necessary ingredients we are using in the following are explained in
apps. C and D. Following refs. \cite{geo,ross}, we denote the basis vectors
of the 32--dimensional space ${\bf C}^2 \otimes {\bf C}^2 \otimes
{\bf C}^2 \otimes {\bf C}^2 \otimes {\bf C}^2$ by
$$
|s_1 s_2 s_3 s_4 s_5 \rangle,
$$
where $s_j = \pm 1$ or, abbreviated, $\pm$, and
$$
|s_1 s_2 s_3 s_4 s_5\rangle = {\bf e}_{s_1} \otimes {\bf e}_{s_2}
\otimes {\bf e}_{s_3} \otimes {\bf e}_{s_4} \otimes {\bf e}_{s_5}
$$
with
\beq
{\bf e}_+ = \left( \ba{c} 1 \\ 0 \ea \right), \qquad
{\bf e}_- = \left( \ba{c} 0 \\ 1 \ea \right) .
\eeq
The 16--dimensional subspaces for the spinor irreps $\{16\}$ and
$\{ \ol{16}\}$ are given by the two projectors $P_\pm$ on the
32--dimensional space where
\beq
P_\pm = \frac{{\bf 1} \pm \Gamma_{11}}{2} , \quad
\Gamma_{11} = \sigma_3 \otimes \sigma_3 \otimes \sigma_3 \otimes
\sigma_3 \otimes \sigma_3 .
\eeq
This means we are taking as basis vectors those with
\beq
\prod_{j=1}^5 s_j = +1 \mbox{ or } -1
\eeq
for the $\{ 16\}$ or the $\{ \ol{16}\}$, respectively.

The subalgebra $so(4)$ of $so(10)$ is assumed to be generated by
$M_{ij}$, $1 \leq i < j \leq 4$, and $so(6)$ by $M_{ij}$,
$5 \leq i < j \leq 10$. The CSA \{$F_3^c,
Y_c, B-L$\} of $su(4)_c$ is easily carried over to the spinor
representation of $so(10)$ by using the Clifford algebra (see app. C)
and the mapping of the Gell--Mann basis into $so(6)$ as given
in app. D.\footnote{The $M_{ij}$ in eq. (D.4) have to be replaced
by $M_{4+i \, 4+j}$ to comply with the embedding of $so(6)$ into
$so(10)$ chosen above.} Thus we obtain
\beqa
\lefteqn{F_3^c = \frac{1}{2} \lambda_3 =
\frac{1}{2} \mbox{diag }(1,-1,0,0)} \no\\
& & \ra (F_3^c)_s = \frac{1}{4}(- {\bf 1} \otimes {\bf 1} \otimes \sigma_3 +
{\bf 1} \otimes \sigma_3 \otimes {\bf 1}) \otimes {\bf 1}^{(2)} \no\\
\lefteqn{Y_c = \frac{1}{\sqrt{3}} \lambda_8 = \frac{1}{3} \mbox{diag }
(1,1,-2,0)} \no\\
& &  \ra (Y_c)_s = \frac{1}{6} (- {\bf 1} \otimes {\bf 1} \otimes \sigma_3
- {\bf 1} \otimes \sigma_3 \otimes {\bf 1} +
2\sigma_3 \otimes {\bf 1} \otimes {\bf 1}) \otimes {\bf 1}^{(2)}\\
\lefteqn{B-L = \sqrt{2/3} \lambda_{15} =
\frac{1}{3} \mbox{diag }(1,1,1,-3)} \no\\
& & \ra (B-L)_s = \frac{1}{3}({\bf 1} \otimes {\bf 1} \otimes \sigma_3 +
{\bf 1} \otimes \sigma_3 \otimes {\bf 1} +
\sigma_3 \otimes {\bf 1} \otimes {\bf 1}) \otimes {\bf 1}^{(2)} . \no
\eeqa
Here, $B$ and $L$ are baryon and lepton number, respectively,
and \{$F_3^c, Y_c$\} generate
the CSA of $su(3)_c$. The symbol ${\bf 1}^{(p)}$ denotes
the p-fold tensor product of the $2 \times 2$ unit matrix
${\bf 1}$ and the subscript $s$ refers to the spinor rep.

By a completely analogous procedure the two $su(2)$ subalgebras of
$so(4)$ can be obtained:
\beqa
a_1 &=& \frac{i}{4} {\bf 1}^{(3)} \otimes (\sigma_2 \otimes \sigma_2 +
\sigma_1 \otimes \sigma_1) \no \\
a_2 &=& \frac{i}{4} {\bf 1}^{(3)} \otimes (\sigma_2 \otimes \sigma_1 -
\sigma_1 \otimes \sigma_2) \no \\
a_3 &=& \frac{i}{4} {\bf 1}^{(3)} \otimes ({\bf 1} \otimes \sigma_3 -
\sigma_3 \otimes {\bf 1}) \no \\
b_1 &=& -\frac{i}{4} {\bf 1}^{(3)} \otimes (\sigma_2 \otimes \sigma_2 -
\sigma_1 \otimes \sigma_1) \no \\
b_2 &=& \frac{i}{4} {\bf 1}^{(3)} \otimes (\sigma_2 \otimes \sigma_1 +
\sigma_1 \otimes \sigma_2) \no \\
b_3 &=& -\frac{i}{4} {\bf 1}^{(3)} \otimes ({\bf 1} \otimes \sigma_3 +
\sigma_3 \otimes {\bf 1}) .
\eeqa
Comparing eq. (2.19) with eq. (D.1), we see the correspondences
$A_j \leftrightarrow a_j$ and $-B_j^T \leftrightarrow b_j$.
The advantage of the choice $\{-B^T_j \}$ compared with the equivalent
$\{ B_j \}$ will become clear at the end of this section. The
algebra $\{ a_j \}$ generates $SU(2)_L$ whereas $\{ b_j \}$ generates
$SU(2)_R$.

The remaining two elements of the Cartan subalgebra in the spinor rep
of $so(10)$ are given by
\beqa
(I_{3L})_s = ia_3 &=& \frac{1}{4} {\bf 1}^{(3)} \otimes (\sigma_3 \otimes
{\bf 1} - {\bf 1} \otimes \sigma_3) \no \\
(I_{3R})_s = ib_3 &=& \frac{1}{4} {\bf 1}^{(3)} \otimes (\sigma_3 \otimes
{\bf 1} + {\bf 1} \otimes \sigma_3) .
\eeqa
This completes the definition of the CSA of $so(10)$ in the 32--dimensional
spinor rep. We can now easily construct $SU(2)_L \times
SU(2)_R$ multiplets by means of the lowering operators
\beqa
a_- &=& i(a_1 - i a_2) = - {\bf 1}^{(3)} \otimes \sigma_- \otimes \sigma_+
\no \\
b_- &=& i(b_1 - i b_2) = - {\bf 1}^{(3)} \otimes \sigma_- \otimes \sigma_-
\eeqa
with $\sigma_\pm = \frac{1}{2} (\sigma_1 \pm i \sigma_2)$
and find the basis vectors, omitting the $so(6)$ part,
\beqa
(2,1) : && |+-\rangle, - |-+\rangle \no \\
(1,2): &&  |++\rangle, - |--\rangle.
\eeqa

Calling the quarks with $(F_3^c,Y_c)$ quantum numbers $(1/2,1/3)$,
$(-1/2,1/3)$, $(0,-2/3)$ red, yellow and blue, respectively, we can
start with a right--handed red up--quark state $u_r = |++-++\rangle$.
Then, applying $b_-$ we obtain $d_r = - |++---\rangle$. The $su(4)$
multiplet is completed by the raising and lowering operators
\beqa
\frac{1}{2}(\lambda_1 \pm i \lambda_2) &\ra&
- {\bf 1} \otimes \sigma_\pm \otimes \sigma_\mp \otimes {\bf 1}^{(2)} \no \\
\frac{1}{2}(\lambda_4 \pm i \lambda_5) &\ra&
\sigma_\pm \otimes \sigma_3 \otimes \sigma_\mp \otimes {\bf 1}^{(2)} \no \\
\frac{1}{2}(\lambda_6 \pm i \lambda_7) &\ra&
- \sigma_\pm \otimes \sigma_\mp \otimes {\bf 1} \otimes {\bf 1}^{(2)} \no \\
\frac{1}{2}(\lambda_9 \pm i \lambda_{10}) &\ra&
 \sigma_\pm \otimes \sigma_\pm \otimes {\bf 1} \otimes {\bf 1}^{(2)} \no \\
\frac{1}{2}(\lambda_{11} \pm i \lambda_{12}) &\ra&
\sigma_\pm \otimes \sigma_3 \otimes \sigma_\pm \otimes {\bf 1}^{(2)} \no \\
\frac{1}{2}(\lambda_{13} \pm i \lambda_{14}) &\ra&
 {\bf 1} \otimes \sigma_\pm \otimes \sigma_\pm \otimes {\bf 1}^{(2)} .
\eeqa
Thus, having embedded $SU(4)_c \times SU(2)_L \times SU(2)_R$
in $Spin(10)$ we find for its
$(4,1,2)$ multiplet of right--handed quarks
\beqa
\lefteqn{\left( \ba{cccc} u_r & u_y & u_b & N \\ d_r & d_y & d_b & e \ea
\right) = } \no \\
&& \left( \ba{rrrr} |++-++\rangle & -|+-+++\rangle & |-++++\rangle &
|---++\rangle \\
- |++---\rangle & |+-+--\rangle & -|-++--\rangle & -|-----\rangle
\ea \right).
\eeqa
The electric charges of the states can easily be checked by means of the
charge operator
\beqa
Q_{em} &=& I_{3L} + I_{3R} + \frac{1}{2}(B-L) \ra \no \\
(Q_{em})_s & = &
\frac{1}{2} {\bf 1}^{(3)} \otimes \sigma_3
\otimes {\bf 1} + \frac{1}{6} ({\bf 1} \otimes {\bf 1} \otimes \sigma_3
+ {\bf 1} \otimes \sigma_3 \otimes {\bf 1} + \sigma_3 \otimes {\bf 1}
\otimes {\bf 1}) \otimes {\bf 1}^{(2)}. \no \\
\eeqa
{}From this and using $F_3^c$, $Y_c$ (see eq. (2.18)) we find that $N$ has
the quantum numbers of a right--handed SM singlet neutrino,
whereas $e$ is the right--handed electron state. The state $|--+-+\rangle$
has the quantum numbers of an anti--red quark belonging to the (2,1)
multiplet of $SU(2)_L \times SU(2)_R$ with eigenvalue $-1/2$ of $(I_{3L})_s$.
This leads to the $(\bar 4,2,1)$ multiplets
\beqa
\lefteqn{\left( \ba{cccc}
u_r^c & u_y^c & u_b^c & \nu^c \\ d_r^c & d_y^c & d_b^ c & e^c \ea
\right) = } \no \\
&& \left( \ba{rrrr}
-|--+-+\rangle & -|-+--+\rangle & -|+---+\rangle & |+++-+\rangle\\
- |--++-\rangle & -|-+-+-\rangle & -|+--+-\rangle & |++++-\rangle
\ea \right).
\eeqa
The choice of signs in eq. (2.26) relative to eq. (2.24)
will become clear from the discussion of parity. We
see that with our states (2.24) and (2.26) we have projected out the
space corresponding to $\prod_j s_j = -1$. All the quantum numbers
correspond to the choice of right--handed fields in the $\{\ol{16}\}$
irrep.
The construction of the gauge theory based on $so(10)$ and the spinor
representation $\{\ol{16}\}$ for the multiplet $\chi_R$, the physical
content of which has been described above, is now straightforward
using the hermitian generators $T_{ij} = i \sigma_{ij}/2$,
$(1 \leq i < j \leq 10)$ with $\sigma_{ij} = (M_{ij})_s$
(see eqs. (C.3) and (C.4)).

Anticipating sect. 3 and trying
\beq
CP : \chi_R(x) \ra - C \chi_R (\wh x)^*
\eeq
we merely have to check that invariance of the Lagrangian is achieved
by an appropriate transformation of the gauge fields. In our case it
is convenient to number the 45 gauge fields by $W_\mu^{ij}(x)$,
corresponding to the generators $T_{ij}$. Analogously to
the case of QCD it is easy to see that it is sufficient for CP invariance
to transform the gauge fields as
\beq
CP : W_\mu^{ij}(x) \ra \ve(\mu) \eta_{ij} W_\mu^{ij}(\wh x)
\eeq
where the signs $\eta_{ij}$ are obtained through
\beq
- \sigma_{ij}^T = \eta_{ij} \sigma_{ij} .
\eeq
It is easy to calculate the $\eta_{ij}$ recalling the definition of
$\sigma_{ij}$ in terms of the elements of the Clifford algebra (see
app. C). We have
\beq
\Gamma_i^T = \xi_i \Gamma_i, \qquad \xi_i = (-1)^{i+1},
\eeq
therefore
\beq
- \sigma_{ij}^T = - \frac{1}{2} [\Gamma_i,\Gamma_j]^T =
\xi_i \xi_j \sigma_{ij}
\eeq
and
\beq
\eta_{ij} = (-1)^{i+j} = \xi_i \xi_j .
\eeq
In sect. 5 it will be shown that one can always choose a basis in
representation space such that $T_a^T = \pm T_a$ is valid for the
generators. Eq. (2.29) tells us that
$$
M_{ij} \ra \eta_{ij} M_{ij}
$$
is an automorphism of $so(10)$ (see app. B and sect. 3). We will see
later that this guarantees a consistent choice of signs in eq. (2.28) such
that the pure gauge part without fermions is also invariant under our
CP transformation.

Turning to parity one is led to presume that this gauge theory is
also parity invariant because the states (2.26) emerge from eq. (2.24) by
a kind of ``antiparticle formation''. Whereas in QCD the states of
$\{ 3\}$ and $\{\bar 3\}$ correspond to each other, we now have such a
correspondence within one $so(10)$ multiplet. In sect. 6 we will call the first
kind of parity ``external'' and the second one ``internal''. One can
indeed formulate a parity transformation by
\beq
P : \chi_R(x) \ra - U_P C \chi_R (\wh x)^*
\eeq
with
\beq
U_P = \sigma_1 \otimes \sigma_2 \otimes \sigma_1 \otimes \sigma_2
\otimes {\bf 1}
\eeq
and
\beq
P : W_\mu^{ij}(x) \ra \ve(\mu) \rho_{ij} W_\mu^{ij}(\wh x)
\eeq
where the signs $\rho_{ij}$ are now obtained by (to be derived in sect. 3)
\beq
U_P(- \sigma_{ij}^T) U_P^{-1} = \rho_{ij} \sigma_{ij}
\eeq
or else, using eqs. (2.31) and (2.32), by
\beq
U_P \sigma_{ij} U_P^{-1} = \eta_{ij} \rho_{ij} \sigma_{ij} .
\eeq
Again, one can check that the pure gauge terms of the Lagrangian
are invariant under the transformation (2.35).
Note that the element $\sigma_{12}$ of the CSA commutes with $U_P$,
whereas $\sigma_{34}$, $\sigma_{56}$, $\sigma_{78}$, $\sigma_{9\,10}$
anticommute. In other words,
$U_P$ transforms eigenvectors of $(I_{3L} - I_{3R})_s$ into states with the
same eigenvalue, whereas the eigenvalues of
$(I_{3L} + I_{3R})_s$, $(B-L)_s$, $(F_3^c)_s$ and $(Y_c)_s$ are reversed.

Our sign conventions are such that $U_P$ verifies
\beq
U_P \left( \ba{cccc} u_r & u_y & u_b & N \\ d_r & d_y & d_b & e \ea
\right) = \left( \ba{cccc}
u_r^c & u_y^c & u_b^c & \nu^c \\ d_r^c & d_y^c & d_b^ c & e^c \ea
\right)
\eeq
where this notation means that $U_P$ is applied to all the states in
the parentheses. Taking into account eq. (2.19) and $\rho_{ij} = 1$ for
$5 \leq i < j \leq 10$ we obtain
\beqa
U_P \sigma_{ij} U_P^{-1} &=& - \sigma_{ij}^T \qquad \mbox{for }
5 \leq i < j \leq 10, \no \\
U_P a_j U_P^{-1} &=& - b_j^T, \\
U_P b_j U_P^{-1} &=& - a_j^T .\no
\eeqa
{}From this it follows that the fields $\chi_{R1}$ and $\chi_{R2}$
corresponding to (4,1,2) and
$(\ol{4},2,1)$, respectively, in the irrep $\{\ol{16}\}$ transform with
$SU(4)$ matrices which are exactly complex conjugate to each other.
The same is true for the ``diagonal $SU(2)$'' generated by
$\{ a_j + b_j\}$. Thus $\chi_{R1}$ and $(\chi_{R2})^c$ transform alike
under $SU(4)_c \times SU(2)_{\rm diag}$ and correspond to $e_R$ and
$e_L$ in QED, or $q_R$ and $q_L$ in QCD. Thus our conventions eqs. (2.19)
and (2.38) were suggested to exhibit the common features of the three
examples.

\section{Conditions for CP--type transformations: gauge bosons and
fermions}
\subsection{The pure gauge sector}

\renewcommand{\theequation}{\arabic{section}.\arabic{equation}}
\setcounter{equation}{0}
To have a starting point for the discussion of CP--type transformations
it is appropriate to repeat shortly the construction of gauge theories.

Let $\{ T_a \}$ be the hermitian generators in the fermionic representation
such that
\beq
[T_a,T_b] = i f_{abc} T_c
\eeq
and
\beq
\mbox{Tr }(T_a T_b) = k \delta_{ab}
\eeq
with totally antisymmetric coefficients $f_{abc}$ (see app. B, first
subsection). Defining
\beq
W_\mu \equiv T_a W_\mu^a
\eeq
the pure gauge Lagrangian is given by (see, e.g., ref. \cite{abers})
\beq
\cL_G = - \frac{1}{4k} \mbox{ Tr }(G_{\mu\nu} G^{\mu\nu})
\eeq
with
\beq
G_{\mu\nu} = \partial_\mu W_\nu - \partial_\nu W_\mu + ig[W_\mu,W_\nu]
\equiv T_a G^a_{\mu\nu}.
\eeq
Under a gauge transformation, the fermionic multiplet $\omega_R$ and
the fields $W_\mu$ transform as
\beqa
\omega_R(x) &\ra& U(x) \omega_R(x) , \\
W_\mu(x) &\ra& U(x) W_\mu(x) U(x)^\dg + \frac{i}{g} (\partial_\mu U(x))
U(x)^\dg
\eeqa
with $U(x) = \exp \{ -i T_a \alpha_a(x)\}$. As a result the field
strength tensor $G_{\mu\nu}$ transforms according to the adjoint
representation written as
\beq
G_{\mu\nu}(x) \ra U(x) G_{\mu\nu}(x) U(x)^\dg
\eeq
therefore leaving $\cL_G$ invariant.

Having fixed our notation we will now examine in detail the effect of a
CP--type transformation in the gauge sector. The general form of such a
transformation acting on the gauge boson multiplet is given by
\beq
W_\mu^a(x) \ra \ve(\mu) R_{ab} W_\mu^b(\wh x) \quad \mbox{with }
R \in O(n_G)
\eeq
where $n_G$ is the number of gauge bosons and thus
equal to the number of group generators. The fields $W_\mu^a$ are
real and therefore $R$ is a real matrix. It will shortly become clear
that the orthogonality condition (3.2) requires $R$ to be orthogonal.
Let us now consider the effect of the transformation
(3.9) on the field strength tensor.
For the terms with derivatives we have
\beq
\partial_\mu W_\nu^a(x) \ra \ve(\nu) R_{ad} \partial_\mu
(W_\nu^d(\wh x)) =
\ve(\mu) \ve(\nu) R_{ad} \wh\partial_\mu W_\nu^d(\wh x)
\eeq
where $\wh \partial_\mu$ is the derivative with respect to $\wh x$.
The commutator transforms according to
\beqa
- g f_{abc} W_\mu^b(x) W_\nu^c(x) &\ra&
- g f_{ab'c'} \ve(\mu) \ve(\nu) R_{b'b} R_{c'c} W_\mu^b(\wh x)
W_\nu^c(\wh x) \no \\
&=& -g R_{ad} f_{a'b'c'} R_{a'd} R_{b'b} R_{c'c} \ve(\mu) \ve(\nu)
W_\mu^b(\wh x) W_\nu^c(\wh x).
\eeqa
Consequently, under a CP--type transformation $G^a_{\mu\nu}$ behaves as
\beq
G^a_{\mu\nu}(x) \ra \ve(\mu) \ve(\nu) R_{ad} (\partial_\mu W_\nu^d
- \partial_\nu W_\mu^d - g \wh f_{dbc} W_\mu^b W_\nu^c)
(\wh x)
\eeq
with
\beq
\wh f_{dbc} = f_{a'b'c'} R_{a'd} R_{b'b} R_{c'c}.
\eeq

This leads us to the first condition for invariance of $\cL$ under a
CP--type transformation:
\beqa
\mbox{Condition A:} & f_{abc} = f_{a'b'c'} R_{a'a} R_{b'b} R_{c'c}.
\eeqa
In what follows eq. (3.14) will be referred to as Condition A.
If it is fulfilled we get
\beq
G^a_{\mu\nu}(x) \ra \ve(\mu) \ve(\nu) R_{ad} G^d_{\mu\nu}(\wh x)
\eeq
and $\int d^4x \;\cL_G$ is clearly invariant under such a transformation.
Note that eq. (3.10) together with eq. (3.2) already leads to $R$ orthogonal
in order to get invariance of the quadratic part of $\cL_G$ under
the transformation (3.9).

\subsection{Fermions and gauge interactions}
As mentioned before we choose to represent all the fermionic degrees of
freedom by a single right--handed Weyl field vector $\omega_R$
transforming according to the rep $\{ T_a\}$ (see app. A). Hence the
fermionic Lagrangian is given by
\beq
\cL_F = \ol{\omega_R} i \gamma^\mu (\partial_\mu + i g T_a W_\mu{}^a)
\omega_R.
\eeq
The general form of a CP transformation acting on the fermionic multiplet
$\omega_R$ is given by
\beq
\omega_R(x) \ra U \gamma^0 C \ol{\omega_R}(\wh x)^T =
- U C \omega_R(\wh x)^*
\eeq
where $U$, here, is a constant unitary matrix in representation space,
i.e., in the same space on which the rep $\{ T_a\}$ operates. It can be
easily checked that the kinetic part of eq. (3.16) transforms as
$\cL_{F {\rm kin}}(x) \ra \cL_{F {\rm kin}}(\wh x)$ under eq. (3.17)
whilst the interaction term leads to the invariance condition
$$
-(U^\dg T_b R_{ba} U)^T = T_a
$$
which can readily be cast into the form

\beqa
\mbox{Condition B:} & U(- T_b{}^T R_{ab}) U^\dg = T_a.
\eeqa
In what follows we will refer to this equation as Condition B. It is
easily verified that $\{ - T_b{}^T R_{ab}\}$ fulfills the commutation
relations (3.1) for $R$ satisfying Condition A. This fact will
be further exploited. Note that every CP--type transformation acting in
a gauge theory with fermions can be described by a pair of matrices
$(R,U)$ defined above.

As we have seen $R$ is an orthogonal $n_G \times n_G$ matrix and $U$ a
unitary matrix with the dimension of $\{ T_a \}$. In general the rep
$\{ T_a \}$ will not be irreducible and we can therefore perform a
decomposition into irreps
\beq
T_a = i \bigoplus_r ({\bf 1}_{m_r} \otimes D_r(X_a)), \qquad
\dim D_r = d_r
\eeq
where $\{ X_a \}$ is an ON basis of the real compact Lie algebra $\cL_c$
(see app. B); $m_r$ is the multiplicity of the irrep $D_r$ in
$\{ T_a\}$ of dimension $d_r$; the direct sum runs over all irreps
included in $\{ T_a\}$ and its total dimension is $\sum_r m_r d_r$.
We will call the degeneracy spaces ``horizontal spaces'' and the indices
associated with them ``horizontal indices''. The decomposition (3.19)
leads immediately to the following statement:

\paragraph{Theorem I:} Let $(R,U_0)$ be a solution of Conditions
A and B and let $(R,U_1 U_0)$ be another solution. Then
\beq
U_1 = \bigoplus_r (u_r \otimes {\bf 1}_{d_r})
\eeq
where the $u_r$ are unitary $m_r \times m_r$ matrices.

\paragraph{Proof:} Inserting $U_1 U_0$ into Condition B we obtain
$$
T_a = U_1 [U_0 (-T_b{}^T R_{ab}) U_0^\dg] U_1^\dg = U_1 T_a U_1^\dg.
$$
Then Schur's lemma forces $U_1$ to be of the form (3.20). Since $U_0$ and
$U_1U_0$ are both unitary the matrices $u_r$ are unitary as well.
\hfill $\Box$

This simple theorem together with Theorem II of subsect. 6.1 will prove
very useful in the discussion of solutions of Conditions A and B.
In fact Theorem I shows that in order to solve Condition B one can
concentrate on the determination of the group theoretical aspects of
$U$ with the freedom in the horizontal component simply given by eq. (3.20).

Defining a linear operator on $\cL_c$ by (see app. B)
\beq
\psi_R : X_a \ra R_{ba} X_b
\eeq
we infer from Condition A that
\beq
[\psi_R(X_a),\psi_R(X_b)] = f_{abc} \psi_R(X_c).
\eeq
Therefore $\psi_R$ is an automorphism of $\cL_c$ as well as
$\psi_{R^{-1}}$ since the set of automorphisms of $\cL_c$ forms a group
(this can also be verified by examining Condition A). As a result Condition A
can be formulated as
\beq
\psi_R \in \mbox{Aut }(\cL_c).
\eeq

For every irrep $D_r$ the complex conjugate irrep is given by $-D_r^T$.
Clearly, $D_r \circ \psi_R$ defined by
$(D_r \circ \psi_R)(X) \equiv D_r (\psi_R(X))$ is also an irrep.
Thus Condition B can be read as
\beq
\bigoplus_r \left( {\bf 1}_{m_r} \otimes (- D_r^T \circ \psi_{R^{-1}})
\right) \sim \bigoplus_r \left( {\bf 1}_{m_r} \otimes D_r\right).
\eeq
Eqs. (3.23) and (3.24) are purely Lie algebra theoretical conditions
and will be discussed in sect. 4 and sects. 5 and 6, respectively.

Finally, let us write down the effect of changing the basis of the
fermionic multiplet on a CP--type transformation. Let $\omega'_R$
be the field vector in the new basis and
\beq
\omega_R = Z \omega'_R.
\eeq
As a result the new matrix $U'$ in eq. (3.17) is given by
\beq
U' = Z^\dg U Z^*.
\eeq
It is important to note the complex conjugation on the right--hand side of
eq. (3.26). This prevents the use of the well--known theorems for normal
matrices (see subsect. 8.3 for a further discussion on basis transformations).

\section{Automorphisms of $\cL_c$}
\renewcommand{\theequation}{\arabic{section}.\arabic{equation}}
\setcounter{equation}{0}

\subsection{Types of automorphisms}
The reformulation of Conditions A and B into eqs. (3.23) and (3.24),
respectively, makes plain that the group of automorphisms of $\cL_c$,
Aut~$(\cL_c)$, plays an important r\^ole in our discussion. Therefore
we want to set forth in this section all the details we will need in
the following. The basic notions of Lie algebras necessary for this
section can be found in app. B.

\paragraph{Inner and outer automorphisms} \cite{corn,sam}: For each
element $Y$ of $\cL_c$ we can define an automorphism $\psi_Y$ of the
form
\beq
\psi_Y(X) \equiv (\exp \mbox{ ad }Y)(X) = e^Y X e^{-Y} \qquad
\mbox{with} \qquad (\mbox{ad }Y)X \equiv [Y,X].
\eeq
Automorphisms of this kind are called inner. The representation of
$\exp \:\mbox{ad}$ by exponentials, second equality of eq.
(4.1), comes about due to the fact that all Lie algebras can be seen
as matrix Lie algebras (Theorem of Ado \cite{jac}). It is clear from eq. (4.1)
that the inner automorphisms of $\cL_c$ from a group denoted by
Int~$(\cL_c)$. For a connected Lie group $G$ it is isomorphic to $G/Z$ where
$Z$ is the centre of $G$ \cite{var}. Moreover, Int~$(\cL_c)$ is a normal
subgroup of Aut~$(\cL_c)$ since if $\psi \in \mbox{Aut }(\cL_c)$ then
$\psi(\mbox{ad }Y)\psi^{-1} = \mbox{ad }\psi(Y)$ and therefore
$\psi(\exp \mbox{ ad }Y)\psi^{-1} = \exp \mbox{ ad }\psi(Y)$.

Outer automorphisms are automorphisms which are not inner. Clearly, it
is sufficient to consider one representative of each coset in
Aut~$(\cL_c)/\mbox{Int }(\cL_c)$ in order to make a complete study of
outer automorphisms.

\paragraph{Root rotations} \cite{corn}: The group of root rotations
Aut~$(\Delta)$ is defined as a mapping of $\Delta$, the set of (non--zero)
roots of $\wt \cL$ (complexification of $\cL_c$), onto itself which
fulfills
\begin{enumerate}
\item[a)] $\tau(\alpha + \beta) = \tau(\alpha) + \tau(\beta)$ $\forall \;
\alpha,\beta \in \Delta$ such that $\alpha + \beta \in \Delta$ and
\item[b)] $\tau(-\alpha) = - \tau(\alpha)$.
\end{enumerate}
Aut~$(\Delta)$ has an important normal subgroup, the Weyl group
$\W$. It is the group generated by the elements $S_\alpha$
$(\alpha \in \Delta)$ which act on $\Delta$ as
\beq
S_\alpha \beta = \beta - 2 \frac{\langle \beta,\alpha\rangle}
{\langle \alpha,\alpha\rangle} \alpha.
\eeq
One can show that $S_\alpha \in \mbox{Aut }(\Delta)$ $\forall \; \alpha
\in \Delta$.

Associated with each root rotation $\tau \in \mbox{Aut }(\Delta)$ there is
an automorphism $\psi_\tau$ of the corresponding algebra $\wt \cL$, which
is also an automorphism of $\cL_c$. The connection between Aut~$(\Delta)$
and Aut~$(\cL_c)$ is given by the following theorem.

\paragraph{Theorem:} For every $\tau \in \mbox{Aut }(\Delta)$ there is a
mapping $\psi_\tau$ of $\wt \cL$ onto itself defined by
\beq
\psi_\tau(h_\alpha) = h_{\tau(\alpha)} \qquad \mbox{and} \qquad
\psi_\tau(e_\alpha) = \chi_\alpha e_{\tau(\alpha)}
\eeq
where $\chi_\alpha = \pm 1$ $\forall \; \alpha \in \Delta$ such that
$$
\chi_\alpha = 1 \mbox{ for all simple roots,}
$$
\beq
\chi_{\alpha+\beta} = \frac{N_{\tau(\alpha) \tau(\beta)}}{N_{\alpha\beta}}
\chi_\alpha \chi_\beta \qquad \mbox{for } \alpha,\beta, \; \alpha + \beta
\in \Delta^+
\eeq
$$
\mbox{and } \chi_{-\alpha} = \chi_\alpha.
$$

Then $\psi_\tau$ is an automorphism of $\wt \cL$ and, moreover, restricted
to $\cL_c$ we also have $\psi_\tau \in \mbox{Aut }(\cL_c)$. See app. B
for the notation in eqs. (4.3) and (4.4).

In what follows we will need the automorphism induced by root reflexion
\beq
\psi^\triangle \equiv \psi_{\tau_r} \qquad \mbox{with} \qquad
\tau_r(\alpha) = - \alpha .
\eeq
It is clear that $\tau_r \in \mbox{Aut }(\Delta)$ and $\chi_\alpha = 1$
$\forall \; \alpha \in \Delta$. The index $\triangle$ stands
for contragredient.

Another important class of root rotations in the case of simple Lie
algebras is defined by those which are symmetries of the Dynkin diagram
or, equivalently, of the Cartan matrix $A_{jk}$ (see app. B). The group
of these rotations is thus given by
Aut~$(DD) = \{\tau \in \mbox{Aut }(\Delta)|\: A_{\tau(j)\tau(k)} =
A_{jk}$ $\forall \; j,k = 1,\ldots,\ell\}$.
{}From the set of Dynkin diagrams depicted in fig.~1 it is clear that
Aut~$(DD)$ can only be isomorphic to $\{e\}$ or ${\bf Z}_2$ or $S_3$.
The latter case is only possible for $D_4$, the complexification of
$so(8)$, where the three outer simple roots $\alpha_1$, $\alpha_3$,
$\alpha_4$ can be permutated leading to the permutation group
$S_3$.
For $A_1$, $B_\ell$, $C_\ell$ and all exceptional algebras except
$E_6$ there is no trivial diagram symmetry and therefore Aut~$(DD)$ is
reduced to $\{e\}$. For $A_\ell$ $(\ell \geq 2)$ there is symmetry
under the inversion of the order of the simple roots, i.e.
$\tau(\alpha_j) = \alpha_{\ell + 1 -j}$, $j = 1,2,\ldots,\ell$.
For $D_\ell$ $(\ell \geq 5)$ exchange of $\alpha_{\ell -1}$ and
$\alpha_\ell$ is the unique Dynkin diagram symmetry, i.e.,
$\tau(\alpha_j) = \alpha_j$ $(j = 1,2,\ldots,\ell-2)$,
$\tau(\alpha_{\ell-1}) = \alpha_\ell$, $\tau(\alpha_\ell) =
\alpha_{\ell-1}$. In $E_6$ we can reverse the order of the roots in the
line containing the five roots $\alpha_1,\ldots,\alpha_5$, i.e., there
the Dynkin diagram is symmetric under $\tau(\alpha_j) =
\alpha_{6-j}$ $(j = 1,2,\ldots,5)$,
$\tau(\alpha_6) = \alpha_6$. The automorphisms associated with non--trivial
elements of Aut~$(DD)$ will be called diagram automorphisms and denoted
by $\psi_d$.

The essence of the above discussion is related to the facts that for a
simple Lie algebra $\cL_c$, $\psi_\tau$ is an inner automorphism if and
only if the root rotation $\tau$ is an element of the Weyl groyp $\W$
of $\wt \cL$ and diagram automorphisms are always outer automorphisms
so that we have \cite{corn,sam}
\beq
\mbox{Aut }(\cL_c)/\mbox{Int }(\cL_c) \cong \mbox{Aut }(\Delta)/\W
\cong \mbox{Aut }(DD).
\eeq

For our purposes it will be sufficient to use as representations of
the cosets in Aut~$(\cL_c)/\mbox{Int }(\cL_c)$ the automorphisms
$\psi_d$, $\psi^\triangle$ or id (see table 1) depending on the algebra $\cL_c$
under consideration. This covers the automorphism structure of simple
Lie algebras.

\subsection{Some examples of non--simple groups}
We want to go a little beyond simple Lie algebras to include the most
frequent cases occurring in model building. To determine Aut~$(\cL_c)$ in
more complicated cases we need two trivial observations:
\begin{enumerate}
\item[i)] Let $I$ be an ideal of $\cL$ and $\psi \in \mbox{Aut }(\cL)$.
Then $\psi(I)$ is again an ideal.
\item[ii)] Let $I_1$, $I_2$ be ideals of $\cL$ then also $I_1 \cap I_2$
is an ideal.
\end{enumerate}

Let us now consider a few cases:
\begin{enumerate}
\item[i)] $\cL_c = \cL'_c \oplus \cL''_c$ with $\cL'_c$, $\cL''_c$ simple
and non--isomorphic:

Here we have
\beq
\mbox{Aut }(\cL'_c \oplus \cL''_c) \cong \mbox{Aut }(\cL'_c) \times
\mbox{Aut }(\cL''_c).
\eeq
\end{enumerate}
\paragraph{Proof:} $\cL'_c \oplus 0$ is an ideal of $\cL_c$. Then
$(\cL'_c \oplus 0) \cap \psi(\cL'_c \oplus 0)$ is an ideal of $\cL_c$
and also of $\cL'_c$. Since $\cL'_c$ is simple and $\cL''_c \not\cong
\cL'_c$ we have $\psi(\cL'_c \oplus 0) = \cL'_c \oplus
0$. The same reasoning is valid for the other summand $\cL''_c$.
\hfill\ $\Box$

In the following we will adopt the physical point of view that even when
mathematically $\cL'_c \cong \cL''_c$, if the associated gauge couplings
are different we will consider both Lie algebras as being different from
each other.
\begin{enumerate}
\item[ii)] $\cL_c = \cL'_c \oplus \cL'_c$, $\cL'_c$ simple and associated
to a group $G$:

In analogy with the previous case we have now
\beq
\mbox{Aut }(\cL'_c \oplus \cL'_c)/(\mbox{Aut }(\cL'_c) \times
\mbox{Aut }(\cL'_c)) \cong {\bf Z}_2
\eeq
with the ${\bf Z}_2$ generated by
\beq
\psi_E((X,Y)) = (Y,X).
\eeq
Now we can imagine that the Lie algebra is actually associated to a group
$G^*$ defined by enlarging $G \times G$ by an element $E$ such that
\beq
G^* = (G \times G) \cup \{E\}, \qquad E^2 = (e,e), \qquad
E(g,g')E = (g',g).
\eeq
The physical background for this construction is that here the reps we
are interested in are actually reps of $G^*$ with the representation of
$E$ giving a symmetry reason for equal coupling constants for both group
factors in $G^*$. In app. E the irreps of $G^*$ are derived. Here
we give the full list. Let $D_r$ denote the irreps of $G$. Then all
irreps of $G^*$ are given by one of the following constructions:
\begin{enumerate}
\item[] $D_r^\pm: (g_1,g_2) \in G \times G$ is represented by $D_r(g_1)
\otimes D_r(g_2)$ and $D(E) v \otimes w = \pm w \otimes v$.
\item[] $D_{rr'} (r \neq r'):(g_1,g_2) \in G \times G$ is represented
by $(D_r(g_1) \otimes D_{r'}(g_2)) \oplus (D_{r'}(g_1) \otimes
D_r(g_2))$ and $D(E)(v \otimes w, x \otimes y) = (y \otimes x,
w \otimes v)$.
\end{enumerate}
These irreps serve as a guideline for the construction of theories with
gauge group $G \times G$ and equal coupling constants. In specific models
$\psi_E$ can be used as the automorphism associated with P (see sect. 6)
or C (see sect. 9) and define in that way a transformation forcing equal
gauge coupling constants.
\item[iii)] $\cL_c = \cL'_c \oplus u(1)$, $\cL'_c$ simple:

Again one can show that
\beq
\mbox{Aut }(\cL'_c \oplus u(1)) \cong \mbox{Aut }(\cL'_c) \times {\bf Z}_2
\eeq
with ${\bf Z}_2$ generated by
\beq
\psi_u((X,X_u)) = (X,- X_u).
\eeq
Actually, since $u(1)$ is abelian any transformation $X_u \ra a X_u$ with
$a \in {\bf R} \setminus \{0\}$ would be an automorphism but the gauge
Lagrangian $\cL_G$ can only be invariant under $a = \pm 1$ since
$R_u = \mbox{diag }(1,\ldots,1,\pm a)$ associated with $\psi_u$ must be
orthogonal.
\end{enumerate}

\subsection{Irreps and automorphisms}
As discussed in subsect. 3.2 composition of irreps with automorphisms plays
a crucial r\^ole in solving Condition B. Confining ourselves to simple
Lie groups we know that any $\psi \in \mbox{Aut }(\cL_c)$ can be written
as $\psi = \psi_Y$ or $\psi_Y \circ \psi_d$. If $D$ is an arbitrary rep then
\beqa
D \circ \psi_Y &=& e^{D(Y)} D e^{-D(Y)}, \no \\
D \circ \psi_Y \circ \psi_d &=& e^{D(Y)} D \circ \psi_d e^{-D(Y)}.
\eeqa
Therefore only outer automorphisms can give non--equivalent reps through
composition.

To explore the effects of diagram automorphisms we consider first the
action of $\tau \in \mbox{Aut }(DD)$ on the fundamental weights (see app. B):
\beqa
\tau(\Lambda_j) &=& \sum_{k=1}^\ell (A^{-1})_{jk} \tau(\alpha_k)
= \sum_{k=1}^\ell (A^{-1})_{jk} \alpha_{\tau(k)} \no \\
&=& \sum_{k=1}^\ell (A^{-1})_{\tau(j)\tau(k)} \alpha_{\tau(k)}
= \sum_{k=1}^\ell (A^{-1})_{\tau(j)k} \alpha_k = \Lambda_{\tau(j)}.
\eeqa
For convenience we did not distinguish between the diagram symmetry
and its ensuing permutation of the indices $1,\ldots,\ell$, i.e.
$\tau(\alpha_k) \equiv \alpha_{\tau(k)}$. Next we find the weights
of the irrep $D \circ \psi_\tau$ by comparing it with $D$. Let
$e(\lambda,q)$ be the eigenvector with the weight $\lambda$ of $D$
and $q = 1,\ldots,m_\lambda$ where $m_\lambda$ is the multiplicity
of $\lambda$. Then
\beqa
D(h_\alpha)e(\lambda,q) &=& \lambda(h_\alpha)e(\lambda,q) =
\langle \lambda,\alpha\rangle e(\lambda,q) \no \\
(D \circ \psi_\tau)(h_\alpha)e(\lambda,q) &=& D(h_{\tau(\alpha)})
e(\lambda,q) = \langle \lambda,\tau(\alpha)\rangle e(\lambda,q).
\eeqa
Since $\tau$ is a root rotation one can show that \cite{corn}
\beq
\langle \tau(\beta),\tau(\gamma)\rangle = \langle \beta,\gamma\rangle
\qquad \forall \; \beta,\gamma \in \Delta.
\eeq
Therefore, if $\lambda$ is a weight of $D$ then $\tau^{-1}(\lambda)$
is a weight of $D \circ \psi_\tau$. This is valid for any root rotation
$\tau$. If in addition $\tau \in \mbox{Aut }(DD)$ we obtain the
highest weight of $D \circ \psi_\tau$ as (see eq. (4.14))
\beq
\tau^{-1}(\Lambda) = n_1 \Lambda_{\tau^{-1}(1)} + \ldots + n_\ell
\Lambda_{\tau^{-1}(\ell)} = n_{\tau(1)} \Lambda_1 + \ldots +
n_{\tau(\ell)} \Lambda_\ell
\eeq
if $\Lambda = n_1\Lambda_1 + \ldots + n_\ell \Lambda_\ell$ is the highest
weight of $D$.

Now we can list the simple complex Lie algebras with non--trivial diagram
symmetries and give the conditions for $D_\Lambda \sim D_\Lambda \circ
\psi_d$ where we indicate the highest weight as a subscript. In all
cases except $D_4$ the automorphism is unique. Thus we have
$D_\Lambda \sim D_\Lambda \circ \psi_d$ if and only if
\beq
\ba{lll}
n_j = n_{\ell+1-j} & j = 1,2,\ldots,\ell & \mbox{ for $A_\ell$
($\ell \geq 2$)} \\[6pt]
n_{\ell -1} = n_\ell && \mbox{ for $D_\ell$ ($\ell \geq 5$)} \\[6pt]
n_1 = n_5, \; n_2 = n_4 && \mbox{ for $E_6$}.
\ea
\eeq
For $D_4$ there are five non--trivial diagram automorphisms $\psi_\tau$
and $D_\Lambda \sim D_\Lambda \circ \psi_\tau$ only if
\beq
n_1 = n_{\tau(1)}, \qquad n_3 = n_{\tau(3)}, \qquad
n_4 = n_{\tau(4)}.
\eeq

In subsect. 4.2 we have also defined the automorphism $\psi^\triangle$
associated
with root reflexion. Clearly, $D_\Lambda \circ \psi^\triangle$ has the weight
$- \lambda$ if $\lambda$ is a weight of $D_\Lambda$. Therefore
$D_\Lambda \circ \psi^\triangle \sim - D_\Lambda^T$ and
$\psi^\triangle$ relates a rep
to its contragredient rep. It is outer for $A_\ell$ $(\ell \geq 2)$,
$D_\ell$ $(\ell = 5,7,9,\ldots)$ and $E_6$. In these cases
$D_\Lambda \circ \psi^\triangle \sim D_\Lambda \circ \psi_d$ and thus the
condition for $D_\Lambda \sim - D_\Lambda^T$ is given by eq. (4.18).
For all other simple Lie algebras $\psi^\triangle$ is inner and
$D_\Lambda \sim - D_\Lambda^T$ for all their irreps (see table 1).

\section{The existence of CP in gauge theories and Condition B}
\renewcommand{\theequation}{\arabic{section}.\arabic{equation}}
\setcounter{equation}{0}

\subsection{The automorphism associated with CP}
In this section we will show that the fermionic and vector boson sectors
of gauge theories (the terms $\cL_F$ and $\cL_G$ of the Lagrangian) are
always CP symmetric. The same is true for the couplings $\cL_H$ of
scalars with the gauge fields (see sect. 7). Imposing CP symmetry to
the full Lagrangian $\cL$ containing the terms $\cL_G$, $\cL_F$, $\cL_H$,
$\cL_Y$, the Yukawa interactions, and $\cL_S$, the scalar potential,
will induce conditions on the only terms that may violate this
symmetry --- $\cL_Y$ and $\cL_S$. In sect. 8 we will discuss the
conditions induced on $\cL_Y$ --- the simplest will be reality of
Yukawa couplings but in general the situation is more complicated (see
full classification in subsect. 8.3).
We will not discuss invariance of $\cL_S$ in this paper.

In our formulation a CP transformation of the fermionic and gauge boson
multiplets is completely specified by the two matrices $R$ and $U$ defined
in sect. 3. Requiring that CP reverses all quantum
numbers of each field we see that $R_{CP}$ corresponds to an automorphism
which induces a reflexion of all the
roots or, in other words, acts as $h \ra -h$ on the CSA
(as required by its physical interpretation). Hence we
identify $R_{CP}$ with the matrix representation of the contragredient
automorphism defined by eq. (4.5) \cite{sla,smo}:
\beq
\psi_{CP} \equiv \psi^\triangle.
\eeq

In what concerns the pure gauge term $\cL_G$ of the Lagrangian we can
conclude from the derivation of Condition A that it is always
invariant under any automorphism of the Lie algebra irrespective of
whether this is an inner or outer automorphism, contragredient or not.

One might think that identifying $\psi_{CP}$ with $\psi^\triangle$ is too
restrictive. This is, however, not the case because any two automorphisms
that act in the same way on the CSA will only differ by an inner automorphism
generated by an element of the CSA itself \cite{sam}. As we
shall see in subsect. 6.1 using $\psi_{CP}$ instead of $\psi^\triangle$
leads to equivalent conditions for CP invariance.

\subsection{Canonical and generalized CP transformations}
We consider now how to define a CP transformation on the fermion fields.
In sect. 3, eq. (3.19) shows how the representation $\{ T_a\}$ can be
decomposed into irreps $D_r$. We know from the discussion of sect. 4
that
\beq
- D_r^T \circ \psi^\triangle \sim D_r
\eeq
because the weights of both irreps are the same. This follows immediately
from
\beq
\psi^\triangle(h_\alpha) = h_{-\alpha} = - h_\alpha .
\eeq
Therefore there exist unitary matrices $V_r$ such that
\beq
V_r(- D_r^T \circ \psi^\triangle)V_r^\dg = D_r \qquad \forall\; r.
\eeq
Thus we define a ``canonical CP transformation'' by \cite{sla,smo}
\beq
(R^\triangle,U_0) \quad \mbox{with} \quad \psi_{R^\triangle}
\equiv \psi^\triangle, \qquad
U_0 = \bigoplus_r {\bf 1}_{m_r} \otimes V_r .
\eeq
Since $(R^\triangle,U_0)$ solves Conditions A and B we have a symmetry of
$\cL_G + \cL_F$. This means that a gauge theory without Yukawa interactions
and a scalar potential is automatically CP invariant (the scalar
couplings to the gauge fields, $\cL_H$, are treated like the fermionic
couplings in sect. 7).

As already mentioned the physical reason to call the above transformation
CP is that it turns around the signs of all members of the CSA in any
rep. Thus it transforms fields into those with opposite quantum numbers.
In order to see that this is indeed the case let $\U$ and $\D_r(g)$ be
the operator implementations of CP and $D_r(g)$, respectively,
on the Hilbert space of
states and $\chi_R$ be a second quantized irreducible field multiplet.
If CP is conserved we have
\beq
\U \chi_R(x) \U^{-1} = - V_r C \chi_R(\wh x)^* \quad \mbox{and} \quad
\D_r(g) \chi_R(x) \D_r(g)^{-1} = D_r(g) \chi_R(x)
\eeq
and, together with eq. (5.4), we obtain
\beq
\D_r(g) \U \chi_R(x) \U^{-1} \D_r(g)^{-1} = - V_r D_r(g)^* C \chi_R(\wh x)^*
= (D_r \circ \psi^\triangle)(g)(- V_r C \chi_R(\wh x)^*).
\eeq
As we know, in the irrep $D_r \circ \psi^\triangle$ the CSA is represented with
opposite signs compared to $D_r$ thus verifying the above statement.

Theorem I (sect. 3) tells us that given $R = R^\triangle$ the most general
solution of Condition B is obtained by $U = U_1U_0$ where $U_1$ is a
horizontal transformation:
\beq
U_1 = \bigoplus_r u_r \otimes {\bf 1}_{d_r}.
\eeq
In the following we will call such transformations associated with
$(R^\triangle,U_1U_0)$ ``generalized CP transformations'' (see refs.
\cite{eck81,eck84}
for the context of left--right symmetric models and refs. \cite{ber,eck87}
for general considerations).

Considering the example of QCD in sect. 2 it is obvious that
$U_0$ can be regarded as the identity matrix. Clearly this result can
be generalized to the defining rep of $SU(N)$ for arbitrary $N$ by
generalizing the Gell--Mann matrices. The reason for $U_0 = {\bf 1}$ is
that these matrices are either symmetric or antisymmetric, therefore
$- \lambda_a^T = \eta_a \lambda_a$ and the signs $\eta_a$ are compensated
by the transformation of the gauge bosons $W_\mu^a \ra \eta_a W_\mu^a$.
Thus we have $R^\triangle = \mbox{diag }(\eta_1,\ldots,\eta_{n_G})$. It turns
out that this is a general feature for any irrep of any semisimple
compact Lie group if we choose the standard basis of $\cL_c$ (see eq.
(B.23)) and a suitable basis in representation space. To be specific,
in app. F the following theorem is proved.

\paragraph{Theorem:} For every irrep $D$ of $\cL_C$ there is an
ON basis of ${\bf C}^d$ ($d = \dim D$) such that
\beq
D(X_a)^T = - \eta_a D(X_a), \qquad \eta_a^2 = 1 \qquad
(a = 1,\ldots,n_G)
\eeq
for the antihermitian generators of $\cL_C$ in $D$. Furthermore the
generators $\{X_a\}$ are those of the compact normal form of $\cL_C$
and therefore the root reflexion $\psi^\triangle$ is given by
\beq
\psi^\triangle(X_a) = \eta_a X_a .
\eeq

In other words, the matrices $D(X_a)$ are either imaginary and symmetric
$(\eta_a = -1)$ or real and antisymmetric $(\eta_a = 1)$ and therefore
a generalization of $-i \sigma_a/2$ (Pauli matrices) for $SU(2)$ and
$- i \lambda_a/2$ (Gell--Mann matrices) for $SU(3)$ to arbitrary irreps
of semisimple compact Lie algebras. The above basis will be called
``CP basis'' in the following. In this basis a canonical CP transformation,
(CP)$_0$, can be simply represented by
\beq
(CP)_0 \ra (R^\triangle,{\bf 1}) \qquad \mbox{with } R^\triangle = \mbox{diag }
(\eta_1,\ldots,\eta_{n_G}).
\eeq
To be more explicit (see app. F) we have the following situation in a CP basis:
$$
D(-i H_j) \quad (j = 1,\ldots,\ell), \quad
D\left( \frac{e_\alpha - e_{-\alpha}}{i \sqrt{2}} \right) \quad
(\alpha \in \Delta) : \eta_a = -1
$$
(imaginary and symmetric)
\beq
D\left( \frac{e_\alpha + e_{-\alpha}}{\sqrt{2}} \right) \quad
(\alpha \in \Delta) : \eta_a = 1
\eeq
(real and antisymmetric).

Therefore we get
\beq
\frac{1}{\sqrt{2}} \left[
D \left( \frac{e_\alpha + e_{-\alpha}}{\sqrt{2}} \right)
+ iD \left( \frac{e_\alpha - e_{-\alpha}}{i\sqrt{2}} \right) \right]
= D(e_\alpha) \qquad \mbox{real } \forall\; \alpha \in \Delta.
\eeq
These results will be of importance when we consider the Yukawa
couplings in sect. 8.

As a conclusion we have seen that a fermionic gauge theory (without
scalars) is always CP invariant. The pure gauge term by itself is
always invariant under any automorphism of the Lie algebra. In what
concerns $\cL_F$, the reason for invariance is due to the fact that
the fermionic multiplet transforms in such a way that the CP
transformed fermion fields are associated to the complex conjugate
rep whilst the same transformation in the gauge boson sector will
reverse this effect through $R^\triangle$ (this is obvious from the
derivation of Condition B in sect. 3).
This is the reason why invariance under a canonical
CP transformation is verified irrespective of whether the
automorphism is inner or outer. In all the examples of sect. 2,
QED, QCD and the $SO(10)$--GUT, $\psi^\triangle$ is outer
(see table 1).

\section{On the general solution of Condition B}
\renewcommand{\theequation}{\arabic{section}.\arabic{equation}}
\setcounter{equation}{0}

\subsection{Introduction}
In sect. 5 we discussed the solution of Condition B together with the
requirement that $R$ be such that it represents a CP transformation.
Yet as we have seen in the examples of sect. 2 the form of CP and P
transformations is the same in a formalism with fermion fields of
a single chirality so that the difference between CP and P lies in
the properties of $R$ and $U$. This justifies that we study now
Condition B with a general $R$. Therefore in this section we will discuss
invariance of $\cL_F$, the fermionic part of the Lagrangian of a gauge
theory, under a CP--type transformation without constraining $R$ to
be the contragredient automorphism. The following theorem will simplify
our discussion.

\paragraph{Theorem II:} Let $(R,U)$ be a pair of matrices that verify
Conditions A and B and let $R_I$ represent an inner automorphism
of $\cL_c$. Then the pairs
\beq
(R_IR, e^{iy_aT_a} U) \qquad \mbox{and} \qquad
(RR_I, U(e^{iy_aT_a})^*)
\eeq
where
\beq
\psi_{R_I}(X) = e^{-y_aX_a} X e^{y_aX_a}
\eeq
are also solutions of Conditions A and B.

\paragraph{Proof:} $R_IR$ and $RR_I$ are solutions of Condition A
because the set of automorphisms form a group. Since $\psi_{R_I}$ is
inner there are real numbers $y_a$ $(a = 1,\ldots,n_G)$ such that
eq. (6.2) is fulfilled. Translating eq. (6.2) into the fermion
representation we get
\beq
e^{i y_cT_c} T_a e^{-i y_c T_c} = R_{Iba} T_b.
\eeq
Writing Condition B in terms of the second pair of eq. (6.1) we get
\beqan
\lefteqn{U(e^{iy_cT_c})^*(-T_b^T(RR_I)_{ab})(e^{-iy_cT_c})^* U^\dg =}\\
&=& -U(e^{iy_cT_c} T_b e^{-iy_cT_c}(RR_I)_{ab})^* U^\dg =\\
&=&U(-T_c^T R_{Icb} (RR_I)_{ab})U^\dg =
U(-T_b^T R_{ab})U^\dg = T_a.
\eeqan
The other pair can be dealt with in an analogous way. \hfill\ $\Box$

Theorem II means for the discussion of invariance of $\cL_F$ under
CP--type transformations that if for a given automorphism there is a
solution of Condition B (i.e., there is a matrix $U$ that verifies
the equality) then for any other automorphism differing from this one
by an inner automorphism there will also be a solution of Condition
B. The theorem also tells us how to relate the new matrix $U$ to the
initial one. In mathematical terms this means that it is only the
quotient group Aut~$(\cL_c)/\mbox{Int }(\cL_c)$ that needs to be
considered so that we can confine ourselves to a particular
representative of each coset.

Let us consider now the simple Lie algebras $\cL_c$ in the light of
Theorem II and distinguish three classes of Lie algebras (see table 1)
\cite{sla}:
\begin{enumerate}
\item[a)] $\cL_c = su(2)$, $so(2\ell +1)$ $(\ell \geq 2)$, $sp(2\ell)$
$(\ell \geq 3)$, $cE_7$, $cE_8$, $cF_4$, $cG_2$

Here no outer automorphisms exist and we take $\psi_R = id$ as
representative. Since $\psi^\triangle$ is inner $-D^T \sim D$ is valid for all
irreps $D$ and therefore there is a matrix $W$ such that
\beq
W(-D^T)W^\dg = D.
\eeq
Consequently Condition B is solvable without restrictions on the irrep
content of $\{T_a\}$.
\item[b)] $\cL_c = su(\ell + 1)$ $(\ell \geq 2)$, $so(2\ell)$
$(\ell = 5,7,9,\ldots)$, $cE_6$

Here we have both inner and outer automorphisms. For the inner
automorphisms we can consider again $\psi_R = id$ as
representative, for the outer automorphisms we can choose the
contragredient one, $\psi_R = \psi^\triangle$,
since in this case $\psi^\triangle$ is
outer. In these Lie algebras any outer automorphism will be a
composition of $\psi^\triangle$ with an inner automorphism

$\psi_R = id$: In general an irrep is not
equivalent to its complex conjugate. As a result Condition B requires
that for any irrep $D$ also its complex conjugate $-D^T$ must be
contained in $\{T_a\}$.

$\psi_R = \psi^\triangle$: Choosing a basis where the CSA in the irrep $D$ is
diagonal it is clear that the weights of $- D^T \circ \psi^\triangle$ are
identical with those of $D$. Therefore $- D^T \circ \psi^\triangle \sim D$
and Condition B is solvable without restriction on $\{T_a\}$.
\item[c)] $\cL_c = so(2\ell)$ $(\ell = 4,6,8,\ldots)$

In this case $\psi^\triangle$ is an inner automorphism, therefore all irreps
$D$ are equivalent to $- D^T$. Hence it is appropriate to represent
Aut~$(\cL_c)$ by the two cases $\psi_R = id$ and $\psi_R = \psi_d$,
the unique diagram automorphism for $\ell = 6,8,10,\ldots$. As we
know for $so(8)$ $(\ell = 4)$ there are five non--trivial
diagram automorphisms.

$\psi_R = id$: Analogously to case a) Condition B is solvable without
restriction on $\{T_a\}$.

$\psi_R = \psi_d$: Condition B is solvable only if for any irrep $D$
also $D \circ \psi_d$ is contained in $\{T_a\}$. The relationship
between $D$ and $D \circ \psi_d$ was discussed in subsect. 4.3.
\end{enumerate}
We may also consider non--simple Lie groups in the light of Theorem II.
For that purpose we take into account the results of sect. 4 and split
the discussion into three generic classes.
\begin{enumerate}
\item[i)] $\cL_c = \cL'_c \oplus u(1)$, $\cL'_c$ simple: We know that there
is the automorphism $\psi_u : X_u \ra - X_u$ eq. (4.12) on $u(1)$ in addition
to Aut~$(\cL'_c)$. Thus we can represent Aut~$(\cL_c)/\mbox{Int }(\cL_c)$
by $\psi = (\psi',id_u)$ or $(\psi',\psi_u)$ where
$\psi' \in \{id,\psi^\triangle,\psi_d\} \subset \mbox{Aut }(\cL'_c)$. Each
irrep $D^u_k$ of $u(1)$ is characterized by the generator $ik$,
$k \in {\bf Z}$. Denoting by $D'$ an irrep of $\cL'_c$ then the irreps
of $\cL_c$ are given by $D' \otimes D_k^u$. For $\psi = (\psi',id_u)$
also $(- D'{}^T \circ \psi') \otimes D^u_{-k}$ must be contained in
$\{T_a\}$ whereas for $\psi = (\psi',\psi_u)$ it is sufficient to
consider $\cL'_c$.
\item[ii)] $\cL_c = \cL'_c \oplus \cL''_c$, $\cL'_c$, $\cL''_c$ simple:
This case leads to the discussion of each simple summand. Note that the
irreps of $\cL_c$ are given by the tensor products of the irreps of
each summand \cite{shaw} as for i).
One should remember that here we have in mind
different gauge coupling constants for $\cL'_c$ and $\cL''_c$ even
for $\cL'_c \cong \cL''_c$ and therefore no automorphisms other than
Aut~$(\cL'_c) \times \mbox{Aut }(\cL''_c)$ exist.
\item[iii)] $\cL_c = \cL(G^*)$ with $G^*$ defined in subsect. 4.2:
The only additional automorphism is $\psi_E$ given by eq. (4.9). It is easy
to check that
\beq
D \circ \psi_E = D(E) D D(E)
\eeq
for all irreps $D = D_r^\pm, D_{r,r'}$. Consequently $\psi_E$ is as good
as inner and one only has to consider a single summand.
\end{enumerate}

\subsection{On the definition of parity}
In the definition of a CP transformation we had the automorphism
$\psi^\triangle$ (changing the sign of all elements of the
CSA $\Ha$) with $(\psi^\triangle)^2 = id$.
This suggests a mathematical definition of a parity transformation
via an involutive automorphism $\psi_P$ (i.e., verifying $\psi^2_P = id$)
which maps the CSA into itself and does not change the sign of the whole
algebra. Therefore we can find an ON basis of $\Ha$ where \cite{sla}
\beq
\psi_P(H_j) = \mu_j H_j, \qquad
\mu_j = \left\{ \ba{rl} 1, & j = 1,\ldots,p \\
-1, & j = p+1,\ldots,\ell . \ea \right.
\eeq
{}From the discussion in subsect. 6.1 it is obvious that if parity is a
symmetry of the theory the matrix $U_P$ associated with this
transformation either maps a given irrep into itself or connects the
irrep $D$ to $(- D^T) \circ \psi_P \sim D \circ \psi^\triangle \circ \psi_P$.
Both irreps have to be present in $\{T_a\}$ with the same multiplicity
to fulfill Condition B. In the first case we call such a parity transformation
{\em internal} with respect to $D$ and in the second {\em external}.
In sect. 2 we had the external cases of QED and QCD and the $\{\ol{16}\}$
of $so(10)$ as an internal case.

Let us for the time being consider the internal case with a single
irrep $D$ with dimension $d$. Then we can define a mapping:
\beqa
f : x &\ra& U_P x^* \no \\
 {\bf C}^d &\ra& {\bf C}^d.
\eeqa
Given an ON basis $e(\lambda,q)$ of weight vectors (see app. B) in the
Hilbert space ${\bf C}^d$ associated with the irrep $D$,
where $q = 1,\ldots,m(\lambda)$
and $m(\lambda)$ denotes the multiplicity of $\lambda$, we have
\beq
D(H_j)[f(e(\lambda,q))] = -U_P D(\psi_P(H_j))^* e(\lambda,q)^* =
- \mu_j \lambda(H_j) f(e(\lambda,q)).
\eeq
To derive this relation we have used eq. (6.6) together with the fact that
$D(H_j)$ is hermitian and that Condition B can be rewritten as
\beq
- U_P [D(\psi_P(iX))]^* U_P^\dg = D(iX) \qquad \forall \; X \in \cL_c.
\eeq
As a result eq. (6.8) shows that $f$ relates states characterized in general
by different weights $\lambda$ and $\lambda_P$ given by
\beqa
\lambda &\leftrightarrow& (\lambda(H_1),\ldots,\lambda(H_p),
\lambda(H_{p+1}),\ldots,\lambda(H_\ell)), \no \\
\lambda_P &\leftrightarrow& (-\lambda(H_1),\ldots,-\lambda(H_p),
\lambda(H_{p+1}),\ldots,\lambda(H_\ell)).
\eeqa
Thus weight vectors associated with $\lambda$ are transformed into weight
vectors associated with $\lambda_P$. Only vectors with
$\lambda(H_1) = \ldots = \lambda(H_p) = 0$ have no partner. Since $D(H_j)$
is hermitian $e(\lambda,q)$ is orthogonal to $e(\lambda_P,q)$ for
$\lambda \neq \lambda_P$. Obviously the contragredient automorphism
$\psi^\triangle$ does not establish such a relationship between weight vectors
because for $\psi_P = \psi^\triangle$ we would have
$\lambda = \lambda_P$ for all weights.

In the external case restricting oneself to single copies of $D$ and
$-D^T \circ \psi_P$, respectively, the above discussion goes through
without change, but now $e(\lambda,q)$ and $e(\lambda_P,q)$ lie in
different spaces, namely in the spaces associated with $D$ and
$-D^T \circ \psi_P$, respectively. Taking into account non--trivial
multiplicities in both cases, internal and external, does not alter the
essence of our discussion but for simplicity we stick to single copies
of irreps for the rest of this section.

We can make an analysis which has some resemblance to the one
for CP in eqs. (5.6) and (5.7). As just
before it will be valid for internal and external parity. Defining
\beq
\omega_R^{\lambda,q}(x) \equiv e(\lambda,q)^\dg \omega_R(x)
\eeq
we get
$$
\D(e^{-isH_j}) \omega_R^{\lambda,q} \D(e^{isH_j}) =
e^{-is\lambda(H_j)} \omega_R^{\lambda,q}
$$
and
\beq
\D(e^{-isH_j}) \omega_R^{\lambda_P,q} \D(e^{isH_j}) =
e^{-is\lambda_P(H_j)} \omega_R^{\lambda_P,q} \qquad \forall \; s \in
{\bf R}.
\eeq
This allows to define
\beq
\omega_L^{\lambda,q}(x) \equiv C \gamma_0^T(\omega_R^{\lambda_P,q}(x))^*
\eeq
for $\lambda \neq \lambda_P$ which has the same quantum numbers
$\lambda(H_j)$ $(j = 1,\ldots,p)$ as $\omega_R^{\lambda,q}(x)$.
Then P transforms right into left and vice versa:
\beqan
\U_P \omega_R^{\lambda,q}(x) \U_P^{-1} &=& - e(\lambda,q)^\dg U_P C
\omega_R(\wh x)^* \\
&=& - (U_P^* e(\lambda,q))^\dg U_P^* U_P C \omega_R(\wh x)^*
= - e^{i\delta} C(\omega_R^{\lambda_P,q}(\wh x))^* \\
&=& e^{i\delta} \gamma_0 \omega_L^{\lambda,q}(\wh x)
\eeqan
and, similarly,
\beq
\U_P \omega_L^{\lambda,q}(x) \U_P^{-1} = - \gamma_0 \omega_R^{\lambda,q}
(x).
\eeq
To derive this equation we have taken into account eq. (6.8) and
\beq
U_P U_P^* = e^{-i\delta} {\bf 1}_d \qquad \mbox{or} \qquad
\left( \ba{cc} e^{i\delta} {\bf 1}_d & 0 \\
                             0       & e^{-i\delta} {\bf 1}_d \ea
\right)
\eeq
for internal and external parity, respectively\footnote{$\delta$ is the
phase of $U_P^* U_P$ in $D$ or $- D^T \circ \psi_P$ for the internal
or external case, respectively.}. Eq. (6.15) can be cast into a more
general form:

If $(R,U)$ is a CP--type transformation with $R^2 = {\bf 1}$ then
$$
[UU^*,T_a] = 0 \qquad \forall \; a = 1,\ldots,n_G.
$$
\paragraph{Proof:}
$$
(UU^*)T_a (UU^*)^\dg = U(U T_a^* U^\dg)^* U^\dg
= - U T_b^* R_{ba} U^\dg = U(-T_b^T R_{ab})U^\dg = T_a.
$$
\hfill\ $\Box$

In contrast to CP there is no canonical way to define P in general. With
eq. (6.6) a mathematical definition was given but what is considered as parity
in a model also depends on the physical situation. A plausible
requirement would be that the signs of the electromagnetic charge and the
two colour charges are not changed under $\psi_P$. For a deeper connection
of P and C with group theory the reader is referred to the discussion of
charge conjugation in ref. \cite{sla} (see also sect. 9).

The external case of parity with $\psi_P = id$, $D \not\sim D^*$ is the type
of parity we are used to (QED, QCD). In a basis where the matrices of
$D^*$ are just the complex conjugate matrices of $D$ we simply form the
Dirac field $\chi_D \equiv \chi_R + (\chi'_R)^c$ where $\chi_R$ and
$\chi'_R$ transform according to $D$ and $D^*$, respectively. In the
$so(10)$ example (sect. 2) $\psi_P$ is an inner automorphism.

\section{CP--type transformations in the Higgs sector}

\renewcommand{\theequation}{\arabic{section}.\arabic{equation}}
\setcounter{equation}{0}

\subsection{Pseudoreal scalars and CP transformations}
Considering a pseudoreal irrep $D$ \cite{corn} there are two unitary
matrices associated with it which are important in our context. The
first one is given by the equivalence of the irrep and its complex
conjugate. In Lie algebra form it is written as
\beq
W(- D^T) W^\dg = D, \qquad W^T = - W
\eeq
where antisymmetry of $W$ follows from pseudoreality \cite{corn,ha}.
The other unitary matrix is associated with the (CP)$_0$ symmetry:
\beq
U (- D^T \circ \psi^\triangle) U^\dg = D.
\eeq
The matrix $U$ is symmetric:
\beq
U^T = U.
\eeq
\paragraph{Proof:} The symmetry of $U$ is actually contained in the proof
of app. F. There it is first demonstrated that $U^T = \pm U$. Then
antisymmetry is excluded for irreps of semisimple compact Lie algebras.
\hfill\ $\Box$

Eq. (7.3) is, of course, valid for arbitrary irreps. Clearly, in a CP
basis $U \sim {\bf 1}$ and part of the discussion here would be
superfluous. However, we think that it is very instructive not
to restrict our discussion to a CP basis but
instead to have the general setting.

The matrix $W$ allows to define
\beq
\wt \phi \equiv W \phi^* \qquad \mbox{with} \qquad
\wt{\wt \phi} = - \phi
\eeq
which transforms in exactly the same way as the scalar multiplet
$\phi$ under the gauge group.
It is more difficult to see that both fields also transform alike under CP:
\beq
CP: \phi \ra U \phi^*, \qquad \wt \phi \ra U {\wt \phi}^*.
\eeq
This follows from the fact that $W$ and $U$ are related by
\beq
W U^\dg W = - U
\eeq
with an appropriate choice of phase, e.g., for $W$.

\paragraph{Proof of eq. (7.6):} Putting $\psi^\triangle$ on
the right--hand side of eq. (7.2) and using eq. (7.1) one obtains
$$
U(- D^T) U^\dg = (UW^\dg) D (UW^\dg)^\dg = D \circ \psi^\triangle .
$$
Taking advantage of $(\psi^\triangle)^2 = id$ one further derives
$$
[(UW^\dg)^2,D] = 0 \qquad \mbox{and thus} \qquad
(UW^\dg)^2 = \mu {\bf 1}.
$$
Choosing the phase of $W$ such that $\mu = -1$ and making some algebraic
manipulations one arrives at eq. (7.6). \hfill\ $\Box$

\paragraph{Proof of eq. (7.5):} Under CP $\wt \phi$ transforms as
$$
\wt \phi \ra W U^* \phi = W U^* W^T {\wt \phi}^* =
- W U^\dg W {\wt \phi}^* = U {\wt \phi}^*
$$
where in the last step eq. (7.6) has been taken into account. \hfill\ $\Box$

Let us assume now that there are $m$ copies $\phi_1,\ldots,\phi_m$ all
in the same irrep $D$. Putting the $m$ multiplets into a vector $\chi$
then $\cL_H$, the scalar kinetic and gauge Lagrangian, can be rewritten
as \cite{bin,eck81}
\beq
\cL_H (\chi) = \frac{1}{2} \cL_H(\chi) + \frac{1}{2} \cL_H(\wt \chi).
\eeq
This allows for a wider class of generalized CP transformations for
pseudoreal scalars leaving $\cL_H$ invariant defined by
\beq
CP : \left( \ba{c} \chi \\ \wt \chi \ea \right) \ra HU
\left( \ba{c} \chi \\ \wt \chi \ea \right)^*
\eeq
where $H$ is a unitary ``horizontal'' $2m \times 2m$ matrix. One has to
take into account, however, that $\wt \chi$ is related to $\chi$ via
definition (7.4) which gives a restriction on $H$. With
\beq
H = \left( \ba{cc} A & B \\ C & D \ea \right)
\eeq
this restriction is derived by the requirement
\beq
W(A U \chi^* + B U \wt \chi^*)^* = C U \chi^* + D U \wt \chi^*.
\eeq
Exploiting this condition by using eq. (7.6) in the form
$W U^* = U W^*$ we finally obtain $D = A^*$, $C = - B^*$ and therefore
\beq
H = \left( \ba{cc} A & B \\ -B^* & A^* \ea \right) \in Sp(2m)
\eeq
where $Sp(2m) = \{H \in U(2m)|\: H^T J_m H = J_m\}$ is the unitary symplectic
group \cite{corn}. $J_m$ is defined by
$$
J_m = \left( \ba{rc} 0 & {\bf 1}_m \\ -{\bf 1}_m & 0 \ea \right).
$$

In a CP basis one can say more about the matrices $U$ and $W$. With
the phase convention of eq. (7.6) we get the following result:
\beq
\mbox{CP basis} \Ra U = {\bf 1}, \qquad W = W^* = - W^T
\eeq
where $U = {\bf 1}$ is the convention of the (CP)$_0$ transformation
in the CP basis.

\subsection{Real scalars in complex disguise and CP transformations}
Scalar multiplets belonging to potentially real irreps $D$ \cite{corn}
can also be represented by complex fields. In general, in a CP basis,
matrices of a potentially real irrep are complex. When these matrices
are rotated into a real rep the scalar multiplets are also transformed
and there are two possibilities -- either the scalar multiplets also
become real or else they remain complex in which case they can be split
into two real multiplets belonging to the same real irrep. The second
case will be illustrated by a familiar example in the gauge theory
$SU(2)_L \times SU(2)_R \times U(1)_{B-L}$. The treatment of two real
multiplets collected in a complex multiplet $\phi$ can proceed along
the same lines as in the case of pseudoreal scalars by the introduction
of the corresponding field $\wt \phi$, instead of splitting it into its
real components. When $\phi$ represents a single real irrep it coincides
with $\wt \phi$ in the real basis so that its introduction in this case
would be meaningless.

For real scalars one has now three unitary matrices associated with the
irrep. $U$ is given as in eq. (7.2), $W$ is now symmetric \cite{corn,ha}
\beq
W(- D^T)W^\dg = D, \qquad W^T = W
\eeq
and $V$ transforms $D_R$, an explicitly real realization of the
potentially real rep $D$, into $D$:
\beq
D = V D_R V^\dg.
\eeq

Let us first have a look on the relationship between the (CP)$_0$
transformation in $D_R$ and $D$. To do this we note that
\beq
W U^\dg W = U
\eeq
and
\beq
V^\dg W V^* = {\bf 1}
\eeq
with appropriate phase factors for $W$ and $V$.

\paragraph{Proof:} Eq. (7.15) is proved as before in the pseudoreal case
but now we choose a phase of $W$ such that we have a plus sign instead
of minus in eq. (7.6). The reason for this choice will become clear when
we later go to a CP basis. The second relation derives from inserting
eq. (7.14) into eq. (7.13) giving
$$
W V^* D_R V^T W^\dg = V D_R V^\dg
$$
or
$$
[V^\dg W V^*,D_R] = 0.
$$
With Schur's lemma and a phase choice for $V$ we obtain eq. (7.16).
\hfill\ $\Box$

Let $\vp = V^\dg \phi$ be the real scalar field. Then the CP transformation
on $\vp$ is given by
\beq
CP : \vp \ra V^\dg U V^* \vp
\eeq
and
\beq
V^\dg U V^* \mbox{ real}.
\eeq

\paragraph{Proof:} To prove eq. (7.18) we rewrite eq. (7.15) as
$$
U^* W = W^* U \qquad \mbox{or} \qquad U^* V V^T = V^* V^\dg U.
$$
Therefore
$$
(V^\dg U V^*)^* = V^T(U^* VV^T)V^* = V^T(V^*V^\dg U)V^* = V^\dg UV^*.
$$
\hfill\ $\Box$ \\
Thus the general CP formalism is consistent with real fields.
Choosing the CP basis we now have
\beq
\mbox{CP basis} \Ra U = {\bf 1}, \qquad W = W^* = W^T.
\eeq
As before $W$ is real. This is a consequence of the phase choice in
eq. (7.15).

Now we come to the second topic, namely to the discussion of a complex
multiplet comprising two real multiplets transforming under the same
irrep. As in the pseudoreal case we define
\beq
\wt \phi \equiv W \phi^* \qquad \mbox{with} \qquad
\wt{\wt \phi} = \phi .
\eeq
Note that now we have a plus sign in the second relation (compare to eq. (7.4))
because $WW^* = {\bf 1}$ for real irreps. As before $\phi$ and $\wt \phi$
transform alike under CP, eq. (7.5). If $D$ has multiplicity $m$ we can form
the vectors $\chi$, $\wt \chi$. Then eq. (7.7) is valid and CP can be defined
as in eq. (7.8). But now the restriction on $H$ differs from eq. (7.11).
Imposing the same condition as in the pseudoreal case (see eq. (7.10))
but taking into account the properties of $U$ and $W$ in the real
case we obtain \cite{eck81}
\beq
H = \left( \ba{cc} A & B \\ B^* & A^* \ea \right).
\eeq
It is easy to check that the matrices $H$ of the form eq. (7.21)
are exactly those unitary matrices for which
\beq
H^T I H = I \qquad \mbox{with} \qquad
I = \left( \ba{cc} 0 & {\bf 1} \\ {\bf 1} & 0 \ea \right)
\eeq
is valid. Therefore they form a group. This group is identical with $O(2m)$
up to the basis transformation
\beq
Z^\dg H Z = H' \in O(2m) \qquad \mbox{with} \qquad
Z = \frac{1}{\sqrt{2}} \left( \ba{cr} {\bf 1} & i{\bf 1} \\
{\bf 1} & - i{\bf 1} \ea \right).
\eeq
This can readily be seen by inserting eq. (7.23) into eq. (7.22) where we
obtain
\beq
H'{}^T(Z^T I Z)H' = Z^T I Z = {\bf 1}_{2m}.
\eeq
Defining real fields $\chi_1$, $\chi_2$ via
\beq
V^\dg \chi = \frac{\chi_1 + i \chi_2}{\sqrt{2}}
\eeq
the CP transformation (7.8) for the potentially real irrep $D$ is
now of the form
\beq
CP: \left( \ba{c} \chi_1 \\ \chi_2 \ea \right) \ra
(Z^\dg H Z)(V^\dg U V^*) \left( \ba{r} \chi_1 \\ -\chi_2 \ea \right)
\eeq
after some calculation. Both matrix products embraced by the parentheses
are real. This again shows the consistency of the formalism. Clearly,
both $\phi$ and $\wt \phi$ have to be coupled in the Yukawa sector to
get the most general interaction.

It may look strange that real scalars are packed together in complex
fields. But we will see now by an example from $SU(2)_L \times SU(2)_R
\times U(1)_{B-L}$ \cite{LR,sen1,sen} that such cases are not uncommon.
There a multiplet $\phi_m$ transforming as (2,2,0) exists which gives
masses to the quarks. It is commonly written as a $2 \times 2$ matrix
of complex fields and transforms as
\beq
\phi_m = \left( \ba{cc} \phi_{11} & \phi_{12} \\ \phi_{21} & \phi_{22}
\ea \right) \ra U_L \phi_m U_R^\dg, \qquad
U_{L,R} \in SU(2)
\eeq
under the gauge group. The index $m$ denotes the $2 \times 2$
matrix version of the field. $\wt \phi_m$ is given by
\beq
\wt \phi_m = \tau_2 \phi_m^* \tau_2 =
\left( \ba{rr} \phi_{22}^* & - \phi_{21}^* \\ - \phi_{12}^* &
\phi_{11}^* \ea \right)
\eeq
where $\tau_2$ is the second Pauli matrix. Switching to a vector
notation we read off $W$ from eq. (7.27):
\beq
\phi \equiv \left( \ba{c} \phi_{11} \\ \phi_{21} \\ \phi_{12} \\
\phi_{22} \ea \right), \qquad
\wt \phi = W \phi^* \qquad \mbox{with} \qquad
W = \left( \ba{crrc} 0 & 0 & 0 & 1 \\ 0 & 0 & -1 & 0 \\
0 & -1 & 0 & 0 \\ 1 & 0 & 0 & 0 \ea \right).
\eeq
Since $W^T = W$ we suspect immediately that (2,2,0) is a real rep. This
can indeed be confirmed by recalling the following theorem.

\paragraph{Theorem:} The irreps of a direct product of groups $G \times
G'$ are exactly the tensor products of irreps $D$ and $D'$ of $G$ and
$G'$, respectively \cite{shaw}. If $D$ and $D'$ are both real or both
pseudoreal then $D \otimes D'$ is real \cite{urb}.

In our case the defining irrep of $SU(2)$ is pseudoreal having spin
$j = 1/2$ and therefore the rep (7.27) actually decays into two real
irreps. Since this is an instructive example we want to show this
explicitly.

With
\beq
U_L = \left( \ba{cl} a & b \\ -b^* & a^* \ea \right), \quad
|a|^2 + |b|^2 = 1 \quad \mbox{and} \quad
U_R = \left( \ba{cl} c & d \\ - d^* & c^* \ea \right), \quad
|c|^2 + |d|^2 = 1
\eeq
we find the corresponding transformations for the vector $\phi$
\beqa
U_L \phi_m &\ra& \left( \ba{cc} U_L & 0 \\ 0 & U_L \ea \right) \phi
\equiv \wh U_L \phi, \no \\
\phi_m U_R^\dg &\ra& \left( \ba{rrcl} c^* & 0 & d^* & 0 \\
0 & c^* & 0 & d^* \\ -d & 0 & c & 0 \\ 0 & -d & 0 & c \ea \right) \phi
\equiv \wh U_R \phi.
\eeqa
Of course, now we have $W \wh U^*_{L,R} W^\dg = \wh U_{L,R}$.
If we perform a basis transformation we get
\beq
\phi = Z \phi' \Ra \wh U'_{L,R} = Z^\dg \wh U_{L,R} Z \qquad \mbox{and}
\qquad W' = Z^\dg W Z^*.
\eeq
The last relation can easily be obtained from eq. (7.13). It is obvious
that in the basis where $W' = {\bf 1}$ we have real representation
matrices. With
\beq
V \equiv Z = \frac{1}{\sqrt{2}} \left( \ba{crrr} 1 & i & 0 & 0 \\
0 & 0 & -i & 1 \\ 0 & 0 & -i & -1 \\ 1 & -i & 0 & 0 \ea \right)
\eeq
we have indeed $W' = {\bf 1}$ and
\beqa
\wh U'_L &\ra& \left( \ba{rrrc} a_1 & - a_2 & b_2 & b_1 \\
a_2 & a_1 & -b_1 & b_2 \\ - b_2 & b_1 & a_1 & a_2 \\
- b_1 & - b_2 & - a_2 & a_1 \ea \right), \qquad
\ba{ll}
a_1 = \mbox{Re } a, & a_2 = \mbox{Im } a \\
b_1 = \mbox{Re } b, & b_2 = \mbox{Im } b
\ea  \no \\
\wh U'_R &\ra& \left( \ba{rrrr} c_1 & c_2 & -d_2 & -d_1 \\
- c_2 & c_1 & -d_1 & d_2 \\  d_2 & d_1 & c_1 & c_2 \\
 d_1 & - d_2 & - c_2 & c_1 \ea \right), \qquad
\ba{ll}
c_1 = \mbox{Re } c, & c_2 = \mbox{Im } c \\
d_1 = \mbox{Re } d, & d_2 = \mbox{Im } d. \ea
\eeqa
Clearly, the new field multiplet $\phi'$ $(\wt \phi' = {\phi'}^*)$ is
still complex and we can split it into two real multiplets as in eq. (7.25).

This concludes our discussion of real scalars. The left--right symmetric
example will be taken up once more in sect. 9 where CP will be discussed
together with C and P in this context.

\subsection{The general case}
In the general case the scalar multiplets may contain real, pseudoreal
and complex irreps. We put them together into a vector
\beq
\Phi = \left( \ba{c} \phi_R \\ \phi_P \\ \wt \phi_P \\ \phi_C \\ \phi^*_C
\ea \right) \qquad \mbox{and} \qquad
D_\mu \Phi = (\partial_\mu + ig \T_a W_\mu^a) \Phi
\eeq
where the indices $R$, $P$ and $C$ denote real, pseudoreal and complex,
respectively. $D_\mu$ is the covariant derivative. Therefore we can write
the scalar kinetic and gauge Lagrangian as
\beq
\cL_H = \frac{1}{2} (D_\mu \Phi)^\dg (D^\mu \Phi).
\eeq
In this way we summarize the gauge interactions into one formula with
automatically correct factors 1/2 for $\phi_R$ and 1 for $\phi_P$ and
$\phi_C$ since the contribution of $\phi_P$ and $\wt \phi_P$ are equal
in $\cL_H$ (see eq. (7.7)) and the same holds true for $\phi_C$ and $\phi^*_C$
(of course, if in the $\phi_C$ sector $\T_a$ is given by $\T_a^C$ then
for $\phi^*_C$ it has to be $- (\T_a^C)^T)$.

In analogy to the fermionic sector we define a CP--type transformation by
\beq
\Phi(x) \ra U_H \Phi(\wh x)^*
\eeq
leading to
\beqa
\mbox{Condition B}_{\rm H}: &  U_H(- \T_b^T R_{ab})U_H^\dg = \T_a .
\eeqa
Clearly, Condition B$_{\rm H}$ is connected to Condition B
by the same automorphism
represented by $R$. For the total Lagrangian a CP--type transformation
is characterized by the triple $(R,U,U_H)$.

It is clear from eq. (7.37) that it is always possible to define a
CP transformation in $\cL_H$ in the same way as is done in the fermionic
sector. Therefore as for CP--type transformations it remains to
consider the automorphisms $\psi_R = id$ and $\psi_R = \psi_d$ in
the relevant cases.

When considering the algebras $su(\ell+1) \: (\ell \geq 2)$, $so(2\ell) \:
(\ell = 5,7,9,\ldots)$ and $cE_6$ in the general discussion of a CP--type
transformation the automorphism $\psi_R = id$ is relevant in addition to
$\psi^\triangle$. Obviously, if we consider the scalars $\phi_C$ in
complex irreps we automatically have their complex conjugate irreps in
in $\cL_H$ eq. (7.35) so that Condition B$_{\rm H}$ can be solved
for $\psi_R = id$ and complex reps. For real irreps and $\psi_R = id$
Condition B$_{\rm H}$ becomes trivial because in this case $-D^T = D$
in a basis where the fields are real. In the case of pseudoreal irreps
and $\psi_R = id$ the CP--type transformation has to be defined as
$\phi \ra W \phi^*$ with $W$ specified by eq. (7.1). A simple calculation
reveals that now eqs. (7.5), (7.8) and (7.11) are valid with $U$ replaced
by $W$. This shows that for all simple algebras except $so(2\ell)$ with
$\ell = 4,6,8,\ldots$ (see table 1) arbitrary CP--type transformations
are always symmetries of $\cL_H$ and in this sense the existence of
a CP--type symmetry for $\cL_H$ is more likely than for $\cL_F$.

For the simple Lie algebras $so(2\ell)$ with $\ell = 4,6,8,\ldots$
the relevant case is $\psi_R = \psi_d$, a diagram automorphism.
In this case we have $(- D_\Lambda^T) \circ \psi_d \sim D_\Lambda \circ
\psi_d \not\sim
D_\Lambda$ if and only if $n_{\ell-1} \neq n_\ell$ for the highest weight
$\Lambda$ $(\ell \geq 6)$ (see subsect. 4.3). In this special case
a CP--type symmetry may not exist for $\cL_H$ since here it
requires that both irreps $D_\Lambda$ and $D_\Lambda \circ \psi_d$
occur in $\Phi$.
The more complicated
structure of $U_H$ in this case and also for similar cases in $so(8)$
is easily worked out with the methods established in this paper.
If $(- D_\Lambda^T) \circ \psi_d \sim D_\Lambda$ and $\psi_d^2 = id$
again the considerations in subsect. 7.1 can be used with
$\psi^\triangle$ replaced by $\psi_d$.

Finally, we want to mention that basis transformations are performed as
in the fermionic case eqs. (3.25) and (3.26)
but attention has to be paid to two
points. In the real case the horizontal part of $U_H$ is real apart
from situations discussed in subsect. 7.2. Therefore the corresponding
basis transformations have to be performed by orthogonal matrices.
Similarly, in the pseudoreal case the transformation matrix must be
an element of $Sp(2m)$. These points will be of significance in
subsect. 8.3.

\section{Yukawa couplings and CP--type symmetries}
\renewcommand{\theequation}{\arabic{section}.\arabic{equation}}
\setcounter{equation}{0}

\subsection{The condition on Yukawa couplings}

With the vector of scalar field $\Phi$ as defined in eq. (7.35) and all
fermionic degrees of freedom in a right--handed vector $\omega_R$ the
most general form of Yukawa couplings is given by
\beq
\cL_Y = \frac{1}{2} i \omega_R^T C^{-1} \Gamma_j \omega_R \Phi_j + h.c.
\eeq
where Fermi statistics implies $\Gamma_j^T = \Gamma_j$ $\forall \; j$.
A CP--type transformation characterized by $(R,U,U_H)$ leads to
\beqa
\mbox{Condition C:} &
\qquad U^T \Gamma_k U U_{kj}^H = \Gamma_j^* \qquad \forall \; j
\eeqa
if we require invariance of $\cL_Y$.
As a matter of convention we have pulled out the factor
$\frac{1}{2} i$ from the
coupling matrices $\Gamma_j$. The $i$ gives plus signs on both sides
of eq. (8.2) whereas the factor 1/2 is motivated by the fact that
$\chi_R^T C^{-1} \Gamma_j \chi'_R = \chi'{}^T_R C^{-1} \Gamma_j
\chi_R$, i.e. terms with different multiplets $\chi_R$,
$\chi'_R$ occurring in $\omega_R$ appear twice in $\cL_Y$, eq. (8.1).

Before we continue the discussion of Condition C we want to make a few general
remarks on CP--type symmetries which we define as those CP--type
transformations which fulfill simultaneously
Conditions A, B, B$_{\rm H}$ and C.
First we have the following statements:
\beq
(R,U,U_H) \mbox{ CP--type symmetry} \Longleftrightarrow
(R^T,U^T,U_H^T) \mbox{ CP--type symmetry.}
\eeq
If $(R_i,U_i,U_i^H)$ $(i = 1,2)$ are CP--type symmetries then
\beqa
\omega_R(x) &\ra& - U_1 U_2^* \omega_R(x) \no \\
\Phi(x) &\ra& U_1^H U_2^{H*} \Phi(x) \\
W_\mu^a(x) &\ra& (R_1 R_2)_{ab} W_\mu^b(x) \no
\eeqa
is a symmetry of $\cL_G + \cL_F + \cL_H + \cL_Y$. We will see in sect. 9
that eq. (8.4) is a C--type symmetry. As mentioned in the introduction we
will not discuss the Higgs potential the inclusion of which would give
conditions additional to Conditions A -- C. It is clear that theorem II also
applies to Condition C.

The coupling matrices $\Gamma_j$ have two types of indices since they have
to tie together irreps and their multiplicities. Therefore, considering
matrices which couple irreps $D_r$, $D_{r'}$ and $D_{r_\phi}$ with
multiplicities $m_r$, $m_{r'}$ and $m_{r_\phi}$, respectively, where
$D_r$, $D_{r'}$ occur in $\omega_R$ and $D_{r_\phi}$ in $\Phi$ we can
write $\Gamma_j$ as a tensor product
\beq
\Gamma_j = (( \Gamma^a \otimes \gamma^k)_{rr'}^{r_\phi})
\eeq
with $j$ corresponding to the triple $(a,k,r_\phi)$. Note that
$a = 1,\ldots,m_{r_\phi}$, $k = 1, \ldots, d_{r_\phi}$,
$(\gamma^k)_{rr'}^{r_\phi}$ are $d_r \times d_{r'}$ matrices and
$(\Gamma^a)_{rr'}^{r_\phi}$ $m_r \times m_{r'}$ matrices. Of course,
for pseudoreal scalars there are separate coupling matrices for $\chi$
and $\wt \chi$. Clearly, the individual factors of the coupling
matrices (8.5) do not have
to be symmetric, only the total matrix has to be. For $r = r'$, the
diagonal of $(\Gamma^a)_{rr}^{r_\phi}$ couples the same fermion fields
to each other and therefore the irrep $D_{r_\phi}$ has to be contained
in the symmetric tensor product $(D_r \otimes D_r)_{\rm sym}$.

In the further discussion we will confine ourselves to
generalized CP invariance and use the following
approach which we think is most appropriate in the context of this
work. We know already from sect. 5 that CP can be defined in a canonical
way for each irrep in $\omega_R$ and $\Phi$, respectively, and that a
general CP symmetry is composed of the canonical CP transformation
(CP)$_0$ eq. (5.5) followed by a horizontal unitary rotation (5.8).
Therefore we adopt the strategy that out of the set of CP symmetries
of $\cL_G + \cL_F + \cL_H$ we have given a particular one which then
imposes Condition C on the Yukawa couplings, i.e., in our strategy
the symmetry is primary and determines the Lagrangian.

We will see in subsect. 8.2 that in the CP basis
the Clebsch--Gordan coefficients
are real. Therefore, $(\gamma^k)_{rr'}^{r_\phi}$, the group--theoretical
part of $\Gamma_j$, couples the fields in a (CP)$_0$--invariant way and
it remains a purely horizontal condition which will be solved in general
in subsect. 8.3.

\subsection{Real Clebsch--Gordan coefficients and the generalized CP
condition}
It is easy to see that in a tensor product of two irreps $D$, $D'$ the
Clebsch--Gordan coefficients can be chosen real if we take the tensor
product of the respective CP bases. One only has to remember the
general procedure for deducing the Clebsch--Gordan series. We additionally
imagine that we are in a CP basis where all $D(H_j)$ are diagonal. This
is possible because the representation matrices of the CSA can
simultaneously be diagonalized by an orthogonal matrix $(D(H_j)$ is
symmetric and real) without disturbing symmetry or antisymmetry of the
$D(X_a)$ $(a = 1,\ldots,n_G)$. Therefore the basis vectors of
$D \otimes D'$ given by $\{ e_i \otimes e'_j|\: i=1,\ldots,d;j=1,\ldots,d'\}$
are all eigenvectors of $D(H_j) \otimes {\bf 1}_{d'} + {\bf 1}_d \otimes
D'(H_j)$ $(e_i,e'_j$ are the canonical basis vectors of ${\bf C}^d$,
${\bf C}^{d'}$, respectively). We can e.g. choose $e_1 \otimes e'_1$
to be the unique vector with highest weight $\Lambda + \Lambda'$ in
$D \otimes D'$, if $\Lambda$, $\Lambda'$ are the highest weights of
$D$, $D'$, respectively. Applying all $D(e_{-\alpha})$, $\alpha \in
\Delta$ or, equivalently, all $D(e_{-\alpha_j})$ $(j = 1,\ldots,\ell)$
with $\alpha_j$ simple to $e_1 \otimes e'_1$ gives the representation
space associated with $\Lambda + \Lambda'$. Because of eq. (5.13) the basis
of this space is real and therefore also the basis of its orthogonal
complement can be chosen real. The highest weight in the orthogonal
complement can have multiplicity larger than one. We choose a (real) basis
vector and make the same procedure as before. We continue along these
lines until we have exhausted the whole space ${\bf C}^d \otimes
{\bf C}^{d'}$ \cite{corn}. Since all $D(e_{-\alpha})$ are real and we
started with a real basis we finally arrived at a real basis in the
Clebsch--Gordan series. Therefore the Clebsch--Gordan coefficients
associated with this basis are real.

For the rest of this section we assume that we consider fixed irreps
$D_r$, $D_{r'}$, $D_{r_\phi}$ with field multiplets $\chi_R$,
$\chi'_R$, $\phi$, respectively. Then one can readily check that
\beq
(CP)_0 :\; \chi_R \ra - C \chi_R^*, \;
\chi'_R \ra - C \chi'{}^*_R, \; \phi \ra \phi^*
\eeq
is a symmetry of
\beq
i \chi_R^T C^{-1} \gamma^k \chi'_R \phi_k + h.c.
\eeq
for real Clebsch--Gordan matrices $\gamma^k$ since the invariance condition
is just
\beq
i \chi_R^T C^{-1} \gamma^k \chi'_R \phi_k \st{(CP)_0}{\longrightarrow}
- i \chi'{}^\dg_R C \gamma^{kT} \chi_R^* \phi_k^* =
(i \chi_R^T C^{-1} \gamma^k \chi'_R \phi_k)^\dg
\eeq
or $\gamma^{kT} = \gamma^{k \dg}$. Consequently, taking into account the
multiplicities $m_r$, $m_{r'}$, $m_{r_\phi}$ of the respective irreps
a generalized CP transformation given by
\beq
CP: \; \chi_R \ra - U C \chi_R^*, \; \chi'_R \ra - U' C \chi'{}^*_R,
\; \phi \ra H \phi^*
\eeq
leads to the condition
\beq
U^T \Gamma^b U' H_{ba} = \Gamma^{a*}
\eeq
in the horizontal spaces on which $U$, $U'$, $H$ act. Since CP does not
connect different irreps the above discussion is also fully general.
Of course, the canonical CP symmetry $U = {\bf 1}$, $U' = {\bf 1}'$,
$H = {\bf 1}_\phi$ would simply require real matrices $\Gamma^a$. Thus
we have achieved to separate group indices and horizontal indices in
the case of generalized CP invariance.
The classes of solutions of eq. (8.10) will be discussed
in the following subsection.

\subsection{Solutions of the generalized CP condition}
The discussion of eq. (8.10) is greatly simplified by the freedom of choosing
suitable bases in the horizontal spaces. As in the previous subsection
we stick to fixed irreps $D_r$, $D_{r'}$, $D_{r_\phi}$ and therefore
we have three independent basis transformations (3.26) represented by
the unitary matrices $Z$, $Z'$, $Z_\phi$. According to the theorem proved
in ref. \cite{eck87} one can find $Z$ and $Z'$ such that
\beqa
Z^\dg U Z^* &=& \mbox{diag }(O(\Theta_1), \ldots,O(\Theta_k),{\bf 1}_p),
\qquad 0 < \Theta_\nu \leq \frac{\pi}{2} \no \\
Z'{}^\dg U' Z'{}^* &=& \mbox{diag }(O(\Theta'_1),\ldots,O(\Theta'_{k'}),
{\bf 1}_{p'}),
\qquad 0 < \Theta'_\nu \leq \frac{\pi}{2}
\eeqa
with
\beq
O(\vt) \equiv \left( \ba{rc} \cos \vt & \sin \vt \\
- \sin \vt & \cos \vt \ea \right).
\eeq
In the Higgs sector we have to distinguish the cases as in sect. 7. For a
complex scalar multiplet the above theorem applies yielding
\beq
Z^\dg_\phi H Z^*_\phi = \mbox{diag }(O(\Theta^H_1),\ldots,
O(\Theta^H_{k_H}),{\bf 1}_{p_H}), \qquad 0 < \Theta_\nu^H \leq
\frac{\pi}{2}.
\eeq
For a real scalar $H$ is real and $Z_\phi$ has to be orthogonal. Then
the real version of the spectral theorem for normal operators tells
that one can achieve
\beq
Z_\phi^T H Z_\phi = \mbox{diag }(O(\Theta_1^H),\ldots,O(\Theta^H_{k_H}),
- {\bf 1}_{p_-}, {\bf 1}_{p_+}), \qquad
0 < \Theta^H_\nu < \pi.
\eeq
The case of a pseudoreal scalar requires $Z_\phi \in Sp(2m)$. In app. G
we prove that for any $H \in Sp(2m)$ there is such a $Z_\phi$ giving
\beq
Z_\phi^\dg H Z_\phi^* = \left( \ba{cc|cc}
0 & 0 & D & 0 \\
0 & {\bf 1}_{p_H} & 0 & 0 \\ \cline{1-4}
- D^* & 0 & 0 & 0 \\
0 & 0 & 0 & {\bf 1}_{p_H}
\ea \right)
\eeq
with $D = \mbox{diag }(d_1,\ldots,d_{k_H})$, $|d_\nu| = 1$.

These basis choices allow to solve eq. (8.10) in a piecewise manner with
submatrices of $\Gamma^a$ of maximal size $2 \times 2$ and at most two
different $\Gamma^a$ involved at a time. We will denote these
submatrices by $A$ or $A_1$, $A_2$ and indicate by arrows which part
of $Z^\dg U Z^*$, $Z'{}^\dg U' Z'{}^*$, $Z_\phi^\dg H Z_\phi^*$ is
under discussion.

The number of different cases to be discussed is reduced by the following
observations. Solutions of the cases with $H \ra -1$ are obtained from
those with $H \ra 1$ by multiplying $A$ or $A_1$, $A_2$ by $i$. The
solutions of the cases $U \ra 1$, $U' \ra O(\Theta')$ are obtained
from those with $U \ra O(\Theta)$, $U' \ra 1$ by transposition of $A$
or $A_1$, $A_2$. This leaves the following nine generic cases to be
investigated:
\begin{enumerate}
\item[1)] $U \ra 1$, $U' \ra 1 \Ra A,A_1,A_2$ are $1 \times 1$ matrices
\item[1a)] $H \ra 1$: $A = A^*$
\item[1b)] $H \ra O(\Theta_H)$: $\ba{l} A_1 \cos \Theta_H -
A_2 \sin \Theta_H = A_1^* \\
A_1 \sin \Theta_H + A_2 \cos \Theta_H = A_2^* \ea$
\item[1c)] $H \ra h(d)$: $\ba{r} A_1 d = A_2^* \\
 - A_2 d^* = A_1^* \ea$
\item[2)] $U \ra O(\Theta)$, $U' \ra 1 \Ra A, A_1, A_2$ are $2 \times 1$
matrices
\item[2a)] $H \ra 1$: $O(\Theta)^TA = A^*$
\item[2b)] $H \ra O(\Theta_H)$: $\ba{l}
O(\Theta)^T(A_1 \cos \Theta_H - A_2 \sin \Theta_H) = A_1^* \\
O(\Theta)^T(A_1 \sin \Theta_H + A_2 \cos \Theta_H) = A_2^* \ea$
\item[2c)] $H \ra h(d)$: $\ba{r}
 O(\Theta)^T A_1 d = A_2^* \\
-O(\Theta)^T A_2 d^* = A_1^* \ea $
\item[3)] $U \ra O(\Theta)$, $U' \ra O(\Theta') \Ra A, A_1, A_2$ are
$2 \times 2$ matrices
\item[3a)] $H \ra 1$: $O(\Theta)^T A O(\Theta') = A^*$
\item[3b)] $H \ra O(\Theta_H)$: $\ba{l}
O(\Theta)^T (A_1 \cos \Theta_H - A_2 \sin \Theta_H)O(\Theta') = A_1^* \\
O(\Theta)^T (A_1 \sin \Theta_H + A_2 \cos \Theta_H)O(\Theta') = A_2^*
\ea $
\item[3c)] $H \ra h(d)$: $\ba{r}
O(\Theta)^T A_1 O(\Theta')d = A_2^* \\
-O(\Theta)^T A_2 O(\Theta')d^* = A_1^* . \ea $
\beq {} \eeq
\end{enumerate}
In all the cases we have
$$
0 < \Theta, \; \Theta' \leq \frac{\pi}{2}, \qquad
0 < \Theta_H < \pi, \qquad h(d) = \left( \ba{cc} 0 & d \\
-d^* & 0 \ea \right) \mbox{ with } |d| = 1.
$$
In the following the solutions for $A$ or $A_1$, $A_2$ of 1a) -- 3c) are
given as functions of $\Theta$, $\Theta'$, $\Theta_H$ or $d$ according to
the spirit of our strategy outlined at the end of subsect. 8.1. The methods
used thereby can be found in app. H. Some cases in the above list
correspond to each other like 1b) -- 2a) and 2b) -- 3a). They are,
however, kept apart for the sake of clearness. In the following list of
solutions $\ve$ denotes the two options $\pm 1$.

\paragraph{Solutions:}
\begin{enumerate}
\item[1a)] $A \in {\bf R}$
\item[1b)] $A_1 = A_2 = 0$
\item[1c)] $d = i \ve \Ra A_2 = - dA_1^*$, $A_1 \in {\bf C}$

$d^2 \neq -1 \Ra A_1 = A_2 = 0$
\item[2a)] $A = 0$
\item[2b)] $\Theta_H = \Theta = \dfrac{\pi}{2} \Ra A_2 = \left(
\ba{cr} 0 & -1 \\ 1 & 0 \ea \right)A_1^*$, $A_1 \in {\bf C}^2$

$\Theta_H = \Theta \neq \dfrac{\pi}{2} \Ra A_2 = \left(
\ba{cr} 0 & -1 \\ 1 & 0 \ea \right)A_1$, $A_1 \in {\bf R}^2$

$\Theta_H = \pi - \Theta \neq \dfrac{\pi}{2} \Ra A_2 = \left(
\ba{rc} 0 & 1 \\ -1 & 0 \ea \right)A_1$, $iA_1 \in {\bf R}^2$

$\Theta_H \not\in \{ \Theta, \: \pi - \Theta \} \Ra A_1 = A_2 = 0$

\item[2c)] $\Theta = \dfrac{\pi}{2}$, $d = \ve \Ra A_2 =
\ve \left( \ba{cr} 0 & -1 \\ 1 & 0 \ea \right) A_1^*$,
$A_1 \in {\bf C}^2$

$\Theta \not= \dfrac{\pi}{2}$, $d = i\ve  e^{\pm i\Theta} \Ra A_1 =
a \left( \ba{r} 1 \\ \pm i \ea \right)$,
$A_2 = -i \ve A_1^*$, $a \in {\bf C}$

$d^2 \neq - e^{\pm 2i\Theta} \Ra A_1 = A_2 = 0$

\item[3a)] $\Theta = \Theta' = \dfrac{\pi}{2} \Ra A =
\left( \ba{cl} a & b \\ -b^* & a^* \ea \right)$, $a,b \in {\bf C}$

$\Theta = \Theta' \neq \dfrac{\pi}{2} \Ra A =
\left( \ba{rc} a & b \\ -b & a \ea \right)$, $a,b \in {\bf R}$

$\Theta \neq \Theta' \Ra A = 0$

\item[3b)] $\Theta_H = \Theta + \Theta' = \dfrac{\pi}{2} \Ra A_1 =
\left( \ba{cr} a & b \\ b & -a \ea \right)$,
$A_2 = \left( \ba{rc} -b & a \\ a & b \ea \right)^*$, $a,b \in {\bf C}$

$\cos \Theta_H = \ve \sin \Theta'$, $\Theta = \dfrac{\pi}{2}$,
$\Theta' \not= \dfrac{\pi}{2} \Ra A_1 =
\left( \ba{cc} a & b \\ -\ve b^* & \ve a^* \ea \right)$,
$A_2 = \left( \ba{cc} \ve b & - \ve a \\ a^* & b^* \ea \right)$
$a,b \in {\bf C}$

$\cos \Theta_H = \ve \sin \Theta$, $\Theta' = \dfrac{\pi}{2}$,
$\Theta \not= \dfrac{\pi}{2} \Ra A_1 =
\left( \ba{cc} a & b \\ -\ve b^* & \ve a^* \ea \right)$,
$A_2 = \left( \ba{lc} - b^* &  a^* \\ - \ve a & -\ve b \ea \right)$
$a,b \in {\bf C}$

$\cos \Theta_H = - \ve \cos(\Theta + \Theta')$, $\Theta \not= \dfrac{\pi}{2}$,
 $\Theta' \not= \dfrac{\pi}{2}$, $\Theta + \Theta' \neq \dfrac{\pi}{2} \Ra$

$ \left.
\ba{lll} A_1 = i \left( \ba{cr} a & b \\ b & -a \ea \right),
& A_2 = -i \left( \ba{rc} -b & a \\ a & b \ea \right)
& \mbox{ for } \ve = 1 \\
 A_1 =  \left( \ba{cr} a & b \\ b & -a \ea \right),
& A_2 =  \left( \ba{rc} -b & a \\ a & b \ea \right)
& \mbox{ for } \ve = -1
\ea \right\} a,b \in {\bf R}$

$\cos \Theta_H =  \ve \cos(\Theta - \Theta')$, $\Theta \not= \dfrac{\pi}{2}$,
$\Theta' \not= \dfrac{\pi}{2}$, $\Theta + \Theta' \neq \dfrac{\pi}{2} \Ra$

$ \left.
\ba{lll} A_1 =  \left( \ba{rc} a & b \\ -b & a \ea \right),
& A_2 = \eta \left( \ba{rr} -b & a \\ -a & -b \ea \right) &
\mbox{ for } \ve = 1 \\
 A_1 = i \left( \ba{rc} a & b \\ -b & a \ea \right),
& A_2 = - i \eta \left( \ba{rr} -b & a \\ -a & -b \ea \right)
& \mbox{ for } \ve = -1
\ea \right\}
\ba{l} a,b \in {\bf R} \mbox{ and}\\
\eta = \mbox{sgn}(\Theta' - \Theta) \ea $

All other choices of $\Theta$, $\Theta'$, $\Theta_H$ lead to $A_1 = A_2
= 0$.
\item[3c)]
$\Theta = \Theta' = \dfrac{\pi}{2}$, $d = i \ve \Ra
A_1 = \left( \ba{cc} a & b \\ c & d \ea \right)$,
$A_2 = - i \ve \left( \ba{rr} d^* & - c^* \\ -b^* & a^* \ea \right)$,
$a,b,c,d \in {\bf C}$

$\Theta = \Theta' \not= \dfrac{\pi}{2}$, $d = i \ve \Ra
A_1 = \left( \ba{cc} a & b \\ -b & a \ea \right)$,
$A_2 = - i \ve \left( \ba{rr} a^* &  b^* \\ -b^* & a^* \ea \right)$,
$a,b \in {\bf C}$

$\Theta + \Theta' = \dfrac{\pi}{2}$, $d = \ve \Ra
A_1 = \left( \ba{cr} a & b \\ b & -a \ea \right)$,
$A_2 =  \ve \left( \ba{rr} -b^* & a^* \\ a^* & b^* \ea \right)$,
$a,b \in {\bf C}$

$\Theta' = \dfrac{\pi}{2}$, $\Theta \not= \dfrac{\pi}{2}$, $d =
\ve e^{-i\Theta} \Ra
A_1 = \left( \ba{rr} a & ib \\ -ia & b \ea \right)$,
$A_2 = \ve \left( \ba{rr} ib^* & a^* \\ -b^* & ia^* \ea \right)$,
$a,b \in {\bf C}$

$\Theta = \dfrac{\pi}{2}$, $\Theta' \not= \dfrac{\pi}{2}$, $d =
\ve e^{-i\Theta'} \Ra
A_1 = \left( \ba{rr} a & -ia \\ ib & b \ea \right)$,
$A_2 = \ve \left( \ba{rr} ib^* & -b^* \\ a^* & ia^* \ea \right)$,
$a,b \in {\bf C}$

$\Theta = \dfrac{\pi}{2}$, $\Theta' \not= \dfrac{\pi}{2}$, $d =
\ve e^{i\Theta'} \Ra
A_1 = \left( \ba{rr} a & ia \\ -ib & b \ea \right)$,
$A_2 = \ve \left( \ba{rr} -ib^* & -b^* \\ a^* & -ia^* \ea \right)$,
$a,b \in {\bf C}$

$\Theta' = \dfrac{\pi}{2}$, $\Theta \not= \dfrac{\pi}{2}$, $d =
\ve e^{i\Theta} \Ra
A_1 = \left( \ba{rr} a & -ib \\ ia & b \ea \right)$,
$A_2 = \ve \left( \ba{rr} -ib^* & a^* \\ -b^* & -ia^* \ea \right)$,
$a,b \in {\bf C}$

The remaining four non--trivial cases can be uniformly described in the
following way:
$$
\Theta \neq \frac{\pi}{2}, \qquad \Theta' \neq \frac{\pi}{2}, \qquad
d = i \ve e^{-i(r \Theta + s \Theta')} \Ra
 A_1 = c v_r v_s^T, \qquad A_2 = - i \ve A_1^*
$$
with $r,s = \pm$, $v_\pm = \left( \ba{r} 1 \\ \mp i \ea \right)$,
$c \in {\bf C}$.

For all other choices of $\Theta, \Theta'$ and $d$ we have $A_1 = A_2 = 0$.
\end{enumerate}

This concludes the complete discussion of the generalized CP condition
(8.10). It shows which solutions apart from the trivial one with real
couplings 1a) one can expect. These solutions might be helpful for
model building. Of course, they are bound to the bases introduced in
the beginning of this subsection. If one has additional conditions on
the Yukawa couplings from further symmetries it might not be useful
to work in these bases. An example for such a case is given in
subsect. 9.3.

\section{C--type transformations}
\renewcommand{\theequation}{\arabic{section}.\arabic{equation}}
\setcounter{equation}{0}

\subsection{Charge conjugation}
We have seen in eq. (8.4) that if we carry out two CP--type
transformations one after the other we obtain a transformation of the
type \cite{sla}
\beqa
W_\mu^a(x) &\ra& R_{ab} W_\mu^b(x) \no \\
\omega_R(x) &\ra& U \omega_R(x) \\
\Phi(x) &\ra& U_H \Phi(x). \no
\eeqa
It is reasonable to call eq. (9.1) a C--type transformation since composing
CP with P, both of CP--type, should give charge conjugation. Though the
discussion of CP and P in the previous sections contains implicitly also
C--type transformations it is nevertheless useful to consider their
features in a separate section.

If $\cL_G + \cL_F + \cL_H + \cL_Y$ is invariant under the
transformation (9.1) we have
\beqa
U T_b R_{ba} U^\dg &=& T_a \no \\
U_H \T_b R_{ba} U_H^\dg &=& \T_a \no \\
U^T \Gamma_k U U^H_{kj} &=& \Gamma_j.
\eeqa
As we have learned in subsect. 3.2 the first relation can be interpreted
as $\{T_a\}$ composed with the automorphism $\psi_R$ being equivalent to
$\{T_a\}$. The second relation has the analogous interpretation for the
Higgs fields. It is clear from Schur's lemma that for $R = {\bf 1}$
the matrices $U$ and $U_H$ only act horizontally.

To define a charge conjugation we require as for P and CP that $\psi_R
= \psi_C$ is a non--trivial
involution. Having fixed the CSA $\Ha$ which determines the quantum
numbers of the states in the irreps it is reasonable to require
$\psi_C(\Ha) = \Ha$. As in the case of P
the CSA can be split into $\Ha = \Ha_+ \oplus \Ha_-$
with $\psi_C(X) = \pm X$ for $X \in \Ha_\pm$. We assume that the CSA
of the unbroken part of the SM group is contained in $\Ha_-$, i.e.
$Q_{em}$, $F_3^c$, $Y_c \in \Ha_-$ as any physically viable model must have
$U(1)_{em} \times SU(3)_c \subseteq G$. Thus $\psi_C$ has to flip the
sign of at least three generators in $\Ha$. To every $\psi_C$ there is
associated a subgroup of $G$ generated by the elements of $\cL_C$ with
$\psi_C(X) = X$. In refs. \cite{sla,sla1} there is a complete list
of all such ``symmetric subgroups'' $S$ for simple groups $G$. Since in
this paper we are mainly concerned with the symmetry aspect of the
Lagrangians and not with the embedding of the SM group in $G$ and
future symmetry breaking we refer the reader to the extensive discussion
in ref. \cite{sla} of these questions.

Let the automorphism defining charge conjugation be given by
\beq
\psi_C(H_j) = \rho_j H_j = \left\{ \ba{rl} -H_j, & j = 1,\ldots,p \\
H_j, & j = p+1,\ldots,\ell . \ea \right.
\eeq
Then as for P we can distinguish an internal and external case with
respect to an irrep $D$ if $D \sim D \circ \psi_C$ or
$D \not\sim D \circ \psi_C$, respectively. As before we consider
now the representation space of $D$ in the internal case and the direct
sum of the spaces of $D$ and $D \circ \psi_C$ in the external case.
Then
\beq
D(H_j) U_C e(\lambda,q) = \rho_j \lambda(H_j) U_C e(\lambda,q)
\eeq
for the ON basis $\{e(\lambda,q)\}$ with weights $\lambda$ of $D$. The
situation is analogous to P in eqs. (6.8) and (6.10). All states with
$(\lambda(H_1),\ldots,\lambda(H_p)) \neq (0,\ldots,0)$ have a
counterpart with opposite quantum numbers with respect to $H_1,\ldots,
H_p$. Defining
\beq
\omega_L^{\lambda,q}(x) \equiv C \gamma_0^T (\omega_R^{\lambda_C,q}(x))^*
\eeq
with $\lambda_C$ given as in eq. (6.10) we can perform an analysis as in
subsect. 6.2 and obtain
\beq
\U_C \omega_R^{\lambda,q}(x) \U_C^{-1} = e^{i\delta} C \gamma_0^T
(\omega_L^{\lambda,q}(x))^*.
\eeq
Here we have used that $U_C^2$ is a phase in each irrep of $\{T_a\}$ just
as in the analogous situation in eq. (6.15)\footnote{$\delta$ is the
phase of $U_C^2$ in $D$ or $D \circ \psi_C$ for the internal or
external case, respectively.}. In eq. (9.6) charge conjugation has the
familiar form.

As mentioned before there is a close relationship between C and P
via CP. It is clear that invariance under CP and P is equivalent
to invariance under CP and C$=$CP$\circ$P. This relationship is
given by the identifications $\psi_C = \psi^\triangle \circ \psi_P$,
$\lambda_P = \lambda_C$ and $U_C = -U_P U_{CP}^*$ (see eq. (8.4)).
Once we fix $\psi_{CP} = \psi^\triangle$ and choose a CP basis one can
therefore identify $U_P$ with $U_C$ in the internal case of parity
(see, e.g., the $SO(10)$ example in sect.2).

Like parity, charge conjugation cannot be defined in the SM where
$\psi_C$ would either reverse the sign of the electric charge and
colour charges or of all the four elements of the CSA. In any case the
right--handed singlets have no partners with opposite charges as
required by the presence of $D \circ \psi_C$ in the case of C invariance.

In the light of the discussion in this section we also learn that in
the irreps of $G^*$, eq. (4.10), $D(E)$ would be a candidate for $U_C$ if
$\psi_C = \psi_E$.

\subsection{Compatibility of CP and C}
In this subsection we will show that one can construct Yukawa couplings
which are not only invariant under the canonical CP transformation but
also under a ``canonical'' charge conjugation at the same time. We will
make use of the advantages of the CP basis where (CP)$_0$ is simply
given by eq. (8.6) and where the Yukawa couplings are real.

To define a canonical charge conjugation we confine ourselves to $D$ and
$D \circ \psi_C$ for every irrep $D$ contained in $\{T_a\}$. Given the
involutive automorphism $\psi_C$ we will assume in the following that
either $D \circ \psi_C \sim D$ or $D \circ \psi_C$ is included in $\{T_a\}$
to allow for a definition of C. The whole discussion will be
confined to simple Lie algebras.

It is clear that for $\psi_C \equiv \psi_Y$ inner (see eq. (4.1)) the C
transformation
\beq
\chi_R \ra W \chi_R, \qquad \chi'_R \ra W' \chi'_R, \qquad
\phi \ra W_\phi \phi
\eeq
with
\beq
W = e^{-D(Y)}, \qquad W' = e^{-D'(Y)}, \qquad
W_\phi = e^{-D_\phi(Y)}
\eeq
where the irreps $D$, $D'$, $D_\phi$ are coupled together in $\cL_Y$ is
just a gauge transformation and $\cL_Y$ is obviously invariant under
transformation (9.7). This allows to restrict the further discussion to
$\psi = \psi^\triangle$ or $\psi_d$. In the following the matrices acting on
the minimal number of irreps involved in the Yukawa couplings for a
definition of C will always be called $W$, $W'$, $W_\phi$. They are
part of $U_C$ and $U_C^H$ in eq. (9.1).

In a CP basis $-D^T = D \circ \psi^\triangle$ is valid and therefore, if
$D \sim D \circ \psi^\triangle$, we obtain $W$ real, $W^T = \lambda W$ with
$\lambda = -1$ for $D$ pseudoreal and $\lambda = 1$ for $D$ real
(see eqs. (7.12) and (7.19), respectively). For $W'$, $W_\phi$ the
parameters analogous to $\lambda$ will be denoted by $\lambda'$,
$\lambda_\phi$, respectively. If $D \sim D \circ \psi_d$ it is shown
in app. I that $W$ is real and symmetric, i.e. $\lambda = 1$. As
discussed in sect. 4 we choose $\psi_d$ as the relevant outer
automorphism for $so(2\ell)$
$(\ell = 4,6,8,\ldots)$ whereas in all other cases of
non--trivial outer involutive automorphisms, i.e., the Lie
algebras $su(\ell + 1)$ $(\ell \geq 2)$, $so(2\ell)$ $(\ell = 5,7,9,
\ldots)$ and $cE_6$, we take $\psi^\triangle$ as the relevant
automorphism.

We have to distinguish four cases according to the types of irreps
with respect to $\psi_C$ coupled together in $\cL_Y$:
\begin{enumerate}
\item[a)] $D \sim D \circ \psi_C$, $D' \sim D' \circ \psi_C$,
$D_\phi \sim D_\phi \circ \psi_C$

The $\cL_Y$ of eq. (8.7) transforms as
\beq
i \chi_R^T C^{-1} \gamma^k \chi'_R \phi_k \ra i \chi_R^T C^{-1} W^T
\gamma^\ell W' \chi'_R W_{\phi \ell k} \phi_k
\eeq
under the C transformation (9.7). In case a) we always have
\beq
\lambda \lambda' \lambda_\phi = 1.
\eeq
This is so for two reasons. First, for $\psi_d$ the $W$'s are symmetric
and all $\lambda$'s are 1. Second, as for $\psi^\triangle$ one can show that in
a tensor product of real or pseudoreal irreps of connected semisimple
Lie groups pseudoreal irreps do not occur. Also if one factor is real and
the other one pseudoreal then real irreps do not occur \cite{urb} and
thus eq. (9.10) is valid. Consequently, we can replace $\gamma^k$ by
\beq
\gamma^k \ra \gamma^k_\pm \equiv \gamma^k \pm W \gamma^\ell W'{}^T
(W_\phi^T)_{\ell k} \quad \forall k
\eeq
with the C transformation given by
\beq
aW, \; a'W',\; a_\phi W_\phi \qquad \mbox{with} \qquad
aa' a_\phi = \left\{ \ba{rl} 1 & \mbox{ for } + \\
-1 & \mbox{ for } - . \ea \right.
\eeq
Clearly, eq. (9.10) is necessary to have invariance of $\cL_Y$ eq. (8.7) with
$\gamma_\pm^k$ given by eq. (9.11). Since the coupling matrices (9.11) are
real (CP)$_0$ invariance is not spoiled. We do not know if one can take
a specific sign for all simple groups and all irreps. In any case, in any
situation the two sets of coupling matrices $\{\gamma_+^k\}$ and
$\{\gamma_-^k\}$ cannot consist of only zero matrices at the
same time and therefore it is possible to have (CP)$_0$ invariance
{\em and} invariance under the ``canonical'' C transformation given by
eq. (9.12) for $\cL_Y$ redefined with $\{\gamma_+^k\}$ or
$\{\gamma_-^k\}$ in case a).
\item[b)] $D \not\sim D \circ \psi_C$, $D' \not\sim D' \circ \psi_C$,
$D_\phi \not\sim D_\phi \circ \psi_C$

In this case
\beq
W,W',W_\phi = \left( \ba{cc} 0 & {\bf 1} \\ {\bf 1} & 0 \ea \right),
\qquad \chi_R = \left( \ba{c} \chi_{R1} \\ \chi_{R2} \ea \right)
\mbox{ etc.}
\eeq
Then
\beq
i(\chi^T_{R1} C^{-1} \gamma^k \chi'_{R1} \phi^1_k +
\chi^T_{R2} C^{-1} \gamma^k \chi'_{R2} \phi^2_k)
\eeq
is clearly invariant under (CP)$_0$ and C given by eq. (9.13).
\item[c)] $D \sim D \circ \psi_C$, $D' \sim D' \circ \psi_C$,
$D_\phi \not\sim D_\phi \circ \psi_C$

With
\beq
i(\chi^T_R C^{-1} \gamma^k \chi'_R \phi^1_k +
\chi^T_R C^{-1} W \gamma^k W'{}^T \chi'_R \phi^2_k)
\eeq
and
\beq
W_\phi = \left( \ba{cc} 0 & \lambda\lambda'{\bf 1} \\ {\bf 1} & 0 \ea \right),
\qquad \phi = \left( \ba{c} \phi^1 \\ \phi^2 \ea \right)
\eeq
invariance is obtained.
\item[c')] $D' \sim D' \circ \psi_C$, $D_\phi \sim D_\phi \circ
\psi_C$, $D \not\sim D \circ \psi_C$

As before we have
\beq
i(\chi^T_{R1} C^{-1} \gamma^k \chi'_R \phi_k +
\chi^T_{R2} C^{-1} \gamma^\ell W'{}^T \chi'_R (W_\phi^T)_{\ell k}\phi_k)
\eeq
with
\beq
W = \left( \ba{cc} 0 & \lambda'\lambda_\phi{\bf 1} \\ {\bf 1} & 0 \ea \right),
\qquad \chi_R = \left( \ba{c} \chi_{R1} \\ \chi_{R2} \ea \right).
\eeq
\item[d)] $D \not\sim D \circ \psi_C$, $D' \not\sim D' \circ \psi_C$,
$D_\phi \sim D_\phi \circ \psi_C$

Now
\beq
i(\chi^T_{R1} C^{-1} \gamma^k \chi'_{R1} +
\chi^T_{R2} C^{-1} \gamma^\ell(W_\phi^T)_{\ell k} \chi'_{R2})\phi_k
\eeq
is invariant with $(a, a' = \pm 1)$
\beq
W = \left( \ba{cc} 0 & a{\bf 1} \\ {\bf 1} & 0 \ea \right), \qquad
W' = \left( \ba{cc} 0 & a'{\bf 1} \\ {\bf 1} & 0 \ea \right), \qquad
\lambda_\phi = a a'.
\eeq
\end{enumerate}

All remaining cases are to be discussed analogously. Thus we have seen
that for every C transformation such that $D \circ \psi_C$ is
contained in $\{T_a\}$ for every irrep $D$ of $\{T_a\}$ one can write down
Yukawa couplings invariant under canonical C and CP with ``canonical C''
defined in this subsection. This invariance is achieved by a redefinition
of the $\gamma^k$. It is not clear to us whether one could circumvent this
redefinition by directly exploring properties of the Clebsch--Gordan
coefficients.

\subsection{An example with $G = SU(2)_L \times SU(2)_R \times
U(1)_{B-L}$}
We want to close this section with a left--right symmetric example
\cite{LR,sen1,sen} which exhibits a lot of interesting features in the light
of the discussion of discrete symmetries. We consider fermions
$\chi_L$, $\chi_R$ transforming as $(2,1,B-L)$, $(1,2,B-L)$,
respectively, where the numbers indicate the dimension $2j+1$ of an irrep
$D_j$ $(j = 0, 1/2,1,\ldots)$ of $SU(2)$. The scalar $\phi_m$
transforming as (2,2,0) was already introduced in subsect. 7.2 as an example of
two real irreps in complex disguise. For quarks the $U(1)$ charge is
1/3 whereas for leptons it is $-1$. Therefore it is given by baryon
number minus lepton number $(B-L)$. Assuming $n_f$ families in
$\chi_{L,R}$ we have
\beq
\omega_R = \left( \ba{c} (\chi_L)^c \\ \chi_R \ea \right)
\eeq
and
\beq
- \cL_Y = \bar \chi_L \Gamma \phi_m \chi_R +
\bar \chi_L \Delta \wt \phi_m \chi_R + h.c.
\eeq
with $\wt \phi_m$ given by eq. (7.28). $\Gamma$ and $\Delta$ are
$n_f \times n_f$ matrices in family (flavour) space. Note that in eq. (9.22)
both $\phi_m$ and $\wt \phi_m$ have to be coupled to get the most general
Yukawa couplings.

Let us now investigate the effects of the following generalized CP and
C transformations \cite{eck84}:
\beqa
CP &:& \chi_L \ra - C \chi_L^*, \quad \chi_R \ra - i C \chi_R^*, \quad
\phi_m \ra - i \phi_m^* \\
C &: & \chi_L \ra - (\chi_R)^c, \quad \chi_R \ra -i(\chi_L)^c, \quad
\phi_m \ra -i \phi_m^T.
\eeqa
Examining the invariance conditions we obtain
\beqa
CP &\Ra& \Gamma \mbox{ real, $\Delta$ imaginary,} \\
C &\Ra& \Gamma^T = \Gamma, \; \Delta^T = - \Delta .
\eeqa
Taking eqs. (9.25) and (9.26) together we see that $\Gamma$ has to be real
and symmetric and $\Delta$ imaginary and antisymmetric. Here we have
not used the bases in the horizontal space introduced in subsect. 8.3 for CP
because there is the additional C invariance (9.24) whose simple form
would be spoiled in these bases.
Furthermore, eqs. (9.25) and (9.26) together are rather restrictive. It was
shown in ref. \cite{eck84} that the effect of these conditions is similar
to the effect of a horizontal symmetry allowing for relations between
flavour mixing and masses but for $n_f = 3$ the top quark mass comes
out too low.

Interpreting the transformation properties of fermions and $\phi_m$ in
the light of the irreps of $G^*$ in subsect. 4.2 and ignoring
the $U(1)$, $\omega_R$ eq. (9.21) is in the irrep $D_{1/2,0}$ of $SU(2)^*$
and $\phi_m$ contains two $D_{1/2}^+$. Thus $\phi_m$ can be considered
as a tensor product $D_{1/2} \otimes D_{1/2}$ with respect to
$SU(2) \times SU(2)$ with $D(E)$ exchanging
the vectors in the product. Since $D(E)$ is related to the C in
eq. (9.24) it becomes clear that C has the effect of transposition on
$\phi_m$. Combining CP and C we obtain
\beq
P = C \circ CP: \chi_L \ra - \gamma_0 \chi_R, \quad
\chi_R \ra - \gamma_0 \chi_L, \quad \phi_m \ra \phi_m^\dg.
\eeq

Now we want to switch to the formalism discussed in this paper, namely
using $\omega_R$ eq. (9.21) and $\phi$ eq. (7.29) instead of $\phi_m$.
Then the fermion rep has the form
\beq
T_a = \left( \ba{cc} 0 & 0 \\ 0 & \tau_a/2 \ea \right), \quad
T_{3+a} = \left( \ba{cc} -\tau^T_a/2 & 0 \\ 0 & 0 \ea \right),
a = 1,2,3, \quad T_7 = \frac{1}{2}(B-L) \left( \ba{cc} - {\bf 1}_2 & 0 \\
0 & {\bf 1}_2 \ea \right)
\eeq
with Pauli matrices $\tau_a$. The generators $\T_a$ acting on $\phi$ are
obtained from the transformation property of $\phi_m$ eq. (7.27), i.e.,
$\T_a \phi$ $(a = 1,2,3)$ corresponds to $- \phi_m \tau_a/2$ and
$\T_{3+a} \phi$ to $\tau_a \phi_m/2$. Therefore we obtain
\beq
\ba{lll}
\T_1 = - \dfrac{1}{2} \left( \ba{cccc} 0 & 0 & 1 & 0 \\
0 & 0 & 0 & 1 \\ 1 & 0 & 0 & 0 \\ 0 & 1 & 0 & 0 \ea \right), &
\T_2 = - \dfrac{1}{2} \left( \ba{cccc} 0 & 0 & i & 0 \\
0 & 0 & 0 & i \\ -i& 0 & 0 & 0 \\ 0 &-i & 0 & 0 \ea \right), & \\
\T_3 = - \dfrac{1}{2} \left( \ba{cccc} 1 & 0 & 0 & 0 \\
0 & 1 & 0 & 0 \\ 0 & 0 & -1 & 0 \\ 0 & 0 & 0 & -1 \ea \right), &
\T_{3+a} = \frac{1}{2} \left( \ba{cc} \tau_a & 0 \\ 0 & \tau_a \ea
\right), & \T_7 = 0.
\ea
\eeq

Now CP eq. (9.23) is quickly treated.
Acting on $\omega_R$, eq. (3.17) determines
\beq
U_{CP} = \left( \ba{cc} -{\bf 1}_{n_f} & 0 \\
0 & i {\bf 1}_{n_f} \ea \right) .
\eeq
Then the automorphism is fixed by Condition B:
\beq
R_{CP} = \mbox{diag }(-1,1,-1,-1,1-1,-1)
\eeq
is just $\psi^\triangle$ for $SU(2) \times SU(2) \times U(1)$. For $\phi$ we
have $U^H_{CP} = - i {\bf 1}_4$ and a quick look at eq. (9.29) confirms that
$- \T_a^T R_{CPaa} = \T_a$ is fulfilled.

As for charge conjugation eq. (9.24) fixes
\beq
U_C = - \left( \ba{cc} 0 & {\bf 1}_{n_f} \\ i{\bf 1}_{n_f} & 0 \ea \right)
\qquad \mbox{and} \qquad
R_C = \left( \ba{rrrrrrr}
0 & 0 & 0 & -1 & 0 & 0 & 0 \\
0 & 0 & 0 & 0  & 1 & 0 & 0 \\
0 & 0 & 0 & 0  & 0 & -1 & 0 \\
-1& 0 & 0 & 0  & 0 & 0 & 0 \\
0 & 1 & 0 & 0 &  0 & 0 & 0 \\
0 & 0 & -1 & 0 & 0 & 0 & 0 \\
0 & 0 & 0 & 0 & 0 & 0 & -1 \ea \right)
\eeq
is then given by eq. (9.2). Now we have to check consistency in the Higgs
sector where from eq. (9.24) we calculate
\beq
U_C^H = - i \left( \ba{cccc} 1 & 0 & 0 & 0 \\
0 & 0 & 1 & 0 \\ 0 & 1 & 0 & 0 \\ 0 & 0 & 0 & 1 \ea \right).
\eeq
It is then tedious but easy to verify that
$$
U_C^H (\T_b R_{Cba}) U_C^{H\dg} = \T_a.
$$

This was a somewhat unorthodox look at the discrete symmetries CP, C and P
in left--right symmetric models. In this example P has the simple form
coming from the idea of left--right symmetry whereas CP and C are
``generalized'' in the sense of subsect. 5.2 because there is a
phase in one part
of the horizontal space. The general discussion in this paper is quite
nicely illustrated here and though the gauge group is rather small the
structure of the discrete symmetries is more complex than in the cases
of QED and QCD.

\section{Comments and conclusions}
\renewcommand{\theequation}{\arabic{section}.\arabic{equation}}
\setcounter{equation}{0}

In this work we have discussed the possibility of defining CP, P and C
invariance in gauge theories before spontaneous symmetry breaking.
As mentioned in the introduction the breaking of the gauge group
and of the discrete symmetries could be performed at the same time.
For CP this possibility was first envisaged in ref. \cite{lee1} and
for P in refs. \cite{fa,sen1}.

Our first comment concerns the question of when the above scenario really
happens. Let us suppose that the Lagrangian is invariant under a
CP transformation where $U_H$ is the unitary matrix appearing
in the transformation of the scalar field vector $\Phi$ eq. (7.37). Then,
if the vacuum expectation value $\langle \Phi\rangle_0$ fulfills
\beq
U_H \langle \Phi\rangle_0^* = \langle \Phi\rangle_0
\eeq
the CP symmetry is not broken \cite{eck83,gri88,bra}. Furthermore,
if eq. (10.1) is not fulfilled there are two possibilities. On the one
hand, there might exist an element of the total symmetry group $G \times
H$ of the Lagrangian ($G$ is the gauge group and $H$ the group of all
other internal symmetry transformations commuting with $G$) such
that its action on $\Phi$ is given by $S_H$ and
\beq
S_H U_H \langle \Phi\rangle_0^* = \langle \Phi \rangle_0 .
\eeq
Then one can define a new CP symmetry
\beq
\Phi(x) \ra S_H U_H \Phi(\wh x)^*
\eeq
with adequate redefinitions in the fermion and gauge boson sectors and
eq. (10.1) is fulfilled with $U_H$ replaced by $U'_H = S_H U_H$. On the other
hand, only if for a given $U_H$ one cannot find an $S_H$ such that
$U'_H$ complies with eq. (10.1) the CP symmetry associated with the $U_H$
of eq. (10.1) is spontaneously
broken together with the gauge group.

The second comment refers to CPT invariance. One might be tempted to
define a CPT--type transformation in analogy to CP--type transformations
defined in sect. 3 by
\beqa
W_\mu^a(x) &\ra& - R_{ab}  W_\mu^b (-x)\no \\
\omega_R(x) &\ra& - U^T \gamma_5^T \omega_R(- x)^* \\
\Phi(x) &\ra& - U_H^T \Phi(- x)^* . \no
\eeqa
It can easily be checked that the invariance conditions ensuing from
eq. (10.4) and from its antiunitary operator implementation are identical
with the conditions (9.2) following from the C--type transformations
(9.1) with the matrices $R$, $U$, $U_H$ of eq. (10.4).\footnote{We
use the transposed matrices in eq. (10.4)
to get exactly the form of eq. (9.2) for the
invariance conditions.} Clearly, if $R$,
$U$, $U_H$ are identity matrices, eq. (9.2) is always fulfilled
suggesting that CPT should be defined by\footnote{The minus in
the transformation of $\omega_R$ is convention, however, the other
two minus signs are required for invariance of the Lagrangian.}
\beqa
W_\mu^a(x) &\ra& - W_\mu^a(-x) \no \\
CPT: \qquad \omega_R(x) &\ra& - \gamma_5^T \omega_R(-x)^* \\
\Phi(x) &\ra& - \Phi(- x)^* . \no
\eeqa
In other words, invariance of the Lagrangian under the transformation
(10.4) leads to
invariance under the corresponding C--type transformation (9.1) yet
the Lagrangian is anyway invariant under the transformation (10.5).
Thus the concept of
CPT--type transformations is void and there is just one canonical form
(10.5) of CPT. This together with the definition of CP eq. (5.1) via
the contragredient automorphism $\psi^\triangle$ fixes the definition of time
reversal in the generalized sense analogous to CP in sect. 5.

It is interesting to note that with $R = {\bf 1}$, but $U$, $U_H$ non--trivial,
conditions (9.2) tell us that $U$, $U_H$ act on $\omega_R$, $\Phi$,
respectively, as representations of an element of $G \times H$.
If $U$, $U_H$ belong to $H$ alone then the effect of the CPT--type
transformation (10.4) or the corresponding C--type transformation (9.1)
is a horizontal symmetry \cite{gri88}.

The comment on CPT explains why in the discussion of the discrete
symmetries C, P, T we could concentrate on CP and P in this work and
being fully general at the same time.

Let us now summarize the main points of this work. The starting point of
our discussion was the observation that CP and P transformations have
the same structure when formulated with fermion fields of one chirality.
General transformations of that type we have called CP--type
transformations. The crucial point is that if a CP--type transformation
is a symmetry of the Lagrangian its action on the gauge bosons can be
described in terms of automorphisms of the Lie algebra $\cL_c$ of the
gauge group. Consequently, the invariance conditions in the fermion
and scalar sectors also have a straightforward interpretation in terms
of Lie algebra representations. In addition, all possible automorphisms
of simple Lie algebras are known and classified in the literature.

Now what distinguishes CP and P from each other? The automorphism
associated with CP is given by the contragredient automorphism $\psi^\triangle$
which has the property $\psi^\triangle(h) = - h$ for all elements $h$ of the
CSA whereas P is associated with an involutive automorphism $\psi_P$
which reverses the signs of at most part of the CSA. These abstract
mathematical definitions were substantiated by physical considerations.

CP is always a symmetry of the gauge Lagrangian  $\cL_{\rm gauge}$,
the Lagrangian without Yukawa couplings
and the Higgs potential. This statement formulated
in the language of representations is expressed by $-D^T \circ \psi^\triangle
\sim D$, saying that for every irrep $D$ of $\cL_c$ its complex
conjugate irrep $- D^T$ connected with the automorphism $\psi^\triangle$ is
equivalent to $D$. (This follows immediately from the fact that the
weights of both irreps are identical.) Consequently, CP does not
impose conditions on the irrep content of the fermion or scalar
representation. If irreps occur with non--trivial multiplicities a
CP transformation not acting in these ``horizontal'' spaces is called
``canonical CP'' or (CP)$_0$, a special case of the ``generalized''
CP transformations.

Since P is not uniquely associated with an automorphism the choice of
$\psi_P$ is subject to physical boundary conditions. For instance, one
would require that $\psi_P$ does not change the sign of those elements
in the CSA which are associated with the electric and colour charges
in order to get a reasonable definition of parity. Also one could imagine
cases with large gauge groups where several automorphisms lead to viable
definitions of P. In contrast to CP, the gauge interactions of fermions
and scalars are not automatically invariant under parity but, in general,
parity invariance introduces a condition on the irrep content of the
fermion and scalar representations or on irreps themselves,
depending on the gauge group. For simple gauge groups a
common condition is that with every irrep in the fermionic sector also its
complex conjugate irrep occurs. This condition is always trivially
satisfied in the scalar sector. We have also worked out the
connection between parity and the definition of Dirac fields.

In the scalar sector a complication arises for pseudoreal irreps where
the horizontal part of a CP--type transformation in general mixes the
pseudoreal fields $\phi$ and $\wt \phi = W \phi^*$ (7.4) which transform
alike under the gauge group. This leads to unitary symplectic matrices
acting horizontally.

Though generalized CP transformations are automatically symmetries
of $\cL_{\rm gauge}$
this is not the case for the Yukawa interactions. Choosing a particular
transformation one obtains conditions on the Yukawa couplings. For
(CP)$_0$ this amounts to real couplings in some phase conventions. In
the general case horizontal transformations are involved. With suitable
basis choices all possible solutions of these conditions for all
possible generalized CP transformations can be derived. This might be of
interest for models relating fermion masses and mixing angles.

We have also considered C--type transformations defined as the composition
of two CP--type transformations. Such symmetry transformations
associated with the trivial automorphism $\psi_C = id$ are just
horizontal symmetries. Finally we have shown that at least for simple
gauge groups but arbitrary reps CP and P are compatible in the
following sense: given fermion and scalar reps,
real Yukawa couplings, i.e. $\cL_Y$ invariant
under (CP)$_0$, and an automorphism $\psi_C$ associated with C
such that $D \circ \psi_C$ is also contained in the
reps for every irrep $D$ occurring there, then one
can construct Yukawa couplings which are invariant under the ``simplest'' C
transformation pertaining to $\psi_C$.

In conclusion, in this work we have tried to understand
CP and P symmetries in gauge theories by pointing out their intimate
connection with
automorphisms of the Lie algebra of the gauge group. We have furthermore
studied in detail the action of such symmetries in the multiplicity
spaces of the irreps in the fermionic and scalar sectors and how these
rotations imply conditions on the Yukawa couplings.
Finally, we have made extensive use of certain basis transformations:
in the representation spaces of the irreps they lead to symmetric or
antisymmetric generators of $\cL_c$ in arbitrary irreps and real
Clebsch Gordan coefficients in the Yukawa couplings and in the
horizontal spaces they give simple forms of generalized CP transformations.
All these considerations might be useful for the construction of models
beyond the SM.

\section*{Acknowledgements}
We would like to thank H. Stremnitzer for thorough and helpful
discussions on $so(10)$. Furthermore, W. G. acknowledges numerous
enlightening conversations on group theory with H. Urbantke.

\newpage
\section*{Appendices}

\newcounter{zahler}
\renewcommand{\thesection}{\Alph{zahler}}
\renewcommand{\theequation}{\Alph{zahler}.\arabic{equation}}
\setcounter{zahler}{1}
\setcounter{equation}{0}

\addcontentsline{toc}{section}{Appendices}
\section{Notation and conventions}
We follow the conventions of ref. \cite{bjo}. Therefore we use the metric
\beq
(g_{\mu\nu}) = \mbox{diag }(1,-1,-1,-1)
\eeq
and thus the Dirac algebra is given by
\beq
\{ \gamma_\mu,\gamma_\nu\} = 2g_{\mu\nu} {\bf 1}_4.
\eeq
Furthermore, we assume that the $\gamma$ matrices fulfill the hermiticity
conditions
\beq
\gamma_\mu^\dg = \gamma_0 \gamma_\mu \gamma_0 = \ve(\mu) \gamma_\mu
\quad \mbox{with} \quad
\ve(\mu) = \left\{ \ba{rl} 1, & \mu = 0 \\ -1, & \mu = 1,2,3.
\ea \right.
\eeq
Then
\beq
\gamma_5 \equiv i \gamma^0 \gamma^1 \gamma^2 \gamma^3
\eeq
is hermitian. Left and right--handed fermion fields are given by the
conditions
\beq
\frac{{\bf 1} + \gamma_5}{2} \chi_R = \chi_R, \qquad
\frac{{\bf 1} - \gamma_5}{2} \chi_L = \chi_L,
\eeq
respectively. The charge conjugation matrix $C$ is defined by
\beq
C^{-1} \gamma_\mu C = - \gamma_\mu^T .
\eeq
As a consequence of eqs. (A.3) and (A.6) we have\footnote{Actually, it can
only be derived that $C^\dg$ is proportional to $C^{-1}$. For
convenience and without loss of generality we assume that these
matrices are equal.}
\beq
C^\dg = C^{-1}, \qquad C^T = -C \qquad \mbox{and} \qquad
C^{-1} \gamma_5 C = \gamma_5^T.
\eeq

A time reversal transformation requires the definition of a matrix $T$
verifying
\beq
T \gamma_\mu T^{-1} = \gamma_\mu^T .
\eeq
{}From the properties of $C$ it is clear that
\beq
T = e^{i\beta} C^{-1} \gamma_5 = - T^T
\eeq
with an arbitrary phase $\beta$. Then, given a solution $\chi(x)$ of the
Dirac equation we obtain its time--reflected solution
\beq
\chi_T(x^0,\vec x) = T^* \chi(-x^0,\vec x)^*.
\eeq
In the second quantized version this translates into
\beq
\T \chi(x^0,\vec x) \T^{-1} = T \chi(-x^0,\vec x)
\eeq
with the time reversal operator $\T$ acting on the Hilbert space of states
as an antiunitary operator.

In this paper we only use right--handed fermion fields. If one starts in
a theory with fields of both chirality $f_L$, $f_R$ then the fermionic
Lagrangian
\beq
\cL_F = \bar f_R i \gamma^\mu (\partial_\mu + ig T_a^R W_\mu^a)f_R
+ \bar f_L i \gamma^\mu (\partial_\mu + ig T_a^L W_\mu^a)f_L,
\eeq
where $\{T_a^R\}$, $\{T_a^L\}$ are arbitrary, in general different reps
of the gauge group, can easily be rewritten as
\beq
\cL_F = \bar \omega_R i \gamma^\mu (\partial_\mu + ig T_a W_\mu^a)
\omega_R
\eeq
with
$$
\omega_R = \left( \ba{c} (f_L)^c \\ f_R \ea \right), \qquad
(f_L)^c \equiv C \gamma_0^T f_L^* \qquad \mbox{and} \qquad
T_a = \left( \ba{cc} - (T_a^L)^T & 0 \\ 0 & T_a^R \ea \right).
$$

\setcounter{equation}{0}
\addtocounter{zahler}{1}
\section{Facts about semisimple Lie groups}
In this appendix we collect all the facts about Lie algebras (in
particular, semisimple Lie algebras) which are used in this work.
Extensive expositions of this subject can e.g. be found in the books
by Cornwell \cite{corn}, Georgi \cite{geo}, Samelson \cite{sam},
Jacobson \cite{jac}, Varadarajan \cite{var}, Wybourne \cite{wy},
Cahn \cite{cahn} and others.

\paragraph{The Killing form:} Every element $X$ of a Lie algebra
(always assumed to be over a field ${\bf K} = {\bf R}$ or {\bf C})
allows to define a linear mapping
\beq
\mbox{ad} \; X : \ba{rcl} \cL &\ra& \cL \\ Y &\ra& [X,Y]. \ea
\eeq
Then the symmetric bilinear form $\kappa$ obtained by
\beq
\kappa(X,Y) = \mbox{Tr }(ad\; X \; \mbox{ad} \; Y)
\eeq
is called Killing form. Automorphisms $\psi$ of $\cL$ are defined as
linear mappings which respect the Lie algebra product, i.e.
\beq
\psi([X,Y]) = [\psi(X), \psi(Y)] \qquad \forall \; X,Y \in \cL.
\eeq
The set of automorphisms form a group denoted by Aut~$(\cL)$. Note that
the Killing form is invariant under automorphisms:
\beq
\kappa(\psi(X),\psi(Y)) = \kappa(X,Y) \; \forall \; X,Y \in \cL.
\eeq

Semisimple Lie algebras are those which have no non--zero Abelian ideals.
Cartan's second criterion states that a Lie algebra $\cL$ is semisimple if
and only if its dimension is positive and its Killing form non--degenerate.

Given a basis $\{X_a\}$ of $\cL$ one can define structure constants by
\beq
[X_a,X_b] = C_{ab}^c X_c .
\eeq
One can show that a semisimple Lie group $G$ (a group whose (real) Lie algebra
is semisimple) is compact if and only if its Lie algebra $\cL_c$ has a
negative definite Killing form. Therefore on such a Lie algebra a scalar
product is given by $-\kappa$. This allows the definition of ON
bases $\{X_a\}$ with $\kappa(X_a,X_b) = -\delta_{ab}$. In such a case
the structure constants $C_{ab}^c$ are usually denoted by $f_{abc}$. One
can prove that $f_{abc}$ is totally antisymmetric in $a$, $b$, $c$ whereas
in general only $C_{ab}^c = - C_{ba}^c$ is valid.

\paragraph{Automorphisms and bases of $\cL$:} Fixing a basis $\{X_a\}$
of $\cL$ allows to associate a matrix with every linear operator on
$\cL$ and vice versa. Denoting such a matrix by $A$ and the corresponding
operator by $\psi_A$ we can thus write
\beq
\psi_A : \ba{rcl} \cL &\ra& \cL \\ X_a &\ra& A_{ba} X_b. \ea
\eeq
It is then easy to check by analysing eqs. (B.3) and (B.5) that matrices $A$
corresponding to automorphisms $\psi_A$ fulfill the following condition:
\beq
\psi_A \in \mbox{Aut }(\cL) \Longleftrightarrow A_{a'a} A_{b'b}
(A^{-1})_{cc'} C_{a'b'}^{c'} = C_{ab}^c.
\eeq
Furthermore, invariance of the Killing form is expressed by
\beq
A_{a'a} A_{b'b} \kappa(X_{a'},X_{b'}) = \kappa(X_a,X_b).
\eeq

For compact Lie algebras and ON bases this immediately
translates into the conditions that $A$ is an orthogonal matrix and
$f_{abc}$ an invariant tensor with respect to $A$:
\beq
A_{a'a} A_{b'b} A_{c'c} f_{a'b'c'} = f_{abc}.
\eeq

\paragraph{The structure of semisimple Lie algebras:} With every Lie
group $G$ a real Lie algebra $\cL$ is associated and by complexification
of $\cL$ a complex Lie algebra $\wt \cL$. The transition from $\cL$ to
$\wt \cL$ conserves semisimplicity. Complex semisimple Lie algebras
are fully classified. Conversely, given a complex semisimple Lie
algebra $\wt \cL$ there is a well defined procedure to go back to the
real Lie algebras $\cL$ associated with $\wt \cL$, i.e. to find all
inequivalent $\cL$'s whose complexification is $\wt \cL$. The structure
of $\cL$ is closely connected with $\wt \cL$. This motivates the
consideration of complex semisimple Lie algebras though in gauge
theories only real Lie algebras occur. Furthermore, since here we are
concerned only with compact Lie groups it is sufficient to consider the
construction of the unique compact Lie algebra $\cL_c$ associated with
$\wt \cL$ (see eq. (B.23)) and refer the reader to the books quoted at the
beginning of this appendix for the general procedure of ``realification''.

A CSA can be defined and its existence proved for
general Lie algebras (see refs. \cite{sam,jac,var}). In the case of a
semisimple complex Lie algebra a CSA $\Ha$ is a maximal abelian
subalgebra of $\wt \cL$ such that $\mbox{ad} \; h$ is completely reducible
(i.e. diagonalizable) $\forall \; h \in \Ha$. All CSAs are mutually
conjugate which means that given two CSAs $\Ha$, $\Ha'$ of $\wt \cL$
then there is an inner automorphism (see eq. (4.1)) $\psi$ such that
$\psi(\Ha) = \Ha'$. In this sense a CSA is unique for semisimple
complex Lie algebras $\wt \cL$. The dimension of the CSA, $\dim \Ha =
\ell$, is called the rank of $\wt \cL$.

As a vector space $\wt \cL$ can be decomposed into
\beq
\wt \cL = \left( \bigoplus_{\alpha \in \Delta} \wt \cL_\alpha \right)
\oplus \Ha
\eeq
which corresponds to the decomposition of $\wt \cL$ into common
eigenstates of $\mbox{ad} \; h$ $(h \in \Ha)$:
\beq
(\mbox{ad}\; h)(X) = [h,X] = \alpha(h) X
\eeq
with $X \in \wt \cL_\alpha$. One can show that
\beq
\dim \wt \cL_\alpha = 1 \qquad \forall \; \alpha \in \Delta
\eeq
for $\wt \cL$ semisimple. The eigenvalues $\alpha(h)$ depend linearly on
$h$ and can therefore be regarded as linear functionals on $\Ha$. These
non--zero functionals are called roots and the set of roots is denoted
by $\Delta$. Since the spaces $\wt \cL_\alpha$ are one--dimensional it
suffices to choose a non--zero vector $e_\alpha \in \wt \cL_\alpha$ as
a basis $\forall \; \alpha \in \Delta$. Then one can show that
\beq
\kappa(e_\alpha,h) = 0 \quad \forall \; \alpha \in \Delta, \; h \in \Ha
\quad \mbox{and} \quad
\kappa(e_\alpha,e_\beta) = 0 \quad \forall \; \alpha,\beta \in \Delta
\mbox{ with } \beta \neq -\alpha.
\eeq
Therefore, non--degeneracy of $\kappa$ requires
\beq
\alpha \in \Delta \Longleftrightarrow - \alpha \in \Delta .
\eeq

As a matter of fact already $\left.\kappa\right|_\Ha$ is non--degenerate.
Thus $\forall \; \alpha \in \Delta$ there exists a unique $h_\alpha \in
\Ha$ such that
\beq
\alpha(h) = \kappa(h_\alpha,h) \qquad \forall \; h \in \Ha.
\eeq
The elements $h_\alpha$ are called root vectors. If $\alpha,\beta,
\alpha + \beta \in \Delta$ then
\beq
h_{\alpha+\beta} = h_\alpha + h_\beta, \qquad h_{-\alpha} = h_\alpha .
\eeq
The real linear span of $\{ h_\alpha|\alpha \in \Delta\}$ is denoted by
$\Ha_{\bf R}$ and $\Ha$ is the complexification of $\Ha_{\bf R}$.
$\left.\kappa\right|_{\Ha_{\bf R}}$ is a positive definite scalar product
and therefore $\Ha_{\bf R}$ is an $\ell$--dimensional euclidean space.
By
\beq
\langle \alpha,\beta\rangle \equiv \kappa(h_\alpha,h_\beta) \: \forall
\alpha,\beta \in \Delta
\eeq
length of roots and the angle between two roots are defined.
The basis elements $e_\alpha$ and the root vectors are connected through
\beq
[e_\alpha,e_{-\alpha}] = \kappa(e_\alpha,e_{-\alpha}) h_\alpha .
\eeq
A crucial point of the theory is that the numbers
\beq
a_{\beta\alpha} \equiv 2 \frac{\langle \beta,\alpha \rangle}{\langle
\alpha,\alpha\rangle}
\eeq
are integers (the Cartan integers) and that only
$a_{\beta\alpha} = 0,\pm 1,\pm 2,\pm 3$ is allowed.

On $\Delta$ a weak order can be defined by choosing an element
$h_0 \in \Ha_{\bf R}$ such that $\alpha(h_0) \neq 0$ $\forall \; \alpha
\in \Delta$. Then for functionals $\mu, \mu'$ on $\Ha_{\bf R}$ the
relation $\mu > \mu'$ $(\mu \geq \mu')$ is defined by
$\mu(h_0) > \mu'(h_0)$ $(\mu(h_0) \geq \mu'(h_0))$.
Let $\Delta_\pm = \{ \alpha \in \Delta | \alpha(h_0) \; {}^>_< \; 0\}$.
Then $\Delta = \Delta_+ \cup \Delta_-$ and
$\Delta_- = \{ -\alpha|\alpha \in \Delta_+\}$.
A root is called simple if it is positive but not the sum of two
positive roots. The set of simple roots consists of exactly $\ell$
linearly independent elements $\{\alpha_1,\ldots,\alpha_\ell\}$.

\paragraph{The Weyl canonical form of $\wt \cL$:} By choosing suitable
bases the commutation relations characterizing semisimple complex Lie
algebras can be brought to certain standard forms one of which is the
Weyl canonical form (for other standard forms see ref. \cite{corn}). There
the basis elements $e_\alpha$ are normalized to
\beq
\kappa (e_\alpha,e_{-\alpha}) = -1.
\eeq
Then one has the following commutators:
\beqa
[e_\alpha, e_{-\alpha}] &=& -h_\alpha \no \\{}
[h,e_\alpha] &=& \alpha(h)e_\alpha \qquad \mbox{or} \qquad
[h_\beta,e_\alpha] = \langle \beta,\alpha\rangle e_\alpha \no \\{}
[h,h'] &=& 0 \qquad \forall \; h,h' \in \Ha \no \\{}
[e_\alpha,e_\beta] &=& 0 \qquad \mbox{if } \alpha + \beta \notin \Delta,
\; \alpha + \beta \neq 0 \no \\{}
[e_\alpha,e_\beta] &=& N_{\alpha \beta} e_{\alpha+\beta} \qquad
\mbox{for } \alpha + \beta \in \Delta
\eeqa
with
$$
N_{\alpha\beta} \in {\bf R} \setminus \{0\}, \qquad
N_{\alpha\beta} = N_{-\alpha -\beta}.
$$
Given a weak order on $\Delta$ every root can be written as a linear
combination of simple roots such that $\forall \; \alpha \in \Delta_+$
\beq
\alpha = \sum_{j=1}^\ell k_j^\alpha \alpha_j \qquad \mbox{with} \qquad
k_j^\alpha \in {\bf N}_0
\eeq
for rank $\ell = \dim \Ha = \dim_{\bf R} \Ha_{\bf R}$.

\paragraph{The compact real form $\cL_c$ of $\wt \cL$:} Given $\wt \cL$
we can go back to the compact real Lie algebra $\cL_c$ by choosing the
following basis elements $\{X_a\}$:
$$
- i H_j \; (j = 1,\ldots,\ell) \qquad \mbox{with} \qquad
H_j \in \Ha_{\bf R}, \qquad \kappa(H_j,H_k) = \delta_{jk},
$$
\beq
\frac{e_\alpha + e_{-\alpha}}{\sqrt{2}} \qquad \mbox{and} \qquad
\frac{e_\alpha - e_{-\alpha}}{\sqrt{2}\; i} \qquad \forall \;
\alpha \in \Delta_+ .
\eeq
To a given semisimple complex Lie algebra $\wt \cL$ there is a unique
compact real form $\cL$, up to isomorphisms. One can easily check that
the above basis elements fulfill $\kappa(X_a,X_b) = - \delta_{ab}$.

\paragraph{Dynkin diagrams:} With the simple roots the Cartan matrix
\beq
A_{jk} = 2 \frac{\langle \alpha_j,\alpha_k\rangle}
{\langle \alpha_k,\alpha_k\rangle}
\eeq
is associated. Clearly, $A_{jj} = 2$ and one can show that $A_{jk}$ can
only be $0,-1,-2$ or $-3$.

Considering the case $j \neq k$ in more detail we find (no sum implied here)
\beq
A_{jk} A_{kj} = 4 \cos^2 \Theta
\eeq
where $\Theta$ is the angle between $\alpha_j$ and $\alpha_k$. Note that
for $\Theta \neq \pi/2$ either $A_{jk}$ or $A_{kj}$ have to be $-1$.
According to the possible angles $\Theta$ we have to distinguish four
cases (assuming that $A_{jk} = -1$ for $\Theta \neq \pi/2$ and
$\omega_j^2 \equiv \langle \alpha_j,\alpha_j\rangle$):
\begin{enumerate}
\item[a)] $\cos \Theta = 0 \Ra \Theta = \pi/2$ or $90^0$,
$\omega_k/\omega_j$ undetermined
\item[b)] $\cos \Theta = - 1/2 \Ra \Theta = 2\pi/3$ or $120^0$,
$\omega_k/\omega_j = 1$
\item[c)] $\cos \Theta = - 1/\sqrt{2} \Ra \Theta = 3\pi/4$ or $135^0$,
$\omega_k/\omega_j = \sqrt{2}$
\item[d)] $\cos \Theta = - \sqrt{3}/2 \Ra \Theta = 5\pi/6$ or $150^0$,
$\omega_k/\omega_j = \sqrt{3}$.
\end{enumerate}
For $\Theta \neq 0$ and $A_{jk} = -1$ the ratio of the lengths of the
simple roots is given by $\omega_k/\omega_j = - 2 \cos \Theta$.

A Dynkin diagram of a semisimple complex Lie algebra is defined in the
following way:
\begin{enumerate}
\item[i)] To each simple root is associated a point (or vertex) of the
diagram.
\item[ii)] The points associated with $\alpha_j$ and $\alpha_k$ are
connected by $A_{jk}A_{kj}$ lines, i.e. there are zero, one, two or
three lines for the above cases a, b, c, d, respectively.
\item[iii)] If there are two or three lines connecting two points then
a black dot denotes the shorter root. Thus a Dynkin diagram is the
graphical representation of the Cartan matrix.
\end{enumerate}

The connected Dynkin diagrams are exactly those associated with simple
complex Lie algebras and there is a one--to--one correspondence between
connected Dynkin diagrams and simple complex Lie algebras. In fig.~1
all possible connected Dynkin diagrams are depicted with the names of
the $\wt \cL$'s associated with them. In table 1 all $\wt \cL$'s are
listed with their respective compact real forms $\cL_c$. Table 2
contains all isomorphisms of the low--dimensional classical $\cL_c$'s
explaining thus why the series $B_\ell$, $C_\ell$, $D_\ell$ start with
$\ell = 2,3,4$, respectively.

\paragraph{Some facts about irreps of semisimple complex Lie algebras:}
All facts mentioned here are also valid for $\cL_c$.

In a rep of $\wt \cL$ weight vectors are those elements of the vector space
which are eigenvectors of the CSA, i.e.
\beq
D(h) e(\lambda,q) = \lambda(h) e(\lambda,q) \qquad
(q = 1,\ldots,m_\lambda)
\eeq
where $m_\lambda$ is the multiplicity of the weight $\lambda$ which,
like a root,
can be conceived as a functional on $\Ha$. One can show that for any
weight $\lambda$ of a rep of $\wt \cL$ and for any $\alpha \in \Delta$
\beq
\frac{2 \langle \lambda,\alpha\rangle}{\langle \alpha,\alpha\rangle}
\in {\bf Z} \quad \mbox{and} \quad
\lambda = \sum_{j=1}^\ell \mu_j \alpha_j \mbox{ with $\mu_j$ real and
rational}.
\eeq
The fundamental weights are defined by
\beq
\Lambda_j = \sum_{k=1}^\ell (A^{-1})_{jk} \alpha_k
\eeq
and it follows that
\beq
\frac{2 \langle \Lambda_j,\alpha_k\rangle}{\langle \alpha_k,\alpha_k
\rangle} = \delta_{jk}.
\eeq
For any irrep of a semisimple complex Lie algebra $\wt \cL$ there is a
unique highest weight $\Lambda$ (a weight $\Lambda$ is called highest if
$\Lambda + \alpha$ is not a weight $\forall \alpha \in \Delta_+$).
It can be written as
\beq
\Lambda = n_1 \Lambda_1 + \ldots + n_\ell \Lambda_\ell
\eeq
where $n_1,\ldots,n_\ell$ are non--negative integers. Conversely, given
$n_1,\ldots,n_\ell$ with all $n_j \in {\bf N}_0$ there is an irrep $D_\Lambda$
of $\wt \cL$ unique up to equivalence such that
$\Lambda = n_1 \Lambda_1 + \ldots + n_\ell \Lambda_\ell$ is its highest
weight. Thus there is a one--to--one correspondence between functionals
$\Lambda$ of the form (B.30) and irreps $D_\Lambda$ of $\wt \cL$. All
irreps are faithful except the trivial irrep with $\Lambda = 0$.

No real form $\cL$ of $\wt \cL$ admits unitary non--trivial irreps except
the compact real form $\cL_c$ where all irreps are equivalent to unitary
ones. On $\left. D_\Lambda\right|_{\cL_c}$ unitarity is expressed by
\beq
D(X)^\dg = - D(X)
\eeq
corresponding to unitary operators $\exp D(X)$. Then from the basis (B.23)
we derive that
\beq
D(H)^\dg = D(H) \quad \forall \; H \in \Ha_{\bf R}, \qquad
D(e_\alpha)^\dg = - D(e_{-\alpha}) \quad \forall \; \alpha \in \Delta.
\eeq

Finally we want to mention that in physics one usually employs hermitian
generators of a unitary rep
\beq
T_a^\dg = T_a, \qquad [T_a,T_b] = i f_{abc} T_c
\eeq
which are obtained by
\beq
T_a = i D(X_a) \qquad \mbox{with} \qquad
[D(X_a),D(X_b)] = f_{abc} D(X_c)
\eeq
from the ON generators $X_a$ of $\cL_c$. In this paper we often switch
between the two forms $\{T_a\}$ and $\{D(X_a)\}$.

\setcounter{equation}{0}
\addtocounter{zahler}{1}
\section{$so(N)$ and the spinor representations}
A convenient choice of basis in the space of antisymmetric real $N
\times N$ matrices is given by
\beq
(M_{pq})_{jk} = \delta_{pj} \delta_{qk} - \delta_{qj} \delta_{pk},
\qquad 1 \leq p < q \leq N
\eeq
with commutation relations
\beq
[M_{pq},M_{rs}] = \delta_{ps} M_{qr} + \delta_{qr} M_{ps} -
\delta_{pr} M_{qs} - \delta_{qs} M_{pr}.
\eeq

A Clifford algebra with $N$ basis elements is defined by the
anticommutators
\beq
\{ \Gamma_p,\Gamma_q\} = 2 \delta_{pq} {\bf 1}.
\eeq
It is easy to show that the elements
\beq
\frac{1}{2} \sigma_{pq} \equiv \frac{1}{4} [\Gamma_p,\Gamma_q]
\eeq
also verify the commutation relations given by eq. (C.2). Thus a rep
of the Clifford algebra automatically gives also a rep of $so(N)$.
It is straightforward to check that for
$N = 2\ell +1$ a rep of the $\{ \Gamma_p\}$
in the $2^\ell$ dimensional space ${\bf C}^2 \otimes \ldots \otimes
{\bf C}^2$ ($\ell$--fold tensor product) is given, in terms of the
Pauli matrices and the two--dimensional identity matrix, by
\beq
\ba{rcllcl}
\Gamma_1 &=& \sigma_3 \otimes \ldots \otimes \sigma_3 \otimes \sigma_3
\otimes \sigma_1 \qquad &
\Gamma_{2\ell-3} &=& \sigma_3 \otimes \sigma_1 \otimes {\bf 1} \otimes
\ldots \otimes {\bf 1} \\[5pt]
\Gamma_2 &=& \sigma_3 \otimes \ldots \otimes \sigma_3 \otimes \sigma_3
\otimes \sigma_2 \qquad &
\Gamma_{2\ell-2} &=& \sigma_3 \otimes \sigma_2 \otimes {\bf 1} \otimes
\ldots \otimes {\bf 1} \\[5pt]
\Gamma_3 &=& \sigma_3 \otimes \ldots \otimes \sigma_3 \otimes \sigma_1
\otimes {\bf 1} \qquad &
\Gamma_{2\ell-1} &=& \sigma_1 \otimes {\bf 1} \otimes {\bf 1} \otimes
\ldots \otimes {\bf 1} \\[5pt]
\Gamma_4 &=& \sigma_3 \otimes \ldots \otimes \sigma_3 \otimes \sigma_2
\otimes {\bf 1} \qquad &
\Gamma_{2\ell} &=& \sigma_2 \otimes {\bf 1} \otimes {\bf 1} \otimes
\ldots \otimes {\bf 1} \\
&\vdots& & \Gamma_{2\ell+1} &=& \sigma_3 \otimes \sigma_3 \otimes
\sigma_3 \otimes \ldots \otimes \sigma_3.
\ea
\eeq
Then the $\{ \frac{1}{2} \sigma_{pq}\}$ of eq. (C.4)
define the irreducible spinor irrep of
$so(2\ell +1)$ which has dimension $2^\ell$.

For $so(2\ell)$ one simply has to take $\Gamma_1,\ldots,\Gamma_{2\ell}$
of the Clifford algebra of $so(2\ell +1)$. However, the
$2^\ell$--dimensional rep of $so(2\ell)$ is not irreducible because
\beq
[ \sigma_{pq},\Gamma_{2\ell +1}] = 0 \qquad \forall \; p,q=1,\ldots, \ell .
\eeq
It decays into two inequivalent irreps of dimension $2^{\ell -1}$ given
by the projectors $({\bf 1} \pm \Gamma_{2\ell +1})/2$.

\setcounter{equation}{0}
\addtocounter{zahler}{1}
\section{On the isomorphisms
$so(4) \stackrel{\sim}{=} su(2) \oplus su(2)$ and  $so(6)
\stackrel{\sim}{=} su(4)$}

In sect. 2 the above isomorphisms are exploited for the example of the
spinor representation of $so(10)$. Since there one assumes that it is
known how to classify the fields according to $su(4) \oplus
su(2) \oplus su(2)$ an explicit realization of the above isomorphisms
has to be established to transfer this classification to
$so(6) \oplus so(4) \subset so(10)$.

Starting from the basis $\{M_{ij}\}$ as given in app. C we can define
the following new basis of $so(4)$:
\beq
\ba{rcllcl}
A_1 &=& \dfrac{1}{2} (M_{23} - M_{14}) \qquad &
B_1 &=& \dfrac{1}{2} (M_{23} + M_{14}) \\[12pt]
A_2 &=& \dfrac{1}{2} (M_{13} - M_{42}) \qquad &
B_2 &=& \dfrac{1}{2} (M_{13} + M_{42}) \\[12pt]
A_3 &=& \dfrac{1}{2} (M_{12} - M_{34}) \qquad &
B_3 &=& \dfrac{1}{2} (M_{12} + M_{34}) .
\ea
\eeq
Then it is easy to check with eqs. (C.1) and (C.2) that
\beq
[A_i,A_j] = \ve_{ijk} A_k, \qquad [B_i,B_j] = \ve_{ijk} B_k, \qquad
[A_i,B_j] = 0
\eeq
thus proving the first of the isomorphisms.

In order to prove the second isomorphism we will specify a basis $\{E_a\}$
of $so(6)$ such that there is a one--to--one correspondence between the
matrices $\{E_a\}$ and the matrices $\{\lambda_a\}$ of $su(4)$ obtained
by generalizing the Gell--Mann basis of $su(3)$, i.e.
\beq
-i \frac{\lambda_a}{2} \longleftrightarrow E_a \qquad (a = 1,\ldots,15),
\eeq
such that the matrices $E_a$ obey the same commutation relations as
$-i \lambda_a/2$. The result which is unique up to orthogonal basis
transformations is given by\footnote{The first eight matrices listed are
generators of the $su(3)$ subalgebra and can be obtained by a general
procedure to derive the generators of the $su(N)$ subalgebra of
$so(2N)$ as given in ref. \cite{geo}.}
\beq
\ba{lcllcl}
E_1 &=& \dfrac{1}{2} (M_{23} - M_{14}) \qquad &
E_9 &=& \dfrac{1}{2} (M_{45} + M_{36}) \\[12pt]
E_2 &=& \dfrac{1}{2} (M_{13} - M_{42}) \qquad &
E_{10} &=& -\dfrac{1}{2} (M_{35} + M_{64}) \\[12pt]
E_3 &=& \dfrac{1}{2} (M_{12} - M_{34}) \qquad &
E_{11} &=& \dfrac{1}{2} (M_{16} + M_{25}) \\[12pt]
E_4 &=& \dfrac{1}{2} (M_{16} - M_{25}) \qquad &
E_{12} &=& -\dfrac{1}{2} (M_{15} + M_{62}) \\[12pt]
E_5 &=& \dfrac{1}{2} (M_{62} - M_{15}) \qquad &
E_{13} &=& \dfrac{1}{2} (M_{23} + M_{14}) \\[12pt]
E_6 &=& \dfrac{1}{2} (M_{45} - M_{36}) \qquad &
E_{14} &=& -\dfrac{1}{2} (M_{13} + M_{42}) \\[12pt]
E_7 &=& \dfrac{1}{2} (M_{35} - M_{64}) \qquad &
E_{15} &=& -\dfrac{1}{\sqrt{6}} (M_{12} + M_{34} + M_{56}) \\[12pt]
E_8 &=& \dfrac{1}{\sqrt{12}} (M_{12} + M_{34} - 2M_{56}). \qquad &
\ea
\eeq

\setcounter{equation}{0}
\addtocounter{zahler}{1}
\section{Irreps of $G^*$}
For the definition of $G^*$ see eq. (4.10). Given a rep of $G^*$ it can
be decomposed according to the subgroup $G \times G$. Since irreps of
$G \times G$ are given by $D_r(g_1) \otimes D_{r'}(g_2)$
$((g_1,g_2) \in G \times G)$ with $D_r$, $D_{r'}$ being irreps of $G$
\cite{shaw} we infer that an irrep $D$ of $G^*$ decays into
$\bigoplus_{(r,r')} (D_r(g_1) \otimes D_{r'}(g_2))$ under $G \times G$.
Using
\beq
D(E) D((g_1,g_2)) D(E) = D((g_2,g_1))
\eeq
we obtain
\beq
\bigoplus_{(r,r')} (D_r(g_1) \otimes D_{r'}(g_2)) \sim
\bigoplus_{(r,r')} (D_r(g_2) \otimes D_{r'}(g_1)) \sim
\bigoplus_{(r,r')} (D_{r'}(g_1) \otimes D_r(g_2)).
\eeq
Therefore, in the irrep $D$ only summands of the type
\beq
D_r(g_1) \otimes D_r(g_2) \quad \mbox{and} \quad
(D_r(g_1) \otimes D_{r'}(g_2)) \oplus (D_{r'}(g_1) \otimes D_r(g_2))
\quad (r \neq r')
\eeq
appear and thus
\beqa
D((g_1,g_2)) &\sim& \left[ \bigoplus_r m_r(D_r(g_1) \otimes D_r(g_2))
\right] \no \\
&\oplus& \left[ \bigoplus_{(r',r'')} m_{r',r''} ((D_{r'}(g_1)
\otimes D_{r''}(g_2)) \oplus (D_{r''}(g_1) \otimes
D_{r'}(g_2)))\right]
\eeqa
with $r' \neq r''$ and $m_r$, $m_{r',r''}$ being the multiplicities of
the reps (E.3). The associated vector spaces will be denoted by
$\V_r$ and $\V_{r',r''}$, respectively. Now we can define an operator
$S$ by
\beq
S: \ba{cccl}
x \otimes y &\ra& y \otimes x & \mbox{ on } \V_r, \\[5pt]
(v \otimes w,x \otimes y) &\ra& (y \otimes x,w \otimes v) &
\mbox{ on } \V_{r',r''} .
\ea
\eeq
One can easily prove that
\beq
SD((g_1,g_2)) S = D((g_2,g_1))
\eeq
and therefore
\beq
[D(E)S,D((g_1,g_2))] = 0.
\eeq
Schur's lemma guarantees that $D(E)S$ operates within $\V_r^{\oplus m_r}$
and $\V_{r',r''}^{\oplus m_{r',r''}}$ (the superscript $\oplus m$
denotes the $m$--fold direct sum).

Discussing first $\V_r^{\oplus m_r} \cong {\bf C}^{m_r} \otimes \V_r$
we note that
\beq
D(E) = A \otimes S \qquad \mbox{with} \qquad A^2 = {\bf 1}_{m_r}
\eeq
since $S^2 = id$. Therefore we can make a basis transformation in
${\bf C}^{m_r}$ diagonalizing $A$ with eigenvalues $\pm 1$.
Consequently we have a type of irreps given by
\beq
\V_r \qquad \mbox{and} \qquad D(E) = \pm S.
\eeq
This defines the irreps $D_r^\pm$ according to the sign in eq. (E.9).

Writing $\V_{r',r''} = \W_{r',r''} \oplus \W_{r'',r'}$ according to the
two inequivalent irreps of $G \times G$ on $\V_{r',r''}$ and
$\V_{r',r''}^{\oplus m_{r',r''}} \cong ({\bf C}^{m_{r',r''}} \otimes
\W_{r',r''}) \oplus ({\bf C}^{m_{r',r''}} \otimes \W_{r'',r'})$ then
$D(E)$ has the form
\beq
D(E) = (A \otimes id_{r',r''}, B \otimes id_{r'',r'})S.
\eeq
A small calculation reveals that $D(E)^2 = id$ requires
\beq
B = A^{-1} .
\eeq
Performing a basis transformation with
\beq
Z = ({\bf 1}_{m_{r',r''}} \otimes id_{r',r''}, A^{-1} \otimes id_{r'',r'})
\eeq
we obtain
\beq
Z^{-1} D((g_1,g_2))Z = D((g_1,g_2)) \qquad \mbox{and} \qquad
Z^{-1} D(E) Z = S
\eeq
on $\V_{r',r''}^{\oplus m_{r',r''}}$. Therefore the rep of $G^*$ on this
space decays into $m_{r',r''}$ equivalent copies of an irrep denoted by
$D_{r',r''}$ defined via $D(E) = S$ as given in the second line of eq. (E.5).
Thus we have found all irreps of $G^*$ as written down in subsect. 4.2.

\setcounter{equation}{0}
\addtocounter{zahler}{1}
\section{On the existence of a CP basis}
Let $\wt \cL$ be a complex semisimple Lie algebra with a $d$--dimensional
irrep $D$ and $D(X_a) \equiv Y_a$ $(a = 1,\ldots,n_G)$ where $\{X_a\}$ is
an ON basis of the compact real form $\cL_c$ of $\wt \cL$. Thus the
$Y_a$ are $d \times d$ matrices and since $D$ is unitary we have
$Y_a^\dg = - Y_a$ in addition.

In this appendix we will show that one can always choose a basis of
${\bf C}^d$ such that
\beq
Y_a^T = - \eta_a Y_a
\eeq
in this basis and where the signs $\eta_a$ are given by $\psi^\triangle(X_a) =
\eta_a X_a$ with $\psi^\triangle$ being the contragredient automorphism defined
in eq. (4.5). For the connection between the signs $\eta_a$ and the ON
basis $\{X_a\}$ defined in eq. (B.23) see eq. (5.12).
Eq. (F.1) defines the (CP)$_0$
basis used in subsect. 5.2.
\paragraph{Proof:} We have
\beq
- D^T \circ \psi^\triangle \sim D
\eeq
because the weights of both irreps are identical. This means that there
is a unitary matrix $U$ such that
\beq
U(- Y_a^T \eta_a) U^\dg = Y_a
\eeq
and, consequently,
\beq
[UU^*,Y_a] = 0 \qquad \forall \; a = 1,\ldots,n_G.
\eeq
Using Schur's lemma we obtain $UU^* = \lambda {\bf 1}$ and therefore
\beq
U^T = \lambda U \qquad \mbox{with} \qquad \lambda = \pm 1.
\eeq
Thus we have shown that the matrix $U$ must be either symmetric or
antisymmetric. (Exactly the same reasoning is valid for the matrix in
the equivalence $- D^T \sim D$ in which case $\lambda = 1$ corresponds
to real and $\lambda = -1$ to pseudoreal irreps.)
\paragraph{Case 1:} $U = U^T$ \\
We use the following lemma: For every unitary symmetric matrix $U$ there
is a unitary symmetric matrix $\wt U$ such that $U = \wt U^2$. Its proof
given at the end of this appendix.

Inserting $\wt U^2$ into eq. (F.3) we obtain
\beq
- \eta_a \wt U Y_a^T \wt U^\dg = - \eta_a(\wt U^\dg Y_a \wt U)^T =
\wt U^\dg Y_a \wt U
\eeq
which is just the desired result (F.1).
\paragraph{Case 2:} $U^T = - U$ \\
It remains to show that this case is impossible for irreps of semisimple
Lie algebras $\cL_c$.

The irrep $D$ is characterized by its highest weight $\Lambda$
which is simple. We get from eq. (F.3) particularized for $D(-i H_j)$
\beq
U \left( D(-i H_j)^T \right) U^\dg e_\Lambda = \Lambda(-i H_j)e_\Lambda
\eeq
where $e_\Lambda$ is the weight vector corresponding to $\Lambda$.
Using antisymmetry of $U$ and the fact that $D(-i H_j)$ is antihermitian
eq. (F.7) can be rewritten as
\beq
D(-i H_j) U e_\Lambda^* = \Lambda(-i H_j) U e_\Lambda^*.
\eeq
Since $\Lambda$ is simple we must have $U e_\Lambda^* = a e_\Lambda$
with $a \neq 0$. Finally, using $U^T = -U$ we derive the contradiction
\beq
0 = e_\Lambda^\dg U e_\Lambda^* = a e_\Lambda^\dg e_\Lambda \neq 0.
\eeq
Thus $U$ is symmetric and therefore the existence of a CP basis for any
irrep of any compact semisimple Lie algebra is proved.
\hfill\ $\Box$
\paragraph{Proof of the lemma:} Let $A$ be a normal symmetric matrix and
$A_1 \equiv \mbox{Re }A$, $A_2 \equiv \mbox{Im }A$. Then it follows from
$A^\dg A = A A^\dg$ and $A^T = A$ that $[A_1,A_2] = 0$. Thus $A$ can be
diagonalized by an orthogonal matrix $O$.

Let now $U$ be unitary and symmetric. Then
$$
U = O \wh d O^T \qquad \mbox{with} \qquad
\wh d = \mbox{diag }(e^{i\Omega_1}, \ldots,e^{i\Omega_d})
$$
and
$$
\wt U = O \; \sqrt{\wh d} \; O^T \qquad \mbox{with} \qquad
\sqrt{\wh d} = \mbox{diag }(e^{i\Omega_1/2},\ldots,e^{i\Omega_d/2}).
$$
\hfill\ $\Box$

\setcounter{equation}{0}
\addtocounter{zahler}{1}
\section{Basis transformations for pseudoreal scalars}
In a situation with $m$ scalar multiplets
transforming under the group according
to a pseudoreal irrep $D$ we are led to matrices $H \in Sp(2m)$ in the
CP--type transformation (7.8). Basis transformations of the form (3.26) involve
again $Sp(2m)$ matrices. This suggests the question whether for a given
$H \in Sp(2m)$ one can find a $Z \in Sp(2m)$ such that
\beq
H' = Z^\dg H Z^*
\eeq
is as ``simple'' as possible. In eq. (8.15) we have written down such a
normal form. The proof of its existence will be given here.

It is useful to reformulate the problem in terms of antilinear operators
and then take advantage of their properties. Let us define the
antilinear operator
\beq
K : x \ra x^*,
\eeq
the matrix
\beq
J_m = \left( \ba{cc} 0 & {\bf 1}_m \\
                     -{\bf 1}_m & 0 \ea \right)
\eeq
and the antilinear operators
\beq
A \equiv HK, \qquad B \equiv J_m K
\eeq
on ${\bf C}^{2m}$. Then the
symplectic property of $H$ is expressed as
\beq
[A,B] = 0.
\eeq
Further properties of $A$, $B$ are
\beq
A^\dg A = {\bf 1}, \qquad B^\dg B = {\bf 1} \qquad \mbox{and} \qquad
B^2 = - {\bf 1}.
\eeq
This allows the following reformulation of the problem posed above
(see eq. (G.1)):

Given two antilinear operators $A$, $B$ on a unitary space $\V$ with
properties (G.5) and (G.6) can one find canonical forms for $A$, $B$?
The answer is the following.

\paragraph{Theorem:} One can always find an ON basis such that $B$ is
given by $J_m K$ and $A$ by eq. (8.15) in that basis. Since basis
transformations
for matrices corresponding to antilinear operators are performed according
to eq. (G.1) the original problem is solved as well.
\paragraph{Proof:} $A$ is antiunitary and therefore $A^2$ is unitary. Let
$\lambda$ be an eigenvalue of $A^2$ with eigenvector $v$. Then $Av$ has
eigenvalue $\lambda^*$. Therefore the eigenvalues of $A^2$ can be denoted
by
\beq
\lambda_1,\ldots,\lambda_\nu,\lambda_1^*,\ldots,\lambda_\nu^*,1,-1 \qquad
\mbox{with} \qquad |\lambda_i| = 1, \; \lambda_i \neq \pm 1 \;
(i = 1,\ldots,\nu)
\eeq
with degeneracies $m_1,\ldots,m_\nu$
and $m_\pm$, respectively. The decomposition of $\V$ into eigenspaces of
$A^2$ is given by
$$
\V = \bigoplus_{i=1}^\nu (\V(\lambda_i) \oplus \V(\lambda_i^*)) \oplus
\V_+ \oplus \V_-
$$
with
\beq
A \V(\lambda_i) = \V(\lambda_i^*), \qquad
B \V(\lambda_i) = \V(\lambda_i^*), \qquad
A \V_\pm = \V_\pm, \qquad B\V_\pm = \V_\pm .
\eeq
Therefore the proof of the above theorem can be devided into three parts.
\begin{enumerate}
\item[a)] $\V = \V(\lambda) \oplus \V(\lambda^*)$, $\lambda \neq \pm 1$,
$\dim \V(\lambda) = \dim \V(\lambda^*) = m_\lambda$

We choose an ON basis $\{e_1,\ldots,e_{m_\lambda}\}$ in $\V(\lambda)$. Then
$\{ f_i = - B e_i| \: i = 1,\ldots,m_\lambda \}$ defines an
ON basis in $\V(\lambda^*)$.
This allows to write
\beq
A e_i = M_{ji} f_j, \qquad Af_i = \wt M_{ji}e_j .
\eeq
Antiunitarity of $A$ results in
\beq
\delta_{ij} = \langle Ae_i|Ae_j\rangle = M_{ki}^* M_{kj}
\eeq
with an analogous consideration for $\wt M$. Consequently,
$M$, $\wt M$ are unitary $m_\lambda \times m_\lambda$ matrices.
Furthermore, we infer from eq. (G.9) that
\beq
A^2 e_i = \lambda e_i = (\wt M M^*)_{ki} e_k .
\eeq
Exploiting eq. (G.5) we obtain
\beq
AB e_i = - A f_i = - \wt M_{ji} e_j = BAe_i = M_{ji}^* e_j
\eeq
and therefore
\beq
\wt M = - M^*, \qquad M^2 = - \lambda^* {\bf 1}_{m_\lambda} .
\eeq
Now we perform a unitary basis transformation on $\V(\lambda)$
\beq
e'_i = z_{ji} e_j \qquad \mbox{and therefore} \qquad
f'_i \equiv - B e_i' = z_{ji}^* f_j
\eeq
and explore its effect on $M$:
\beq
A e'_i = (z^T M z^*)_{ji} f'_j , \qquad
A f'_i = - (z^T M z^*)_{ji}^* e'_j .
\eeq
Unitarity of $M$ allows to choose a $z$ such that
\beq
z^T M z^* = \mbox{diag }(e^{i\Theta_1},\ldots,e^{i \Theta_{m_\lambda}}).
\eeq
With $- \lambda^* \equiv \mu^2 = e^{2i\Theta_i}$ eq. (G.13),
$e^{i\Theta_i} = \ve_i \mu$, $\ve_i = \pm 1$ and
$\wh \ve = \mbox{diag }(\ve_1,\ldots,\ve_{m_\lambda})$ we finally
see that $A$ and $B$ are represented by
\beq
A \ra \left( \ba{cc} 0 & - \mu^* \wh \ve \\ \mu \wh \ve & 0 \ea \right)K,
\qquad B \ra J_{m_\lambda} K
\eeq
in the basis $\{e'_1,\ldots,e'_{m_\lambda} ,f'_1,\ldots,f'_{m_\lambda} \}$.
\item[b)] $\V = \V_+$

One can easily prove that it is possible to find a vector $e_1 \in \V_+$
with $A e_1 = e_1$ as a consequence of $A^2 = {\bf 1}_{m_+}$. Then
$f_1 \equiv - B e_1$ is orthogonal to $e_1$, $A f_1 = f_1$ and the space
orthogonal to $\{ e_1,f_1\}$ is invariant under $A$, $B$. Therefore we
can repeat the previous steps in $\{ e_1,f_1\}^\perp$ and continue the
process until no dimension is left in $\V_+$. Thus we find that $m_+$
is even and
\beq
A \ra {\bf 1}_{m_+} K, \qquad B \ra J_{m_+/2} K
\eeq
in the basis $\{ e_1,\ldots,e_{m_+/2},f_1,\ldots,f_{m_+/2}\}$.
\item[c)] $\V = \V_-$

Now we have $A^2 = - {\bf 1}_{m_-}$ and $AB$ is a unitary operator with
$(AB)^2 = {\bf 1}_{m_-}$. Therefore we can find an eigenvector $e_1$ of
$AB$ with eigenvalue $\eta_1 = \pm 1$. We define $f_1 \equiv - B e_1$.
Then
\beq
\langle Ae_1|f_1\rangle = \langle e_1|ABe_1\rangle^* = \eta_1
\eeq
and therefore
\beq
A e_1 = \eta_1 f_1 .
\eeq
With the analogous basis as for eq. (G.18) we find that
\beq
A \ra \left( \ba{cc} 0 & - \wh \eta \\ \wh \eta & 0 \ea \right) K,
\qquad \wh \eta = \mbox{diag }(\eta_1,\ldots,\eta_{m_-/2}), \qquad
B \ra J_{m_-/2} K.
\eeq
\end{enumerate}

We have seen that the matrices for $B$ in eqs. (G.17), (G.18) and (G.21) all
have the form $J_m$ eq. (G.3) and $A$ is represented by matrices of the
type (8.15). In general, with all cases a, b, c involved, trivial basis
permutations lead to eq. (8.15). \hfill\ $\Box$

\setcounter{equation}{0}
\addtocounter{zahler}{1}
\section{How to solve the generalized CP conditions}
The methods and strategies to solve eq. (8.10) in the bases (8.11), (8.13),
(8.14) and (8.15) will be discussed in this appendix. Thereby the ranges
of the angles $0 < \Theta$, $\Theta' \leq \pi/2$ and $0 < \Theta_H < \pi$
have to be kept in mind because they will play a crucial r\^ole in the
following.

The first observation is that the matrix $O(\vt)$ (8.12) has eigenvalues
$\exp (\pm i\vt)$:
\beq
O(\vt) \left( \ba{r} 1 \\ \pm i \ea \right) = e^{\pm i\vt}
\left( \ba{r} 1 \\ \pm i \ea \right).
\eeq
If $0 < \vt < \pi$, there are no real eigenvalues. In the cases 1b) and
2a) eigenvectors of such a rotation matrix with eigenvalues $\pm 1$ are
required which do not exist in the ranges of $\Theta$ and $\Theta_H$.
Therefore, we have only the zero solutions.

Case 1c) is trivially solved.

Expressing $A_2$ by $A_1$ in case 2b) gives
\beq
A_2 = \frac{A_1 \cos \Theta_H - O(\Theta) A_1^*}{\sin \Theta_H}.
\eeq
For the second equality gives the condition
\beq
\cos \Theta A_1 = \cos \Theta_H A_1^*.
\eeq
Therefore, $A_1 = A_2 = 0$ for $\Theta \not\in \{ \Theta_H, \pi - \Theta_H\}$.
Taking $\Theta = \Theta_H \neq \pi/2$, we necessarily have
$A_1 \in {\bf R}^2$, $\Theta = \pi - \Theta_H \neq \pi/2$ requires
$A_1$ to be imaginary. For $\Theta = \Theta_H = \pi/2$ eq. (H.3) gives
no restriction and $A_1 \in {\bf C}^2$. $A_2$ is determined by eq. (H.2).

Case 2c) requires
\beq
A_2 = - dO(\Theta)A_1^* \qquad \mbox{and} \qquad
O(2\Theta)A_1 = - d^2 A_1.
\eeq
Thus $- d^2$ must be an eigenvalue of $O(2\Theta)$ of
the form $e^{\pm 2i\Theta}$
or $A_1 = A_2 = 0$ if $d^2 \neq - e^{\pm 2i\Theta}$. For
$\Theta = \pi/2$ and $d = \ve$ the matrix $A_1$ is arbitrary.
On the other hand,
$\Theta < \pi/2$ requires $A_1$ to be an eigenvector of $O(2\Theta)$
from which the solution follows.

For the discussion of the remaining three cases we make the ansatz
$$
A_1 \mbox{ (or }A) = \sum_{r,s=\pm} A_{rs} v_r v_s^T \qquad \mbox{with}
\qquad v_\pm = \left( \ba{r} 1 \\ \mp i \ea \right)
$$
and
\beq
O(\Theta)^T v_\pm = e^{\pm i\Theta} v_\pm, \qquad
v_\pm^T O(\Theta') = e^{\pm i\Theta'} v_\pm^T .
\eeq
Furthermore, after summation we obtain
\beq
A_1 (\mbox{or }A) = \left( \ba{lr}
A_{++} + A_{--} + A_{+-} + A_{-+} & i(-A_{++} + A_{--} + A_{+-} - A_{-+})\\
i(-A_{++} + A_{--} - A_{+-} + A_{-+}) & -A_{++} - A_{--} + A_{+-} +
A_{-+} \ea \right).
\eeq

Application of eq. (H.5) in case 3a) readily gives
\beq
\sum_{r,s} (A_{rs} e^{i(r\Theta + s\Theta')} - A_{-r-s}^*) v_r v_s^T = 0.
\eeq
Evaluation of this equation leads to
\beq
A_{++}(e^{2i(\Theta + \Theta')} -1) = A_{+-} (e^{2i(\Theta - \Theta')} -1)
= 0.
\eeq
The only possibility for $A_{++} \cdot A_{+-} \neq 0$ requires $\Theta =
\Theta' = \pi/2$ and, consequently, $A_{--} = - A_{++}^*$ and
$A_{-+} = A_{+-}^*$. Inserting this into eq. (H.6) gives the first subcase
of the solutions 3a). Going on to $\Theta = \Theta' < \pi/2$ one gets
$A_{++} = A_{--} = 0$ and $A_{-+} = A_{+-}^*$. Inspecting again eq. (H.6)
one sees that $A$ has the same form as before but now its elements are
real. Finally, $\Theta \neq \Theta'$ gives $A_{+-} = 0$ and
$\Theta + \Theta' < \pi$. Therefore also $A_{++} = 0$ and thus
$A = 0$ results.

The last two cases are more complicated. In 3b) we get the following
equation for $A_1$:
\beq
O(\Theta)^T A_1 O(\Theta') + O(\Theta) A_1 O(\Theta')^T =
2 \cos \Theta_H A_1^* .
\eeq
Using the ansatz (H.5) we straightforwardly arrive at
\beq
\sum_{r,s} (A_{rs} \cos(r \Theta + s \Theta') - A_{-r-s}^* \cos \Theta_H)
v_r v_s^T = 0 .
\eeq
Furthermore, $A_2$ is expressed in terms of $A_1$ by
\beq
A_2 = \frac{A_1 \cos \Theta_H - O(\Theta) A_1^* O(\Theta')^T}
{\sin \Theta_H} .
\eeq

Let us first discuss $\Theta_H = \pi/2$. Then $A_{+-} = A_{-+} = 0$
because $\cos (\Theta-\Theta') \neq 0$. On the other hand, it is clear
that $A_{++}$, $A_{--} \neq 0$ is only possible if $\Theta + \Theta' =
\pi/2$. Thus with the help of eq. (H.6) we obtain the $A_1$ of the first
solution of 3b).
Applying eq. (H.11) we get with $\Sigma =
\mbox{diag }(1,-1)$
\beqa
A_2 &=& - O(\Theta) \left( \ba{cr} a & b \\ b & -a \ea \right)^*
O(\Theta')^T = - \Sigma O(-\Theta) \left( \ba{rc} a & b \\ -b & a \ea
\right)^* O(\Theta')^T \no \\
&=& - \Sigma \left( \ba{rc} a & b \\ -b & a \ea \right)^* O(-\Theta-\Theta')
= \left( \ba{cr} a & b \\ b & -a \ea \right)^*
\left( \ba{rc} 0 & 1 \\ -1 & 0 \ea \right) =
\left( \ba{rc} -b & a \\ a & b \ea \right)^*. \no \\
\eeqa

Now we turn to the discussion of $\Theta_H \neq \pi/2$. Our starting
point is
\beq
A_{++} (\cos^2(\Theta + \Theta') - \cos^2\Theta_H) =
A_{+-} (\cos^2(\Theta - \Theta') - \cos^2\Theta_H) = 0.
\eeq
We can only have both $A_{++}$ and $A_{+-}$ non--zero at the same time if
$\cos^2(\Theta + \Theta') = \cos^2(\Theta - \Theta')$. It is easy to
check that in such a case either $\Theta = \pi/2$ or $\Theta' = \pi/2$
(in addition we have $\Theta \neq \Theta'$ to prevent
$\Theta_H = \pi/2$). Let us first
discuss $\Theta = \pi/2$ and $\Theta' < \pi/2$. Then
$\cos(\Theta \pm \Theta') = \mp \sin \Theta'$,
$\cos \Theta_H = \ve \sin \Theta'$ $(\ve = \pm 1)$ and
$A_{--} = - \ve A_{++}^*$, $A_{-+} = \ve A_{+-}^*$.
With eq. (H.6) we recover $A_1$ of the second solution of 3b). Using eq. (H.9)
to replace $A_1 \cos \Theta_H$ in eq. (H.11) we obtain
\beq
A_2 = \frac{1}{2 \sin \Theta_H} (O(\Theta)^T A_1^* O(\Theta') -
O(\Theta) A_1^* O(\Theta')^T).
\eeq
This equation is generally valid in case 3b). Taking $\Theta = \pi/2$ eq.
(H.14) simplifies to
\beq
A_2 = \left( \ba{cr} 0 & -1 \\ 1 & 0 \ea \right) A_1^*.
\eeq
The subcase $\Theta' = \pi/2$, $\Theta < \pi/2$ is dealt with in an
analogous way.

Finally, we consider $\Theta,\Theta' < \pi/2$. Under this condition
$A_{++}$ and $A_{+-}$ cannot be non--zero at the same time. Thus
assuming first $\cos \Theta_H = - \ve \cos(\Theta + \Theta')$ we have
$A_{+-} = A_{-+} = 0$ and $A_{--} = - \ve A_{++}^*$. This gives $A_1$
of the fourth solution of 3b). Now we apply eq. (H.14) and derive
\beqa
A_2 &=& \frac{1}{2 \sin \Theta_H} ( \Sigma O(\Theta) (\Sigma A_1^*)
O(\Theta') - \Sigma O(-\Theta) (\Sigma A_1^*) O(-\Theta')) \no \\
&=& \frac{1}{2 \sin \Theta_H} A_1^* (O(\Theta + \Theta') -
O(-\Theta -\Theta')) = A_1^* \left( \ba{rc} 0 & 1 \\ -1 & 0 \ea \right).
\eeqa
In this way we have obtained the complete fourth solution. For
$\cos \Theta_H = \ve \cos(\Theta - \Theta')$ one can apply the same
procedure.

Finally, let us discuss case 3c). $A_1$ is determined by
\beq
A_1 = -d^2 O(2\Theta)^T A_1 O(2\Theta').
\eeq
Thus with the decomposition (H.5) we get
\beq
A_{rs} (d^2 e^{2i(r\Theta + s\Theta')} +1) = 0 \qquad \forall \;
r,s = \pm .
\eeq
There is exactly one possibility to have an arbitrary $A_1$, namely
$d = i\ve$, $\Theta = \Theta' = \pi/2$. $A_2$ is always computed by
\beq
A_2 = d^* O(\Theta)^T A_1^* O(\Theta')
\eeq
in case 3c).

Next we observe that it is possible to get $A_{++},A_{--} \neq 0$ by
$\Theta + \Theta' = \pi/2$, $d = \ve$. In this case $A_{+-} = A_{-+}
= 0$. The $A_2$ is given by
\beq
A_2 = \ve \Sigma O(\Theta) \Sigma A_1^* O(\Theta') = \ve A_1^*
\left( \ba{rc} 0 & 1 \\ -1 & 0 \ea \right) .
\eeq
There is an analogous case with $A_{++} = A_{--} = 0$ and $A_{+-},
A_{-+} \neq 0$ characterized by
$\Theta = \Theta' \not= \pi / 2$ and $d=i\ve$.

Allowing for $A_{++},A_{+-} \neq 0$ the conditions
\beq
- d^2 e^{2i(\Theta + \Theta')} = - d^2 e^{2i(\Theta - \Theta')} = 1
\eeq
must be fulfilled. In this case $\Theta' = \pi/2$ and $d = \ve
e^{-i\Theta}$. Then eqs. (H.6) and (H.19) determine $A_1$ and $A_2$,
respectively. There are three similar cases characterized by
$A_{++},A_{-+} \neq 0$, $A_{--},A_{+-} \neq 0$ and
$A_{--},A_{-+} \neq 0$.

Finally we are left with cases where only one $A_{rs}$ is non--zero. The
discussion goes along the line presented here.

\setcounter{equation}{0}
\addtocounter{zahler}{1}
\section{On the case $D_\Lambda \sim D_\Lambda \circ \psi_d$ in
$so(2\ell)$}
The non--trivial diagram automorphism of $D_\ell$ $(\ell \geq 5)$
corresponds to exchanging $\alpha_{\ell-1}$ and $\alpha_\ell$ in the
Dynkin diagram (fig.~1), for $D_4$ there are other ones in addition
(see sect. 4). Then $D_\Lambda \sim D_\Lambda \circ \psi_d$ for irreps
with highest weight $\Lambda$ is valid if and only if
$n_{\ell -1} = n_\ell$. Here we want to show that if this is the case
and if $D_\Lambda$ is given in a CP basis (see subsect. 5.2) then the
unitary matrix $W$ establishing the equivalence
\beq
W D_\Lambda W^\dg = D_\Lambda \circ \psi_d
\eeq
can be chosen real and symmetric (see subsect. 9.2 for the use of this
result).

To proceed with the proof we first describe how irreps of the kind
$n_{\ell -1} = n_\ell$ can be built up as tensor products of the
defining fundamental irrep $D_{\Lambda_1}$. For simplicity of notation
we write $D_j \equiv D_{\Lambda_j}$. Then one can show that \cite{sam}
\beq
\wedge^k D_1 \cong D_k \; (k = 1,\ldots,\ell-2) \quad \mbox{and} \quad
\wedge^{\ell-1} D_1 \cong D_{\Lambda'} \quad \mbox{with} \quad
\Lambda' = \Lambda_{\ell -1} + \Lambda_\ell .
\eeq
The symbol $\wedge^k$ denotes the $k$--fold antisymmetric tensor product.
Thus $D_\Lambda$ with $\Lambda = n_1 \Lambda_1 + \ldots + n_{\ell-2}
\Lambda_{\ell-2} + n_\ell (\Lambda_{\ell-1} + \Lambda_\ell)$ can be
obtained as the irrep with highest weight in
\beq
D_1^{\otimes n_1} \otimes \ldots \otimes D_{\ell-2}^{\otimes n_{\ell-2}}
\otimes D_{\Lambda'}^{\otimes n_\ell}
\eeq
where the superscript $\otimes n_j$ indicates the $n_j$--fold tensor
product of $D_j$. $D_\Lambda$ can therefore be constructed from tensor products
of $D_1$.

The r\'esum\'e of this consideration is that if one can show that
$\wt A D_1 \wt A^\dg = D_1 \circ \psi_d$ is achieved with a real
symmetric matrix $\wt A$ for $D_1$ in a CP basis the $W$ for
$D_\Lambda$ with $n_{\ell-1} = n_\ell$ is obtained by suitable tensor
products of $\wt A$ conserving reality and symmetry. Also in these
tensor products $D_\Lambda(e_\alpha)$ will be real and $D_\Lambda(-iH_j)$
imaginary and symmetric if this is so in $D_1$. Then by reversing the
argument leading to eq. (5.13) and by using $D_\Lambda(e_\alpha)^\dg =
- D_\Lambda(e_{-\alpha})$ we easily see that $D_\Lambda$ is given in a
CP basis.

Let us now consider $D_1$ and show first that there is a real and
symmetric matrix $A$ with $A D_1 A^\dg = D_1 \circ \psi_d$ where $D_1$
is given in the natural basis of $so(2\ell)$ with real antisymmetric
matrices. We will closely follow the discussion in app. G of ref.
\cite{corn}. Using the matrices $\{ M_{pq}|\: 1 \leq p < q \leq 2\ell\}$
(see eq. (C.1)) as a basis of $so(2\ell)$ or of its complexification
$D_\ell$ then $\{ h_j = M_{2j-1,2j}|\: j = 1,\ldots,\ell\}$ defines a
basis of the CSA. It is shown in ref. \cite{corn} that with the
functionals $\ve_j$ on the CSA defined by
\beq
\ve_j (h_k) = i \delta_{jk} \qquad (j,k = 1,\ldots,\ell)
\eeq
all basis elements $e_\alpha$ with positive roots are given by
\beqa
e_{\ve_j+\ve_k} &=& \frac{1}{\sqrt{16(\ell - 1)}}
(M_{2j,2k} - i M_{2j,2k-1} - i M_{2j-1,2k} - M_{2j-1,2k-1}) \no \\
e_{\ve_j-\ve_k} &=& \frac{1}{\sqrt{16(\ell - 1)}}
(M_{2j,2k} + i M_{2j,2k-1} - i M_{2j-1,2k} + M_{2j-1,2k-1})
\eeqa
with $1 \leq j < k \leq \ell$. Therefore
\beq
\Delta_+ = \{ \ve_j + \ve_k, \ve_j - \ve_k | \; 1 \leq j < k \leq \ell\}
\eeq
and the simple roots may be defined as
\beq
\alpha_j = \left\{ \ba{ll} \ve_j - \ve_{j+1}, & j = 1,2,\ldots,\ell - 1 \\
\ve_{\ell -1} + \ve_\ell, & j = \ell .\ea\right.
\eeq
Then one can calculate
$$
h_{\alpha_j} = - \frac{i}{4(\ell - 1)} (M_{2j-1,2j} -
M_{2j+1,2j+2}) \qquad (j = 1,2,\ldots,\ell-1)
$$
and
\beq
h_{\alpha_\ell} = - \frac{i}{4(\ell - 1)}
(M_{2\ell-3,2\ell-2} + M_{2\ell -1,2\ell}).
\eeq
With
\beq
t \equiv \left( \ba{rl} 0 & 1 \\ -1 & 0 \ea \right)
\eeq
we have
\beqa
h_{\alpha_\ell} &=& - \frac{i}{4(\ell - 1)} \mbox{diag }
(0_{2\ell - 4},t,t), \no \\
h_{\alpha_{\ell -1}} &=& - \frac{i}{4(\ell - 1)} \mbox{diag }
(0_{2\ell - 4},t,-t), \no \\
h_{\alpha_{\ell -2}} &=& - \frac{i}{4(\ell - 1)} \mbox{diag }
(0_{2\ell - 6},t,-t,0_2), \mbox{ etc.}
\eeqa
In eq. (I.10) the symbol diag means arranging the matrices along the
diagonal and $0_m$ denotes the $m \times m$ zero matrix. The basis
elements corresponding to simple roots are represented by
$$
e_{\alpha_\ell} = \frac{1}{\sqrt{16 (\ell-1)}} \mbox{ diag }
(0_{2\ell - 4}, F')
$$
with
\beq
F' = \left( \ba{crrr} 0 & 0 & -1 & -i \\ 0 & 0 & -i & 1 \\
1 & i & 0 & 0 \\ i & -1 & 0 & 0 \ea \right)
\eeq
and, for $j \not= \ell$,
$$
e_{\alpha_j} = \frac{1}{\sqrt{16 (\ell-1)}} \mbox{ diag }
(0_{2(j - 1)}, F, 0_{2(\ell-1-j)})
$$
with
\beq
F = \left( \ba{rrcr} 0 & 0 & 1 & -i \\ 0 & 0 & i & 1 \\
-1 & -i & 0 & 0 \\ i & -1 & 0 & 0 \ea \right).
\eeq

The diagram automorphism $\psi_d$ is already uniquely determined by
its action on $h_{\alpha_j}$ and $e_{\alpha_j}$:
\beq
\psi_d(h_{\alpha_j}) = \left\{ \ba{lll}
h_{\alpha_j}, & j \leq \ell - 2 & \\
h_{\alpha_\ell}, & j = \ell - 1, & \psi_d(e_{\alpha_j}) \mbox{ analogously}
\\
h_{\alpha_{\ell -1}}, & j = \ell  \ea
\right.
\eeq
Then one quickly confirms with eqs. (I.10), (I.11) and (I.12)
that eq. (I.13) is reproduced by
\beq
\psi_d(X) = A X A \qquad \mbox{with} \qquad
A = \mbox{diag } ({\bf 1}_{2\ell-2},-1,1)
\eeq
for $X \in so(2\ell)$ or its complexification $D_\ell$.
Since $D_1(X) = X$ we have obtained
\beq
(D_1 \circ \psi_d)(X) = \psi_d(X) = AXA = A D_1(X) A^\dg
\eeq
with $A$ real and symmetric.

In the last step we have to switch to a CP basis by conserving the above
properties of $A$. Defining
\beq
T = \mbox{diag }(s,\ldots,s) \qquad \mbox{with} \qquad
s = \frac{1}{\sqrt{2}} \left( \ba{cr} 1 & 1 \\ i & -i \ea \right)
\eeq
we get
\beq
T^\dg h_j T = i \mbox{ diag }(0_{2(j-1)}, \tau_3,0_{2(\ell - j)})
\eeq
with the Pauli matrix $\tau_3$. Furthermore, one readily verifies that
\beq
(T^\dg M_{pq} T)_{rs} = T_{pr}^* T_{qs} - T_{qr}^* T_{ps}.
\eeq
This allows to check easily that
$$
(T^\dg M_{2j,2k} T)_{rs} \qquad \mbox{and} \qquad
(T^\dg M_{2j-1,2k-1} T)_{rs} \qquad \mbox{are real } \forall \; r,s
$$
and
$$
(T^\dg M_{2j,2k-1} T)_{rs} \qquad \mbox{and} \qquad
(T^\dg M_{2j-1,2k} T)_{rs} \qquad \mbox{are imaginary } \forall \; r,s.
$$
Therefore, looking at eq. (I.5) we immediately see that we have
\beq
T^\dg e_{\ve_j \pm \ve_k} T \qquad \mbox{real } \forall \; j,k =
1,\ldots,\ell \mbox{ with } j < k .
\eeq
Thus the matrix $T$ achieves a basis transformation into a CP basis.
Finally, with the Pauli matrix $\tau_1$ we obtain
\beq
\wt A = T^\dg A T = \mbox{diag }({\bf 1}_{2\ell -2}, - \tau_1)
\eeq
which is real and symmetric. This proves the claim made in the beginning
of this appendix.

\newpage
\begin{table}[h]
\addcontentsline{toc}{section}{Tables}
$$
\begin{tabular}{|c|c|c|c|} \hline
$\wt\cL$ & $\cL_c$ & Aut $(\cL_c)/\mbox{Int }(\cL_c)$ & $\psi^\triangle$
\\ \hline
$A_1$ & $su(2)$ & $\{e\}$ & inner \\
$A_\ell$ $(\ell \geq 2)$ & $su(\ell+1)$ & ${\bf Z}_2$ & outer \\
$B_\ell$ $(\ell \geq 2)$ & $so(2\ell +1)$ & $\{e\}$ & inner \\
$C_\ell$ $(\ell \geq 3)$ & $sp(2\ell)$ & $\{e\}$ & inner \\
$D_4$ & $so(8)$ & $S_3$ & inner \\
$D_\ell$ $(\ell = 5,7,9,\ldots)$ & $so(2\ell)$ & ${\bf Z}_2$ & outer \\
$D_\ell$ $(\ell = 6,8,10,\ldots)$ & $so(2\ell)$ & ${\bf Z}_2$ & inner \\
$E_6$ & $cE_6$ & ${\bf Z}_2$ & outer \\
$E_7$ & $cE_7$ & $\{e\}$ & inner \\
$E_8$ & $cE_8$ & $\{e\}$ & inner \\
$F_4$ & $cF_4$ & $\{e\}$ & inner \\
$G_2$ & $cG_2$ & $\{e\}$ & inner \\ \hline
\end{tabular}
$$
\caption{The complete list of simple complex Lie algebras with their
compact real forms} and the structure of their automorphism groups. We
have also indicated in which algebras the contragredient automorphism
$\psi^\triangle$ is inner $(\psi^\triangle \in \mbox{Int }(\cL_c))$ or outer
$(\psi^\triangle \notin \mbox{Int }(\cL_c))$. $S_3$ denotes the group of
permutations of $\{1,2,3\}$ and $sp(2\ell)$ is the Lie algebra
corresponding to the compact symplectic group
$Sp(2\ell) = \{U \in U(2\ell)|\: U^T J U = J\}$ with
$$
J = \left( \ba{cc} 0 & {\bf 1}_\ell \\ -{\bf 1}_\ell & 0 \ea \right).
$$
\end{table}
\begin{table}[h]
\beqan
so(3) &\cong& su(2) \\
sp(2) &\cong& su(2) \\
sp(4) &\cong& so(5) \\
so(2) &\cong& u(1) \\
so(4) &\cong& su(2) \oplus su(2) \\
so(6) &\cong& su(4)
\eeqan
\caption{Complete list of isomorphisms within the four series of classical
compact algebras $su(N)$ $(N \geq 2)$, $so(N)$ $(N \geq 2)$ and
$sp(N)$ $(N = 2,4,6,\ldots)$. Of all these algebras only $so(2)$ and
$so(4)$ are not simple. The last two isomorphisms are proved in apps.
C and D, respectively.}
\end{table}

\setlength{\unitlength}{0.6mm}
\newsavebox{\dynloop}
\savebox{\dynloop}(0,0){\put(2,0){\circle{4}}
\put(4,0){\line(1,0){16}}}
\newsavebox{\dynloopf}
\savebox{\dynloopf}(0,0){\put(2,0){\circle*{4}}
\put(4,0){\line(1,0){16}}}
\newpage
\begin{figure}[h]
\addcontentsline{toc}{section}{Figure}
\begin{picture}(145,360)(-40,0)
\put(30,320){\begin{picture}(120,40)
\multiput(40,20)(20,0){2}{\usebox{\dynloop}}
\put(82,20){\circle{4}}
\put(84,20){\line(1,0){4}}
\put(93,20){\makebox(0,0){$\cdots$}}
\put(102,20){\line(-1,0){4}}
\put(102,20){\usebox{\dynloop}}
\put(124,20){\circle{4}}
\put(-30,20){\makebox(0,0)[l]{${\bf A_\ell} \; (\ell=1,2,3,\ldots)$}}
\put(42,17){\makebox(0,0)[t]{$\alpha_1$}}
\put(62,17){\makebox(0,0)[t]{$\alpha_2$}}
\put(82,17){\makebox(0,0)[t]{$\alpha_3$}}
\put(104,17){\makebox(0,0)[t]{$\alpha_{\ell-1}$}}
\put(124,17){\makebox(0,0)[t]{$\alpha_\ell$}}
\end{picture}}
\put(30,280){\begin{picture}(120,40)
\multiput(40,20)(20,0){2}{\usebox{\dynloop}}
\put(82,20){\circle{4}}
\put(84,20){\line(1,0){4}}
\put(93,20){\makebox(0,0){$\cdots$}}
\put(102,20){\line(-1,0){4}}
\put(104,20){\circle{4}}
\put(104,22){\line(1,0){20}}
\put(104,18){\line(1,0){20}}
\put(124,20){\circle*{4}}
\put(-30,20){\makebox(0,0)[l]{${\bf B_\ell} \; (\ell=2,3,4,\ldots)$}}
\put(42,17){\makebox(0,0)[t]{$\alpha_1$}}
\put(62,17){\makebox(0,0)[t]{$\alpha_2$}}
\put(82,17){\makebox(0,0)[t]{$\alpha_3$}}
\put(104,17){\makebox(0,0)[t]{$\alpha_{\ell-1}$}}
\put(124,17){\makebox(0,0)[t]{$\alpha_\ell$}}
\end{picture}}
\put(30,240){\begin{picture}(120,40)
\multiput(40,20)(20,0){2}{\usebox{\dynloopf}}
\put(82,20){\circle*{4}}
\put(84,20){\line(1,0){4}}
\put(93,20){\makebox(0,0){$\cdots$}}
\put(102,20){\line(-1,0){4}}
\put(104,20){\circle*{4}}
\put(104,22){\line(1,0){20}}
\put(104,18){\line(1,0){20}}
\put(124,20){\circle{4}}
\put(-30,20){\makebox(0,0)[l]{${\bf C_\ell} \; (\ell=3,4,5, \ldots)$}}
\put(42,17){\makebox(0,0)[t]{$\alpha_1$}}
\put(62,17){\makebox(0,0)[t]{$\alpha_2$}}
\put(82,17){\makebox(0,0)[t]{$\alpha_3$}}
\put(104,17){\makebox(0,0)[t]{$\alpha_{\ell-1}$}}
\put(124,17){\makebox(0,0)[t]{$\alpha_\ell$}}
\end{picture}}
\put(30,200){\begin{picture}(120,40)
\multiput(40,20)(20,0){2}{\usebox{\dynloop}}
\put(82,20){\circle{4}}
\put(84,20){\line(1,0){4}}
\put(93,20){\makebox(0,0){$\cdots$}}
\put(102,20){\line(-1,0){4}}
\put(104,20){\circle{4}}
\put(105.029,21.715){\line(3,5){8.1}}
\put(105.029,18.285){\line(3,-5){8.1}}
\put(114.29,37.15){\circle{4}}
\put(114.29,2.85){\circle{4}}
\put(-30,20){\makebox(0,0)[l]{${\bf D_\ell} \; (\ell=4,5,6,\ldots)$}}
\put(42,17){\makebox(0,0)[t]{$\alpha_1$}}
\put(62,17){\makebox(0,0)[t]{$\alpha_2$}}
\put(82,17){\makebox(0,0)[t]{$\alpha_3$}}
\put(104,17){\makebox(0,0)[tr]{$\alpha_{\ell-2}$}}
\put(117.29,37.15){\makebox(0,0)[l]{$\alpha_\ell$}}
\put(117.29,2.85){\makebox(0,0)[l]{$\alpha_{\ell-1}$}}
\end{picture}}
\put(30,160){\begin{picture}(120,40)
\multiput(40,20)(20,0){4}{\usebox{\dynloop}}
\put(122,20){\circle{4}}
\put(82,22){\line(0,1){16}}
\put(82,40){\circle{4}}
\put(-30,20){\makebox(0,0)[l]{${\bf E}_6$}}
\put(42,17){\makebox(0,0)[t]{$\alpha_1$}}
\put(62,17){\makebox(0,0)[t]{$\alpha_2$}}
\put(82,17){\makebox(0,0)[t]{$\alpha_3$}}
\put(102,17){\makebox(0,0)[t]{$\alpha_4$}}
\put(122,17){\makebox(0,0)[t]{$\alpha_5$}}
\put(85,40){\makebox(0,0)[l]{$\alpha_6$}}
\end{picture}}
\put(30,120){\begin{picture}(120,40)
\multiput(40,20)(20,0){5}{\usebox{\dynloop}}
\put(142,20){\circle{4}}
\put(82,22){\line(0,1){16}}
\put(82,40){\circle{4}}
\put(-30,20){\makebox(0,0)[l]{${\bf E}_7$}}
\put(42,17){\makebox(0,0)[t]{$\alpha_1$}}
\put(62,17){\makebox(0,0)[t]{$\alpha_2$}}
\put(82,17){\makebox(0,0)[t]{$\alpha_3$}}
\put(102,17){\makebox(0,0)[t]{$\alpha_4$}}
\put(122,17){\makebox(0,0)[t]{$\alpha_5$}}
\put(142,17){\makebox(0,0)[t]{$\alpha_6$}}
\put(85,40){\makebox(0,0)[l]{$\alpha_7$}}
\end{picture}}
\put(30,80){\begin{picture}(120,40)
\multiput(40,20)(20,0){6}{\usebox{\dynloop}}
\put(162,20){\circle{4}}
\put(82,22){\line(0,1){16}}
\put(82,40){\circle{4}}
\put(-30,20){\makebox(0,0)[l]{${\bf E}_8$}}
\put(42,17){\makebox(0,0)[t]{$\alpha_1$}}
\put(62,17){\makebox(0,0)[t]{$\alpha_2$}}
\put(82,17){\makebox(0,0)[t]{$\alpha_3$}}
\put(102,17){\makebox(0,0)[t]{$\alpha_4$}}
\put(122,17){\makebox(0,0)[t]{$\alpha_5$}}
\put(142,17){\makebox(0,0)[t]{$\alpha_6$}}
\put(162,17){\makebox(0,0)[t]{$\alpha_7$}}
\put(85,40){\makebox(0,0)[l]{$\alpha_8$}}
\end{picture}}
\put(30,40){\begin{picture}(120,40)
\put(40,20){\usebox{\dynloop}}
\put(62,20){\circle{4}}
\put(62,22){\line(1,0){20}}
\put(62,18){\line(1,0){20}}
\put(82,20){\circle*{4}}
\put(84,20){\line(1,0){16}}
\put(102,20){\circle*{4}}
\put(-30,20){\makebox(0,0)[l]{${\bf F}_4$}}
\put(42,17){\makebox(0,0)[t]{$\alpha_1$}}
\put(62,17){\makebox(0,0)[t]{$\alpha_2$}}
\put(82,17){\makebox(0,0)[t]{$\alpha_3$}}
\put(102,17){\makebox(0,0)[t]{$\alpha_4$}}
\end{picture}}
\put(30,0){\begin{picture}(120,40)
\put(40,20){\usebox{\dynloop}}
\put(62,20){\circle*{4}}
\put(42,22){\line(1,0){20}}
\put(42,18){\line(1,0){20}}
\put(-30,20){\makebox(0,0)[l]{${\bf G}_2$}}
\put(42,17){\makebox(0,0)[t]{$\alpha_1$}}
\put(62,17){\makebox(0,0)[t]{$\alpha_2$}}
\end{picture}}
\end{picture}
\caption{The Dynkin diagrams of all simple complex Lie
algebras.}
\end{figure}
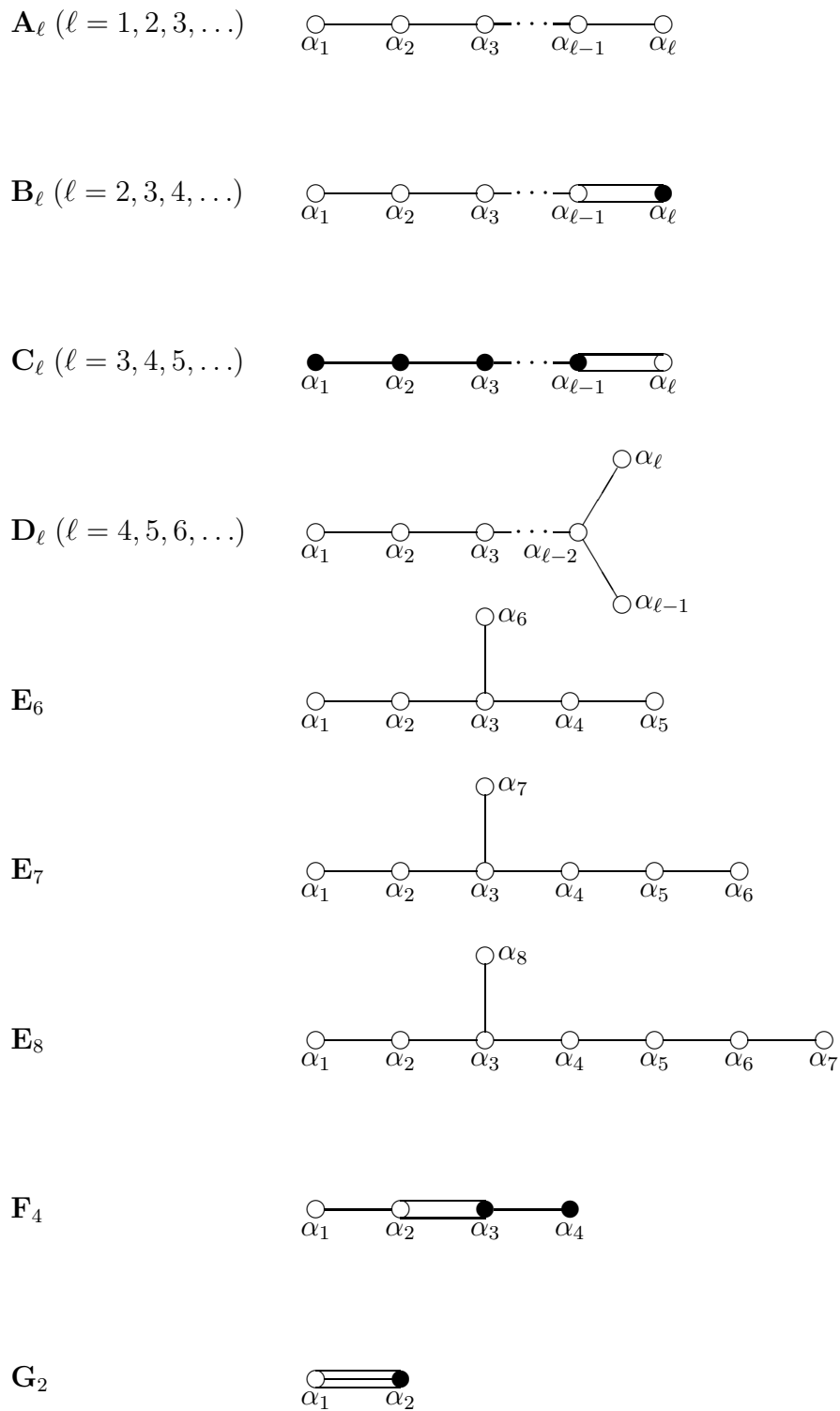


\newpage
\addcontentsline{toc}{section}{References}
\begin{thebibliography}{99}
\bibitem{lee} T.D. Lee and C.N. Yang, {\em Phys. Rev.} {\bf 104} (1956) 254.
\bibitem{amb} E. Ambler, R.W. Hayward, D.D. Hoppes, R.R. Hudson and
C.S. Wu, {\em Phys. Rev.} {\bf 105} (1957) 1413.
\bibitem{chris} J.H. Christenson, J.W. Cronin, V.L. Fitch and R. Turlay,
{\em Phys. Rev. Lett.} {\bf 13} (1964) 138.
\bibitem{SM} S.L. Glashow, {\em Nucl. Phys.} {\bf 22} (1961) 579;\\
S. Weinberg, {\em Phys. Rev. Lett.} {\bf 19} (1967) 1264;\\
A. Salam, in ``Proc. of 8th Nobel Symposium'', Aspen\"asg\aa rden,
1968, ed. N. Svartholm (Almqvist and Wiksell, Stockholm, 1968).
\bibitem{KM} M. Kobayashi and T. Maskawa, {\em Prog. Theor. Phys.}
{\bf 49} (1973) 652.
\bibitem{GUT} J.C. Pati and A. Salam, {\em Phys. Rev. D} {\bf 8} (1973)
1240;\\
H. Georgi and S.L. Glashow, {\em Phys. Rev. Lett.} {\bf 32} (1974) 438.
\bibitem{sla} R. Slansky, {\em Phys. Rep.} {\bf 79} (1981) 1.
\bibitem{smo} N.V. Smolyakov, {\em Theor. and Math. Phys.} {\bf 50}
(1982) 225.
\bibitem{bin} P. Binetruy, M.K. Gaillard and Z. Kunszt, {\em Nucl. Phys. B}
{\bf 144} (1978) 141.
\bibitem{eck81} G. Ecker, W. Grimus and W. Konetschny, {\em Nucl. Phys. B}
{\bf 191} (1981) 465.
\bibitem{comp} L. O'Raifeartaigh, {\em Phys. Rev. B} {\bf 139} (1965)
1050; \\
S. Coleman and J. Mandula, {\em Phys. Rev.} {\bf 159} (1967) 1251;\\
W. Grimus, {\em J. Phys. A} {\bf 26} (1993) L435.
\bibitem{pati} J.C. Pati and A. Salam, {\em Phys. Rev. D} {\bf 10} (1975)
275.
\bibitem{LR} R.N. Mohapatra and J.C. Pati, {\em Phys. Rev. D} {\bf 11} (1975)
566 and 2558.
\bibitem{CPT} J. Schwinger, {\em Phys. Rev.} {\bf 82} (1951) 914;\\
G. L\"uders, {\em Dansk Mat. Fys. Medd.} {\bf 28} (1954) 5; \\
W. Pauli, in ``Niels Bohr and the Development of Physics'', ed. W. Pauli
(Pergamon Press, New York, 1955).
\bibitem{bjo} J.D. Bjorken and S.D. Drell, ``Relativistic Quantum Fields''
(McGraw--Hill, New York, 1965).
\bibitem{corn} J.F. Cornwell, ``Group Theory in Physics'' (Academic Press,
London, 1984).
\bibitem{gri93} W. Grimus, in ``Proc. of 4th Hellenic School on Elementary
Particle Physics'', Corfu, 1992, eds. E.N. Gazis et al. (National
Technical University, Zografou Campus, Athens).
\bibitem{gell} M. Gell--Mann, {\em Phys. Rev.} {\bf 125} (1962) 1067.
\bibitem{SO(10)} H. Georgi, in ``Proc. of Particles and Fields'', A.I.P.,
ed. C.E. Carlson (New York, 1975); \\
H. Fritzsch and P. Minkowski, {\em Ann. Phys.} {\bf 93} (1975) 193.
\bibitem{geo} H. Georgi, ``Lie Algebras in Particle Physics''
(Benjamin/Cummings, Menlo Park, 1982).
\bibitem{ross} G.G. Ross, ``Grand Unified Theories'' (Benjamin/Cummings,
Menlo Park, 1985).
\bibitem{moh} R.N. Mohapatra, ``Unification and Supersymmetry'' (Springer,
New York, 1986).
\bibitem{abers} E.S. Abers and B. Lee, {\em Phys. Rep. C} {\bf 9} (1973) 1.
\bibitem{sam} H. Samelson, ``Notes on Lie Algebras'' (Springer, New York,
1990).
\bibitem{jac} N. Jacobson, ``Lie Algebras'' (Interscience Publishers,
New York, 1962).
\bibitem{var} V.S. Varadarajan, ``Lie Groups, Lie Algebras, and Their
Representations'' (Springer, New York, 1984).
\bibitem{eck84} G. Ecker, W. Grimus and H. Neufeld, {\em Nucl. Phys. B}
{\bf 247} (1984) 70.
\bibitem{ber} J. Bernab\'eu, G.C. Branco and M. Gronau, {\em Phys. Lett.
B} {\bf 169} (1986) 243.
\bibitem{eck87} G. Ecker, W. Grimus and H. Neufeld, {\em J. Phys. A}
{\bf 20} (1987) L807.
\bibitem{shaw} R. Shaw, ``Linear Algebra and Group Representations''
(Academic Press, London, 1982).
\bibitem{ha} M. Hamermesh, ``Group Theory and its Application to Physical
Problems''\\ (Addison--Wesley, Reading, MA, 1962).
\bibitem{sen1} G. Senjanovi\'c and R.N. Mohapatra, {\em Phys. Rev. D}
{\bf 12} (1975) 1502.
\bibitem{sen} R.N. Mohapatra and G. Senjanovi\'c, {\em Phys. Rev. Lett.}
{\bf 44} (1980) 912; {\em Phys. Rev. D} {\bf 23} (1981) 165.
\bibitem{urb} W. Grimus and H. Urbantke, ``Antilinear Operators and Group
Representations'', preprint UWThPh-1994-30.
\bibitem{sla1} R. Slansky, in ``Proc. of First Workshop on Grand
Unification'', eds. P.H. Frampton, S.L. Glashow and A. Yildiz (Math. Sci.
Press, Brookline, 1980).
\bibitem{lee1} T.D. Lee, {\em Phys. Rev. D} {\bf 8} (1973) 1226.
\bibitem{fa} P. Fayet, {\em Nucl. Phys. B} {\bf 78} (1974) 14.
\bibitem{eck83} G. Ecker, W. Grimus and H. Neufeld, preprint
TH.3780--CERN (1983) (this is a longer version of ref. [27]).
\bibitem{gri88} W. Grimus, {\em Fortschr. Phys.} {\bf 36} (1988) 201.
\bibitem{bra} G.C. Branco, J.--M. G\'erard and W. Grimus, {\em Phys. Lett.
B} {\bf 136} (1984) 383.
\bibitem{wy} B.G. Wybourne, ``Classical Groups for Physicists''
(Wiley--Interscience, New York, 1974).
\bibitem{cahn} R.N. Cahn, ``Semi--Simple Lie Algebras and Their
Representations'' (Ben\-jamin/ Cummings, Menlo Park, 1984).
\end{thebibliography}
\end{document}